\documentclass[10pt,notitlepage,twoside]{report}
\usepackage{graphicx}
\usepackage{amsmath}
\usepackage{psfrag}
\usepackage[T1]{fontenc}
\usepackage[cp1250]{inputenc}

\newcommand{\bq}{\begin{equation}}
\newcommand{\eq}{\end{equation}}
\newcommand{\ba}{\begin{eqnarray}}
\newcommand{\ea}{\end{eqnarray}}
\newcommand{\dd}{{\rm d}}
\newcommand{\p}{\mathbf{p}}
\newcommand{\pp}{\mathbf{P}}
\newcommand{\A}{\mathbf{A}}

\newcommand{\tr}{\tau}
\newcommand{\eps}{\epsilon}
\newcommand{\Tr}{\mbox{Tr}}

\begin{document}
\pagenumbering{roman}

\thispagestyle{empty}
\setcounter{page}{0}
\begin{center}
\mbox{}

\vspace{4cm}
{\huge \bf Statistical mechanics of complex networks} \\
\vspace{1cm}
{\large \bf Bartłomiej Wacław} \\
\vspace{1cm}
\today \\

\vspace{2cm}
{\bf Abstract}

\end{center}

\noindent
The science of complex networks is a new interdisciplinary branch of science which has arisen recently on the interface of physics, biology, social and computer sciences, and others. Its main goal is to discover general laws governing the creation and growth as well as processes taking place on networks, like e.g. the Internet, transportation or neural networks. It turned out that most real-world networks cannot be simply reduced to a compound of some individual components. Fortunately, the statistical mechanics, being one of pillars of modern physics, provides us with a very powerful set of tools and methods for describing and understanding these systems. In this thesis, we would like to present a consistent approach to complex networks based on statistical mechanics, with the central role played by the concept of statistical ensemble of networks. We show how to construct such a theory and present some practical problems where it can be applied. Among them, we pay attention to the problem of finite-size corrections and the dynamics of a simple model of mass transport on networks.
In particular, we calculate the cutoff function for finite growing networks in the generalized Barab\'{a}si-Albert model and show how the maximal degree observed in such a network depends on its size and on the exponent $\gamma$ in the power-law degree distribution. We show that this structural cutoff is gaussian only for $\gamma=3$, and is never exponential for $2<\gamma<4$. In parallel, we present numerical results for equilibrated networks, that is networks obtained in a sort of ``thermalization'' (randomization) process. We discuss also similarities and differences between growing and equilibrated networks.
Concerning dynamics on networks, we study so called zero-range process being a system of particles hopping between sites of a network. We discuss known results for its static and dynamical properties on homogeneous networks, where all nodes have the same degrees, and derive new predictions for inhomogeneous graphs. We show that when the density of particles passes a certain threshold, a condensate emerges at the most inhomogeneous node. Its life-time grows exponentially with the size of the system, contrary to homogeneous graphs where it grows only like a power law. We find also a special type of an inhomogeneous network, for which the average distribution of balls is scale-free at the critical point.

\newpage

\thispagestyle{empty}

\mbox{}

\vspace{4cm}
\noindent
{\bf \large Preface}\\

\noindent
This is a slightly modified text of the PhD thesis written as a part of the author's PhD studies in theoretical physics under the supervision of Prof. Z. Burda, and defended on April 5th, 2007, at the Faculty of Physics, Astronomy and Applied Computer Science, Jagellonian University in Cracow, Poland. 
In comparison to the officially accepted doctoral dissertation, available from Jagellonian University Library, this version has been changed according to some critical remarks of referees and other people who read it before and after the defense. In particular, some typos and errors in formulas have been corrected, and some references added or updated. There are also some minor changes. For instance, in the original text we used the word ``homogeneous'' to refer
to a certain type of networks. We replaced it here by the word
``equilibrated'' which, as we realized, better relfects
the structure of these networks and does not lead to a confusion with another, commonly accepted meaning.
The contents is, however, almost unchanged, so is the order of all chapters, sections etc.

\newpage

\tableofcontents

\newpage
\noindent
{\huge \bf List of publications}\\

\vspace{1cm}

\noindent
Some results presented in this thesis have been published in the following papers:\\

\begin{itemize}
\item L. Bogacz, Z. Burda, and B. Waclaw, {\em Homogeneous complex networks}, Phys.~A {\bf 366} (2006), 587; {\tt arXiv:cond-mat/0502124}.
\item P. Bialas, Z. Burda, and B. Waclaw, {\em Causal and homogeneous networks}, AIP Conf. Proc. {\bf 776} (2005), 14; {\tt arXiv:cond-mat/0503548}.
\item L. Bogacz, Z. Burda, W. Janke, and B. Waclaw, {\em A program generating random graphs with given weights}, Comp. Phys. Comm. {\bf 173} (2005), 162; {\tt arXiv:cond-mat/0506330}.
\item B. Waclaw and I. M. Sokolov, {\em Finite size effects is Barab\'{a}si-Albert growing networks}, accepted for publication in Phys. Rev. E; {\tt arXiv:cond-mat/0609728}.
\item L. Bogacz, Z. Burda, W. Janke, and B. Waclaw, {\em Balls-in-boxes condensation on networks}, to appear in the ``Chaos'' focus issue on ``Optimization in Networks''; {\tt arXiv:cond-mat/0701553}.
\item B. Waclaw, L. Bogacz, Z. Burda, and W. Janke, {\em Condensation in zero-range processes on inhomogeneous networks}, submitted to Phys. Rev. E; {\tt arXiv:cond-mat/0703243}.

\end{itemize}

\noindent
Other papers to which author has contributed during his PhD studies:\\
\begin{itemize}
\item Z. Burda, A. G\"{o}rlich, B. Waclaw, {\em Spectral properties of empirical covariance matrices for data with power-law tails}, Phys. Rev. E {\bf 74} (2006), 041129; {\tt arXiv:physics/0603186}.
\item Z. Burda, J. Jurkiewicz, B. Waclaw, {\em Eigenvalue density of empirical covariance matrix for correlated samples}, Acta Phys. Pol. B {\bf 36} (2005), 2641; {\tt arXiv:cond-mat/0508451}.
\item Z. Burda, A. G\"{o}rlich, J. Jurkiewicz, B. Waclaw, {\em Correlated Wishart matrices and critical horizons}, Eur. Phys. J. B {\bf 49} (2006), 319; {\tt arXiv:cond-mat/0508341}.
\item Z. Burda, J. Jurkiewicz, B. Waclaw, {\em Spectral moments of correlated Wishart matrices}, Phys. Rev. E {\bf 71} (2005), 026111; {\tt arXiv:cond-mat/0405263}.
\end{itemize}

\newpage
\noindent
{\huge \bf Acknowledgments}\\

\vspace{1cm}
\noindent
Many people helped me during my PhD studies. Here I would like to thank them. I hope that they will be able to decipher correctly enigmatic sentences written below.

\vspace{5mm}
\noindent
I would like to thank my supervisor Prof. Z. Burda, who has been and still is doing much more for me than a supervisor should.
%I owe him endless stimulating discussions

\vspace{5mm}
\noindent
The work presented in this thesis was partially done during my several stays abroad. 
I am indebted to my hosts Prof. I. M. Sokolov from the Department of Physics, Humbold-Universit\"{a}t zu Berlin and Prof. W. Janke from the Institute of Theoretical Physics, Universit\"{a}t Leipzig, for a warm welcome and many useful discussions. 

\vspace{5mm}
\noindent
I would like to thank Prof. S. N. Dorogovtsev, Prof. J. F. F. Mendes, Dr. A. Goltsev and the University of Aveiro for hospitality. This was an unforgettable stay. I am especially grateful to Prof. Dorogovtsev for many interesting conversations.

\vspace{5mm}
\noindent
I am grateful to Dr. R. Xulvi-Brunet for the time we spent talking about dynamically rewired complex networks.

\vspace{5mm}
\noindent
I thank Dr. L. Bogacz for his hospitality during my first stay in Leipzig, for many things he taught me about programming and for long discussions about physics and computer science.

\vspace{5mm}
\noindent
I am indebted to Dr. P. Bialas for pointing me out the importance of correlations in networks.

\vspace{5mm}
\noindent
I would like to thank the German Academic Exchange Service (DAAD) for a research fellowship at Universit\"{a}t Leipzig, Germany, and F. Kogutowska Foundation for Jagellonian University and EU grant MTKD-CT-2004-517186 (COCOS) for supporting my visit to Humbold-Universit\"{a}t zu Berlin. 

\vspace{5mm}
\noindent
Last but not least I thank my wife Justyna who had always deep understanding for the way of life I have chosen.

\newpage
\pagenumbering{arabic}

\chapter{Introduction}

\section{Science of complex networks}

We live in the world dominated by networks, either in technological or social sense. Who can now imagine our existence without electric power transmission lines, organized in a kind of network with nodes being power plants or transformer substations, or without the Internet, the most powerful medium of the 21th century? In fact, networks surround us. We ourselves are also a part of a huge network of interpersonal contacts, where ideas or diseases can spread. Highways, subways, air traffic as well as scientific collaborations or sexual contacts' networks are just a few further examples. Some of them are real physical networks (the Internet), some of them describe non-physical relations between objects (the World Wide Web), being defined in some abstract space. During the last decade, networks became a subject of interest of scientists who want to discover general laws governing their formation and growth. It is a great success that despite an enormous variety of networks and essential differences in their physical structure, it is possible to find such laws, applying to the majority of real-world networks. The most important observation is that these networks are complex, what means that their properties cannot be simply reduced to a compound of individual components.
Instead, a new quality emerges when many objects are linked together forming a network. Therefore, the reductionism - a powerful tool of physics - fails when one tries to examine complex networks. Fortunately, one branch of physics, namely the statistical mechanics, provides us with an ideal set of tools, methods and ideas for describing and understanding these sophisticated systems. The application of these ideas to complex networks uncovers unexpected connections to other areas of physics, as for instance to percolation or Bose-Einstein condensation.

In recent years, many properties of real-world networks have been described. Many models have been proposed. As a result, a new inter-disciplinary science, the science of complex networks, has emerged on the interface of physics, 
chemistry, biology, computer science and other disciplines.
It is not the intention of author to review all important results of the science of complex network in this short introductory chapter. For a review, we refer the reader to excellent papers \cite{ref:barab,ref:cn2,ref:mejn2}, or to a newer one \cite{ref:physrep2006} presenting also some recent developments in the field. However, to give a better comprehension of results presented in this thesis and to make it self-contained we shall describe some ideas which are especially important for our purposes. So in the next section of this chapter we shall discuss some basic concepts of graph theory, which provides a natural framework for description of networks. Then in the subsequent section we shall recall the empirical findings on real-world networks which have motivated the outbreak of interest in the field and then the rapid development of the science of complex networks in recent years. The explanation of the observed real-world properties is still the main objective of many scientific publications. In the last section of this chapter we shall briefly discuss the aim and the scope of the thesis.

\begin{figure}
\center
\includegraphics[width=13cm]{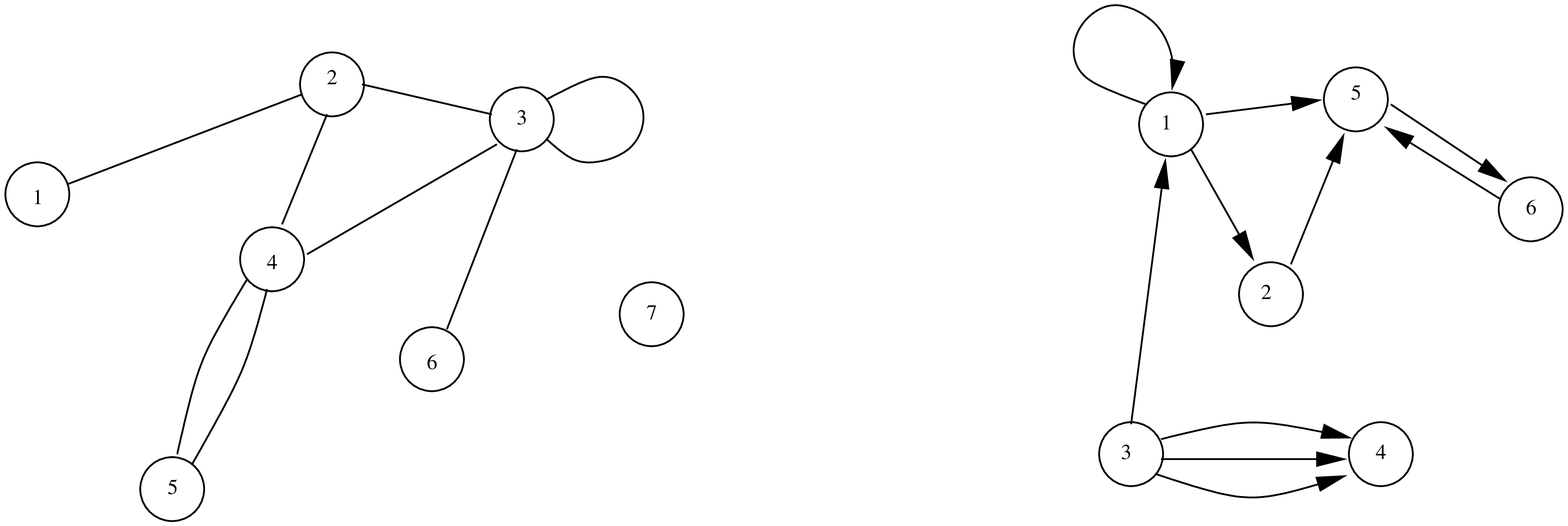}
\caption{\label{fig:some-gr}Left: an example of an undirected pseudograph with seven nodes and eight edges. The nodes are labeled for convenience. The node 7 is disconnected from the main body. The node 3 has a loop that is a self-connecting link. There is a double-link between the nodes 4 and 5.  If the graph were simple, it would have neither multiple- nor self-connections. The degrees of all nodes are $1,3,5,4,2,1,0$, respectively. In addition, there is a triangle on the nodes $2,3,4$. Right: an example of directed pseudograph with $N=6,L=10$.}
\end{figure}

\section{Graphs as models of networks}
The material presented in this section is intended to give the reader a brief introduction to the notation and some basic concepts developed by mathematicians in graph theory and then widely accepted by the community of complex networks. All the definitions given here and also many others can be found for example in the book \cite{ref:gt_book}. The reader familiar with graph theory may skip this section. 

It is probably a trivial statement that a network can be represented as a graph, a mathematical object consisting of a set of nodes (called also vertices or sites) 
and a set of edges (links), which are related by incidence relations.
The nodes are joined by edges and the whole object is usually represented graphically as in Fig.~\ref{fig:some-gr}. For each edge the incidence relation says which nodes are its endpoints. In the thesis we shall denote the total number of nodes and links in the graph by $N$ and $L$, respectively. While referring to graph's size we shall usually mean the number of nodes. 
Nodes shall be denoted by small Latin letters $i,j,\dots$. For simple graphs (see below), each link is uniquely determined by a pair $(i,j)$ of nodes being its endpoints.

For many purposes it is convenient to differentiate between a directed graph, where every link $i\to j$ points only in one direction, and an undirected graph for which links do not have orientation. In Fig.~\ref{fig:some-gr} we show examples of an undirected and a directed graph. Not every edge must connect distinct vertices. An edge which has two identical end-points is called a loop or a self-connection. If two nodes are connected by more than one link, the corresponding links are called a multiple-connection or multiple-links. One is often interested in graphs without self- and multiple-connections, which are called simple graphs or sometimes Mayer graphs, in contrast to graphs with self- or multiple-connections which are called pseudographs or degenerate graphs. In the course of this work we will see however that in some respects pseudographs are more convenient for analytical treatment.  
A graph is fully described by its adjacency matrix $\A$, whose entries $A_{ij}$ give the number of edges between nodes $i,j$. In this thesis we shall mainly consider undirected graphs, for which $\A$ is symmetric: $A_{ij}=A_{ji}$. Because each self-connection can be viewed as two links: one going out and one going in, it is convenient to define diagonal elements of $\A$ to be equal to twice the number of loops incident with the node: $A_{ii}=0,2,4,\dots\;\;$. Alternatively the factor of two for the diagonal elements can be attributed to the fact that each loop is incident with the vertex two times. Of course for simple graphs, all diagonal elements vanish: $A_{ii}=0$ and off-diagonal $A_{ij}$ are either zero or one.

The most important local quantity characterizing a graph is node degree. The degree $k_i$ of node $i$ is just the number of links incident with the node: $k_i = \sum_j A_{ij}$. In case of directed graphs one can define the out- and in- degree separately, for outgoing and incoming links. For a simple undirected graph, the node degree is equal to the number of nearest neighbors of the given node, that is nodes linked to it by an edge. 
The average degree $\bar{k}$ of a graph is the average number of links per one node, that is $\bar{k}=2L/N$, because each link is counted twice in the sum $\sum_i k_i = 2L$. We shall use the notation $\bar{k}$ when $N,L$ are fixed, as for instance for the given network,
or $\left<k\right>$ when $N$ or $L$ may fluctuate, as for instance for networks in
the given statistical ensemble.

A graph is said to be dense if the average degree is of order $O(N)$ for $N\rightarrow \infty$ or to be sparse if $\bar{k}$ approaches a constant in this limit. A special example of a dense graph is a complete graph for which every pair of nodes is connected by an edge, and thus $L=N(N-1)/2$ and $\bar{k}=N-1$, and of a sparse graph is an empty graph with $L=0$ and $\bar{k}=0$. There are more special graphs having their own names, some of which will be mentioned in the next chapters.

A subgraph is a graph defined on a subset of nodes which are connected by links preserving the incidence relation of the whole graph. The simplest subgraphs are a line (a single edge joining two nodes) or a triangle: three nodes joined together by three links, see Fig.~\ref{fig:some-gr}. Small subgraphs are called motifs in the language of complex networks and will be discussed later.

A path joining nodes $i_1$ and $i_n$ is a set of all nodes $i_1,\dots,i_n$, where all intermediate nodes are distinct and every pair $i_k,i_{k+1}$ is connected by a link. In other words, it is a walk which starts from $i_1$, ends in $i_n$ and goes along links through the network, visiting each node no more than once. The length of a path is just its number of links. A shortest path (there may be more than one) between a pair of nodes is called geodesic path, and the length of this path is called geodesic distance. The longest geodesic from all possible paths is called a diameter of graph. The average geodesic distance is sometimes also called diameter, but strictly speaking it is quite a distinct quantity. In this paper we shall however use the latter definition since it is often much simpler to calculate, and moreover for graphs representing complex networks these two quantities are strongly correlated.
A graph is said to be connected if every two nodes can be connected by a path. 
Each subgraph built on all vertices which can be connected by a path is called 
connected component, or just component, of the graph. When the size of a component scales as $O(N)$ it is called a giant component. 
The graph on the left-hand side of Fig.~\ref{fig:some-gr} has two components: one 
has six nodes and the other only one, namely the node 7.

\begin{figure}
\center
\includegraphics*[width=5cm]{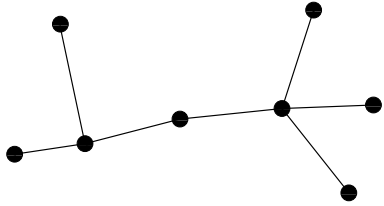}
\caption{\label{fig:atree}A tree graph with $N=8$ nodes and $L=7$ edges. Note that for any tree $L=N-1$.}
\end{figure}

A close path is called cycle. The simplest cycle is the triangle graph.
A connected graph with no cycles is called tree (Fig.~\ref{fig:atree}). 
Trees play an important role because on the one hand many models of complex networks can be exactly solved for trees and on the other hand some important classes of graphs locally look like trees. 

\section{Properties of real-world networks}

All definitions presented in the previous section have been developed by mathematicians long before the science of complex networks received its name and became popular between scientists working in different disciplines. In this section we shall present some of new ideas which have emerged recently, mostly in the last decade, as a result of empirical studies of real-world networks. Some of them have been introduced not as well-defined mathematical concepts but rather as ``operative'' definitions which captured interesting properties of investigated networks. 

Since the works of Milgram, Albert, Barab\'{a}si, Watts, Strogatz, and many others, three concepts have occupied an important place in the science of complex networks. These are power-law (or more generally: heavy tailed) degree distributions, the concept of small-world and the clustering. We shall discuss them shortly and describe how they apply to some real-world networks. All quantities and definitions shall be given for undirected networks if it is not stated otherwise.

{\bf Degree distribution.} Like we said, node's degree is the number of links connected to that node. Let us define now the probability $\Pi(k)$ that a randomly chosen node has exactly $k$ links. $\Pi(k)$ is called the degree distribution and can be obtained for any given network by making a histogram of the degrees for all nodes. By definition, the degree distribution is normalized: $\sum_k \Pi(k)=1$ and its mean $\sum_k k \Pi(k)$ equals to the average degree $\bar{k}$. Investigations of real networks have led to a surprising result that many of them have a power-law tail in the degree distribution:
\bq
	\Pi(k) \sim k^{-\gamma},
\eq
for intermediate values of $1\ll k\ll N$ where $N$ is the number of nodes in the network\footnote{For $k$ of order $N$ there is always some correction, see the next chapter.}. The value of $\gamma$ is typically between 2 and 4. This differs crucially from what one can imagine either for purely random networks, or for regular grids like square or cubic lattices. In the case of regular lattices, all degrees are the same, so $\Pi(k)=\delta_{k,\bar{k}}$, while for random graphs one can argue that since edges are placed randomly, the distribution $\Pi(k)$ should be close to a Poissonian one centered around $\bar{k}$. The network with a power-law degree distribution is called scale-free network (S-F), to emphasize the fact that there is no typical scale in the power-law describing the tail of the node degree distribution. Many models have been proposed to explain this feature, some of them will be presented later.

\begin{figure}
\center
\psfrag{knn}{$\bar{k}_{\rm nn}(k)$} \psfrag{k}{$k$} \psfrag{ass}{$\mathcal{A}>0$} \psfrag{dis}{$\mathcal{A}<0$}
\psfrag{lack}{$\mathcal{A}=0$}
\includegraphics*[width=7cm]{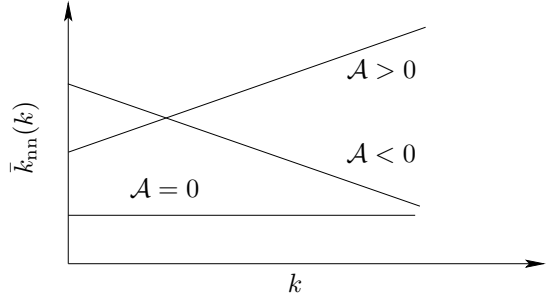}
\caption{\label{fig:assort}The assortativity $\mathcal{A}$ of the network and the average degree of nearest neighbors $\bar{k}_{\rm nn}(k)$.
For an uncorrelated network, the assortativity is zero and $\bar{k}_{\rm nn}(k)$ is a horizontal line. For correlated networks, two scenarios are possible: either $\mathcal{A}<0$ if the network is disassortative, or $\mathcal{A}>0$ if it is assortative.}
\end{figure}

Another quantity related to degrees is a two-point function $\eps(k,q)$ giving the probability that a randomly chosen edge joins two nodes of degrees $k$ and $q$. The values $\eps(k,q)$ form a symmetric matrix: $\eps(k,q)=\eps(q,k)$. The function has the following properties:
\ba
	\sum_{k\leq q} \eps(k,q) &=& 1, \\
	\sum_q \eps(k,q) = \sum_q \eps(q,k) &=& k\Pi(k)/\bar{k}.
\ea
The last equality becomes obvious when one realizes that the sum over $q$ gives the probability that a randomly chosen edge incidents on a vertex with degree $k$. But the fraction of such edges is just $k\Pi(k)$ and the division over $\bar{k}$ gives the correct probabilistic interpretation. If there were no correlations, the probability $\eps(k,q)$ would factorize:
\bq
	\eps_{\rm r}(k,q) = \frac{k\Pi(k)\,q\Pi(q)}{\bar{k}^2}, \label{eq1:uncorr}
\eq
but for almost all networks $\eps(k,q)\neq\eps_{\rm r}(k,q)$. The two-point function is, however, not convenient for examining degree-degree correlations, therefore another alternative quantities based on $\eps(k,q)$ have been introduced, as for instance an average degree $\bar{k}_{\rm nn}(k)$ of nearest neighbors of a node with degree $k$. It can be expressed through the two-point correlations as follows:
\bq
	\bar{k}_{\rm nn}(k) = \frac{\bar{k}}{k\Pi(k)} \sum_q q \,\eps(k,q). \label{eq1:knnk}
\eq
In Fig.~\ref{fig:assort} we sketch three possible behaviors of the correlations in the network, studied by means of $\bar{k}_{\rm nn}(k)$. When this quantity grows with $k$, it means that the higher is degree of a node, the higher is average degree of its neighbors. In order to describe this behavior one uses the term ``assortativity'' which is borrowed from social sciences. If $\bar{k}_{\rm nn}(k)$ decreases with $k$, the network is said to be disassortative. One can easily show that in case of uncorrelated degrees (\ref{eq1:uncorr}), the average degree of nearest neighbors is constant (horizontal line in Fig.~\ref{fig:assort}).
One can go further and reduce assortativity to a single coefficient \cite{ref:xbs-assort}:
\bq
	\mathcal{A} = \frac{\Tr \eps - \Tr \eps_{\rm r}}{1-\Tr \eps_{\rm r}}. \label{eq1:ac}
\eq
This quantity is equal 1 for a totally assortative network: $\eps(k,k)>0, \eps(k,q)=0$ for $k\neq q$, and is negative for disassortative networks.
In the paper \cite{ref:mejn-ass} a slightly different quantity, namely the Pearson correlation coefficient, was measured for real networks. It was found that artificial networks like the WWW or the Internet are mostly disassortative, while the citation network or other networks describing relations between human beings are rather assortative.

{\bf Small-worlds.} The most popular manifestation of the small-world effect is the ``Six degrees of separation'' being also the title of S. Milgram's book. He found that a typical distance in the network of acquaintance among people in the USA is about six. In the language of graph theory, this means that the average path length is six. If relationships between people formed a regular, two-dimensional grid, then for $N=3\times 10^8$ people the average distance $\left<l\right>$ would be of order $10^4$. The experiment made by Milgram showed that $\left<l\right>$ grows rather as $\sim \ln N$. More generally, one speaks about a small-world network when the typical distance or the diameter grows like logarithm of the system size. It is different from the case of a regular lattice in $d$ dimensions where
\bq
 \left<l\right>\sim N^{1/d}, \label{eq1:d}
\eq
but it agrees well with simple models of random graphs. Indeed, one can estimate the number of nodes of a random graph at distance $l$ to some particular node as $\bar{k}^l$. This has to be equal to $N$ for $l$ being the diameter and hence $\left<l\right> \sim \ln N$. If one defines a fractal dimension of the network as $d$ from Eq.~(\ref{eq1:d}), one gets $d=\infty$ for a small-world. We will see later that random graphs of a special type, namely homogeneous random trees (Section 2.3) do not need to be small-worlds, thus randomness {\em per se} is not a sufficient condition to trigger the effect.

{\bf Clustering.} This is a common property of many social networks which describes the tendency to form cliques of acquaintances. It is a rule that friends of our friends are often also our friends. In the language of graphs this means that there are many triangles in the network. Two measures of clustering are most popular. The first one is a global measure or a clustering coefficient $C$:
\bq
	C = \frac{3\times\mbox{number of triangles}}{\mbox{number of connected triples}},
\eq 
where a connected triple is a subgraph consisting of three nodes with at least two links between them. For a complete graph, all nodes are connected and thus $C=1$ which agrees with the intuition that the complete graph forms a big interconnected clique. For trees, $C=0$ because of the absence of any cycles (and hence also triangles) which is also intuitively comprehensible. For any other networks, $C$ lies somewhere between $0$ and $1$. 
Another definition of the clustering coefficient is based on local properties of nodes. Let $i$ be a node with degree $k_i$ and $c_i$ be the number of edges existing between the neighbors of $i$, or in other words, the number of triangles having one vertex at $i$. Then we define a local clustering coefficient:
\bq
	C_i = \frac{c_i}{k_i(k_i-1)/2},
\eq
which is one if all neighbors of the node $i$ are connected. The clustering coefficient for the whole network is the average of all $C_i$'s.
Both definitions of $C$ are qualitatively consistent and give roughly the same values for real networks, being rather high (typically $C>0.1$) in comparison to random graphs of the same size where $C\sim 1/N$ (see Section 2.1.1).  

After this introduction to the most interested observables on complex networks, let us shortly discuss some examples of real-world networks. Let us begin with the World Wide Web (WWW), which represents the largest network for which information about topology is currently available. The nodes are web pages and the edges are hyperlinks pointing from one page to another. This kind of structure can be represented by a directed graph. It is, however, possible to consider undirected networks where nodes are self-contained collections of web pages and the undirected link is formed if there is any hyperlink between pages belonging to different sites.
The Web was probably the first network, for which the power-law degree distribution $\Pi(k)$ was discovered by Albert and Barab\'{a}si, and Kumar et. al. \cite{ref:barab2,ref:kumar}. The total number of nodes in the WWW is of order several billions\footnote{In 2004 Google stated that they indexed over 4,000,000,000 web pages. It is, however, difficult to define, what is a single web page, therefore their number varies by two order of magnitude from definition to definition. Moreover, the WWW grows very quickly, making all precise estimations meaningless.}, but until now it was impossible to search the whole network. For a subset of about $N=300,000$ nodes the exponents $\gamma_{in}$ and $\gamma_{out}$ in power laws for in- and out-degree distribution were estimated to be $2.1$ and $2.45$, respectively. Later these estimations have been corrected to $2.1$ and $2.72$, where both distributions were collected on a network with 200 millions of documents \cite{ref:broder}. The power law is observed in a range of $k$ covering five orders of magnitude.
The network considered in \cite{ref:broder} had the average degree $\bar{k}=7.5$, and the average path length around $16$, which agrees with the conception of small worlds since $\ln N\approx 19$. The clustering coefficient has been found for another subset of the WWW  \cite{ref:adam}, with 1-degree nodes excluded and the size $N\approx 150,000$, to be about $0.11$, being much larger than for randomized graph of the same size.

In contrast to the WWW, the Internet is a physical network of computers (nodes) and wire- or wireless connections (edges). For the Internet treated as undirected network, the power law in the degree distribution has been found to hold over three orders of magnitude with $\gamma$ being in the range $2.1-2.5$ \cite{ref:falou,ref:govin}. Also the high clustering and small-world behavior have been confirmed, indicating a similarity of the Internet to the WWW.

There is also a large class of so called social networks. These are networks describing relationships between humans, either based on physical interpersonal contacts (movie or science collaboration networks, human sexual contacts) or non-physical like the citation network being in fact the graph of citation patterns of scientific publications (for a review see e.g. \cite{ref:barab,ref:physrep2006}). These networks share also some common features. All of them are scale-free with the exponent $\gamma$ varying between $2.1$ for some scientific collaborations to $3.5$ for the web of sexual contacts. Also the clustering is much larger than for an analogous random graph. The very interesting case happens for the citation network of papers published in Phys. Rev. D \cite{ref:redner}, where $\gamma$ seems to be very close to three, indicating that it can evolve due to the preferential attachment \cite{ref:bamodel} described below in Section 2.2.1.

The last class of networks we want to mention are various biological networks. For instance, one can study the metabolism of a living cell and construct a graph with nodes being chemical reagents and links denoting possible chemical reactions. Again, for a network like that \cite{ref:wagner} the usual behavior has been found: the power-law degree distribution, high clustering and small diameter. Another type of networks, showing possible bindings between proteins \cite{ref:jeong}, exhibits the power-law behavior with $\gamma=2.4$ but with an exponential cutoff above $k\approx 20$.

There are many other examples of networks: ecological, neural or linguistic ones or
a network of call phones etc., which have not been cited here. We decided to skip them because  the examples presented above are already a good sample of what one can find in real networks. The inquiring reader is referred to the review articles given in the first section of this chapter.

\section{The aims and the scope of the thesis}

Although many models have been proposed to capture various properties of complex networks, there is a relatively small number of papers which aim on formulating a sort of general theory of complex networks from physicist's point of view. In such a theory one is interested in having a general framework for modeling, calculating, computing or estimating quantities of interest and explaining the existing facts rather than in formulating general theorems or finding formal proofs of statements with idealized assumptions. Of course these two directions should be developed in parallel, since they are complementary. Here we shall concentrate on the former one, that is on practical aspects and on a physical theory of complex networks. The latter direction is covered by the mathematical literature on random graphs.

Therefore, the aim of this thesis is to present a theory of complex networks based on statistical mechanics, where the central role is played by the concept of statistical ensemble of graphs. This approach to complex networks has been continuously developed over the past ten years by many people, including D. ben-Avraham, M. Bauer, J. Berg, D. Bernard, P. Bialas, G. Bianconi, P. Blanchard, M. Bogu\~{n}\'{a}, Z. Burda, G. Caldarelli, J. D. Correia, I. Der\'{e}nyi, S. N. Dorogovtsev, I. Farkas, A. Frączak, K.-I. Goh, A. V. Goltsev, S. Havlin, J. A. Hołyst, B. Kahng, D. Kim, P. Krapivsky, T. Kr\"{u}ger, A. Krzywicki, M. L\"{a}ssig, D.-S. Lee, J. F. F. Mendes, M. E. J. Newman, G. Palla, J. Park, R. Pastor-Satorras, A. M. Povolotsky, S. Redner, A. N. Samukhin, T. Vicsek, and many others. 
In the thesis we discuss ideas and methods developed in this approach as well as a variety of results obtained within this framework. In this general context we present the original contribution of the author. It is partially based on yet unpublished work. The remaining part of the thesis is divided into three chapters, subdivided into sections. Each chapter and each section begin with a short introduction and a summary of most important results derived there.

Chapter 2 is devoted to a general presentation of statistical mechanics of networks. First, the most important models are described and then they are formulated in terms of statistical physics. The approach via statistical ensemble of random graphs is developed and it is shown how to design a very general Monte Carlo algorithm, suitable for generating various networks on a computer. Also the rate equation approach is presented since it is well suited for growing networks, being the vast part of proposed models.
The chapter is ended with a section on the comparison between growing networks and networks obtained by a process of ``thermalization'' (or ``homogenization'') by rewiring links.

Chapter 3 deals with applications of these ideas to complex networks. First we discuss finite-size corrections to power-law degree distributions. Such corrections are always present for finite networks and may significantly affect actual properties of the network. We develop an analytical method to evaluate the corrections and present results of this evaluation for some S-F networks. We compare the analytic results with numerical simulations. In the subsequent section of this chapter we discuss dynamics taking place on networks. We consider a special model called zero-range process. The application of the zero-range process to the description of many important phenomena like mass transport or condensation in homogeneous systems has been widely discussed in the literature. We concentrate here on the behavior of the zero-range process on complex networks, where inhomogeneity in nodes degrees plays an important role. After a preliminary discussion of the model we present our findings concerning static and dynamical properties of the process on inhomogeneous networks.

Chapter 4 contains conclusions and outlook.

\chapter{Statistical mechanics of networks}
For a long time networks were studied by mathematicians as a part of graph theory. In recent years it has been discovered that many concepts and methods of statistical physics can be successfully applied to description of complex networks, and many papers have been dedicated to the problem of formulating principles of statistical mechanics of networks (see e.g. \cite{ref:barab,ref:snd-stat-mech,ref:mejn,ref:bjk1,ref:krzyw}). In this chapter we shall present some of these ideas. Although all of them originate from the same statistical physics, the nature of complex networks leads to distinguishing two classes of methods: those for networks being in a sort of equilibrium to which the concept of statistical ensembles naturally applies, and those for non-equilibrium networks which are most naturally formulated within the rate-equation approach. We shall refer to the two classes of networks as to {\bf equilibrated} and {\bf growing (causal)} networks, respectively. While the second term is commonly accepted in this context, networks belonging to the first class are sometimes called homogeneous networks \cite{ref:homnasz} or maximal entropy random networks \cite{ref:bauer}.
Here we shall use the term ``equilibrated'' since it resembles the way how these networks are generated\footnote{In our earlier papers we often called these networks ``homogeneous''. In this thesis we shall reserve this word for networks with all nodes having the same degree, like $k$-regular graphs, with all degrees being equal to $k$. This is also the most popular meaning in the literature.}.

%the first one is usually used in a different sense. In the literature the word ``homogeneous'' refers mainly to networks with all nodes having almost the same degree, like Erd\"{o}s-R\'{e}nyi random graphs or $k$-regular graphs\footnote{In the $k$-regular graph all degrees are exactly equal to $k$.}. We are aware of this problem but believe that the reader will not be confused.

We have split this chapter into three sections. In the first section we discuss equilibrated networks. We first present the most famous examples of networks of that type. Then we show how to formulate a consistent theory of statistical ensembles of these networks starting from the simplest construction of Erd\"{o}s-R\'{e}nyi random graphs. We show that ascribing non-trivial statistical weights to graphs from this set we can produce networks with any desired features, as for instance networks having the power-law degree distribution, high clustering, degree-degree correlations etc. We present also a dynamical Monte Carlo algorithm, based on a construction of Markov chains, which allows one to generate equilibrated graphs.

At the beginning of the second section we show some famous examples of growing networks. These networks are generated by a growth process in which new nodes are attached to the existing network, so by the construction nodes are causally ordered in time. Therefore the words ``growing'' and ``causal'' are used to name these networks. Their growing character explains why the rate equation approach is so successful in this field. We will see, however, that causal networks can also be described within the formulation via statistical ensembles which in some cases is even more convenient.

In the third section we present differences and similarities between equilibrated and growing complex networks. The causality manifests itself as a very strong constraint that selects a subset of networks from the corresponding set of equilibrated graphs. In effect, ``typical'' networks in this subset usually have quite different properties than ``typical'' networks in the whole ensemble, even if networks in the two ensembles have identical statistical weights.

The ideas presented in this chapter have been introduced earlier by many people. They are scattered in many papers and used in different contexts. Here we want to collect them and comment on their applicability to some problems in the theory of complex networks. This shall form a basis for the considerations presented in the next chapter, where some applications will be discussed.

\section{Equilibrated networks}

As mentioned, we shall use the term ``equilibrated networks'' to refer to networks which are closely related to maximally random graphs. Although they can be constructed by many different methods, their common feature is that nodes, even if labeled, cannot be distinguished by any other attribute. For example, they may not be causally ordered. This means that if one generates a labeled network and repeats the process of generation many times, each node will have statistically the same properties as every other node, e.g. it will have on average the same number of neighbors, the same local clustering etc. We stress here the meaning of the phrase ``statistically the same'', which means that it does not make sense to speak about a single network, but rather about a set of networks, similarly as one speaks about the set of states in classical or quantum physics. In this way a statistical ensemble of graphs naturally arises as a tool for studying ``typical'' properties of networks. 
As we shall see later, equilibrated networks belonging to the given ensemble are in a sort of thermodynamical equilibrium, however it is not an equilibrium in the sense of classical thermodynamics, where the statistical weight of a state is given by the Gibbs measure: $\sim \exp (-\beta E)$, with $E$ being the energy of the state. In the case of complex networks it is convenient to abandon the concept of energy and Gibbs measure and consider a more general form of statistical weights. Therefore such a concept like temperature is often meaningless, although there were some attempts to define this quantity for networks \cite{ref:temp}.

Before we define a statistical ensemble of equilibrated networks, we shall present some examples of networks belonging to this class. They were introduced over the past 50 years. One of them, known as Erd\"{o}s-R\'{e}nyi model (ER model), is a pure mathematical construction. Much is known and can be proved rigorously for that model. In this respect, the ER model is exceptional since other models invented to mimic some features of real networks have not been studied so thoroughly and many results are not so rigorous. 
After a short presentation of the ER model we shall show how to change statistical properties of typical graphs by introducing an additional weight to every graph in the ER ensemble. The resulting ensemble of equilibrated graphs can be flexibly modeled by choosing appropriate weights. For instance, we shall see how to obtain a power-law degree distribution, or how to introduce degree-degree correlations. Towards the end of the section we shall present a quite general Monte Carlo algorithm for generating such equilibrated weighted graphs.

\subsection{Examples of equilibrated networks}
\begin{figure}
\center
\includegraphics*[width=14cm]{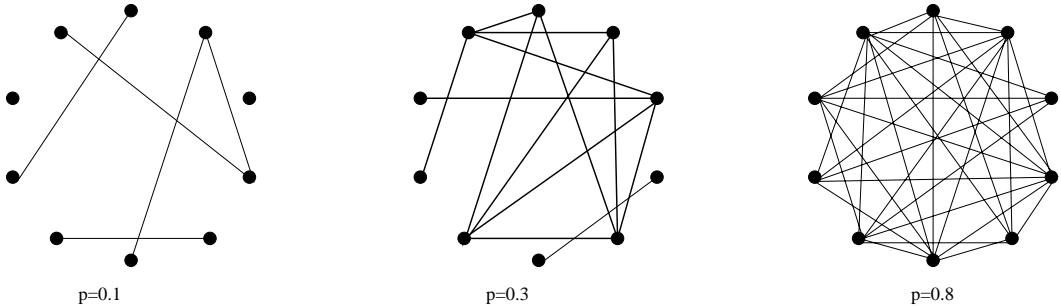}
\caption{\label{fig:e-r}Binomial graphs for $N=10$ and various $p$. For $p=0.1$ the graph consists of separated trees. For $p=0.3$ for which $\bar{k}= 2.7$ we are above the percolation threshold $\bar{k}_c=1$ and the giant component emerges (fat lines in the middle picture). In the limit of $p\to 1$ the graph becomes dense.} 
\end{figure}

As a first example of a network model belonging to the class of equilibrated networks we shall describe the {\bf Erd\"{o}s-R\'{e}nyi model}. In their classical papers \cite{ref:er} in 1950s  Erd\"{o}s and R\'{e}nyi proposed to study a graph obtained from linking $N$ nodes by $L$ edges, chosen uniformly from all $\binom{N}{2}=N(N-1)/2$ possibilities. In this thesis we shall often refer to it as a maximally random graph since it is totally random, that is edges are dropped on pairs of nodes regardless of how many links the nodes have already got. The only constraint is that one cannot connect any pair of nodes by more than one link, so the ER graph is a simple graph. The graph can be constructed in an alternative way by random rewirings. This will be discussed later in section 2.1.5 which is devoted to computer simulations. Beside the ER model there is also a very similar construction called the {\bf binomial model}. Here one starts with $N$ empty nodes and joins every pair of nodes with probability $p$. The name of the model becomes obvious when one realizes that the distribution $P(L)$ of the number of links $L$ is given by the formula:
\bq
	P(L) = \binom{N(N-1)/2}{L} p^L (1-p)^{N(N-1)/2-L}.
\eq
In this model, also introduced by Erd\"{o}s and R\'{e}nyi, the number of nodes is not fixed, but fluctuates around $\langle L\rangle=pN(N-1)/2$. This means that also the average degree $\bar{k}$ is a random variable with the mean $\langle k\rangle \cong p N$. However, because real-world networks have fixed average degree while their size can be very large, in order to compare binomial-graphs to real-world networks one usually scales $p\propto 1/N$. Under this scaling the corresponding graphs have fixed $\bar{k}$ and thus are sparse. If one calculates the variance of $P(L)$ keeping the average degree constant in the limit of $N\to\infty$, one finds that the variance grows only as $\sim N$ and that the distribution $P(L)$ becomes Gaussian with the relative width $\sim 1/\sqrt{N}$ falling to zero. Thus almost all binomial graphs have the same number of links $\left<L\right>$ in the thermodynamical limit and therefore the ER and binomial graphs become equivalent to each other for large $N$. Later on we shall see that the ER model defines a canonical ensemble of graphs while the binomial model - a grand-canonical ensemble, with respect to the number of edges.
In Fig.~\ref{fig:e-r} we show some examples of binomial graphs for different $p$.

Like we have already mentioned, we are interested in properties of the model in the thermodynamical limit, that is for very large graphs. The great discovery of Erd\"{o}s and R\'{e}nyi was that many motifs, like trees of a given size, cycles or the giant component, appear for typical graphs suddenly when $p$ crosses a certain threshold value $p_c$. The thresholds are different for different motifs. For $p$ just below $p_c$ there are almost no motifs of a given type, while for $p$ just above $p_c$ the motifs can be found with probability one. This is similar to the percolation transition on a lattice. For random graphs, however, $p_c$ depends usually on the system size, $N$, and must be properly scaled to get fixed values of critical parameters in the thermodynamical limit. For example, if one scales $p$ as $p=\bar{k}/N$, the desired average degree $\bar{k}$ plays the role of a control parameter. In the limit $N\to\infty$,
\bq
	\left<k\right> = \bar{k} ,
\eq
and the graph is sparse making it comparable to some real networks. One can ask what is the critical value of the control parameter $\bar{k}$, for some motifs to appear on the network. Erd\"{o}s, R\'{e}nyi and their followers were interested in a more general problem. If one assumes that $p$ scales as $p\sim N^{-z}$ for large $N$ with $z$ being an arbitrary real number, what are critical values of $z$ at which some properties appear in the thermodynamic limit? Below we present some important findings.

1) Subgraphs: for binomial graphs one can determine the threshold values of the exponent $z$ when subgraphs of a given type appear. One can argue \cite{ref:bb} that the average number of subgraphs having $n$ nodes and $l$ edges is equal to
\bq
	\binom{N}{n} \frac{n!}{n_I} p^l \approx \frac{N^n p^l}{n_I} \sim N^{n-zl}, \label{eq2:exp}
\eq
because $n$ nodes can be chosen out of $N$ in $\binom{N}{n}$ possible ways and they can be connected by $l$ edges with probability $p^l$. In addition one has to take into account that if one permutes $n$ nodes' labels one obtains $n!/n_I$ different graphs, where $n_I$ is the number of isomorphic graphs. From the formula (\ref{eq2:exp}) one can infer the critical value of the exponent $z$ for having at least $O(1)$ subgraphs of the given type in the limit $N\to\infty$. For instance, the critical value of $z$ for a tree of size $n$ is $z_c=n/(n-1)$, since for trees $l=n-1$. This means that for $z\geq 2$ the only subgraphs present in the graph are empty nodes and separated edges. When $z$ decreases from $2$ to $1$, trees of higher and higher size appear in the graph. Finally for $z\leq 1$ trees of all sizes are present as well as cycles, because for cycles the critical $z_c$ is also $1$.
However, the number of cycles of a given length is always constant for $z=1$, regardless of the size $N$. Thus binomial and ER random graphs are locally tree-like if $p\sim 1/N$. 
Because the clustering coefficient $C$ is proportional to the number of triangles ($n=l=3$) and inversely proportional to the number of connected triples ($n=3,l=2$), one sees from Eq.~(\ref{eq2:exp}) that $C\sim 1/N$ and that it vanishes  for sparse networks in the thermodynamic limit. This is the first property of random graphs that disagrees with empirical results for real networks, for which, like we saw in Sec. 1.3, $C$ is always much greater than zero.

2) Giant component. For the most interesting case of $p\sim 1/N$, there is a critical value of $\bar{k}_c=1$, above which a finite fraction of all nodes forms a connected component, called giant component. For $\bar{k}=1$ it has approximately $N^{2/3}$ nodes but it grows quickly with $\bar{k}$ so that for $\bar{k}$ of order $5$ and large $N$, more than $99\%$ nodes belong to the giant component. All other clusters are relatively small, and most of them are trees. Thus when $\bar{k}$ passes the threshold $\bar{k}_c=1$, the structure of graph changes from a collection of small clusters being trees of size $\sim\ln N$, to a single large cluster of size $\sim N$ containing loops (cycles), and the remaining components being small trees.
This behavior is characteristic not only for ER or binomial graphs, but it is a general feature of random graphs with various degree distributions \cite{ref:barab}. 

3) Degree distribution. For binomial graphs it is extremely simple to obtain the formula for degree distribution $\Pi(k)$, just by observing that a node of degree $k$ has $k$ neighbors chosen out of $N-1$ other nodes, and each of them is linked to the node with probability $p$:
\bq
	\Pi(k) = \binom{N-1}{k} p^k (1-p)^{N-1-k},
\eq
which for large $N$ becomes a Poissonian distribution:
\bq
	\Pi(k) \cong e^{-\bar{k}} \frac{\bar{k}^k}{k!}.	\label{eq2:poiss}
\eq
The same function describes the node-degree distribution for ER graphs in the limit of $N\to\infty$. For large $\bar{k}$ the degree distribution has a peak at $k\approx \bar{k}$. Its width grows as $\sqrt{\bar{k}}$, so random graphs with high average degree are almost homogeneous in the sense of nodes degrees which assume values very close to $\bar{k}$.
This is a second feature that disagrees with real networks, where $\Pi(k)$ has often a heavy tail meaning that there are some nodes with high degrees far from the mean value, called ``hubs''.

4) Diameter. As we mentioned in Sec.~1.2 we will calculate the diameter defined as the average distance  $\bar{l}$ between pairs of nodes in the network rather than the maximal distance $d$. We expect that $\bar{l}$ is roughly proportional to $d$. Certainly it is a good measure of the linear extension of the network. For both definitions it has been found that above the threshold $\bar{k}_c=1$, when a giant component is formed, the diameter grows only logarithmically with the size of the graph:
\bq
	\bar{l} \propto  \frac{\ln N}{\ln \bar{k}}.
\eq
We know that this behavior is called a small-world effect, and is almost always present in real-world networks.\\

\begin{figure}
\center
\includegraphics*[width=14cm]{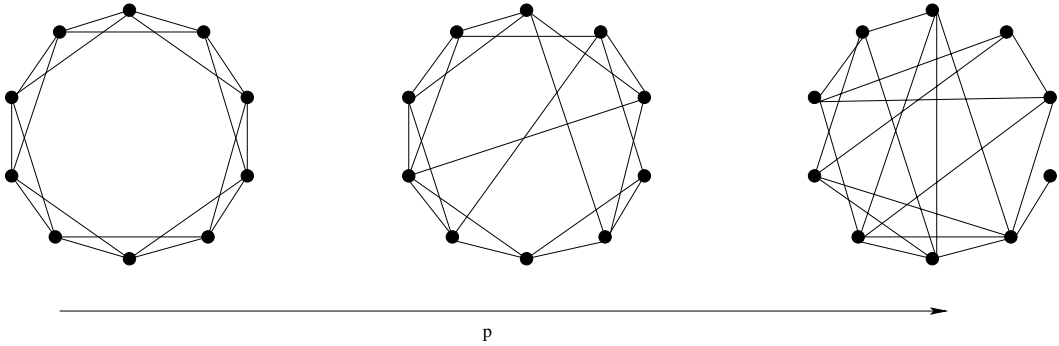}
\caption{\label{fig:ws}Example of the construction of Watts-Strogatz small-world network. Starting from $N=10$ nodes, each with degree $4$, one rewires some edges with probability $p$. As $p$ increases, the graph becomes more random.} 
\end{figure}

A next construction, which we briefly discuss, is the {\bf Watts-Strogatz model} \cite{ref:watts}. Its main feature is that it extrapolates between regular and random graphs. We start with $N$ nodes located on a ring (see figure \ref{fig:ws}). Each node is connected to $K$ of its nearest-neighbors, so all nodes have initially degree $K$. Then one rewires each edge with probability $p$ to randomly chosen nodes, or leave it in place with probability $1-p$. Self- and multiple-connections are excluded. By tuning $p$ one can extrapolate between $K$-regular graph ($p=0$) and the maximally random ER graph ($p=1$). This model originally arose from considerations of social networks, where people have mainly friends from local neighborhood, but sometimes they know someone living away - these cases are represented by rewired long-range edges.

An important feature of this model is that the network can have a small diameter and large clustering coefficient at the same time.
Let us consider first the limit $p=0$. The network is regular and a ring-like. Therefore, the diameter $d_{\rm reg}\sim N$ grows linearly with $N$. The clustering coefficient $C_{\rm reg}$ is constant and larger than zero when $N\to\infty$ because the nearest neighborhood of each node looks the same and there are always some triangles\footnote{We consider only the case of $K>2$ when the network is connected.}. On the other hand, for $p=1$ we have the ER random graph for which $d_{\rm rand}\sim \ln N$ and $C_{\rm rand}\sim 1/N\to 0$.
Watts and Strogatz found \cite{ref:watts} that there is a broad range of $p$, where $d\approx d_{\rm rand}$ and $C\approx C_{\rm reg}$.
This is the result of a rapid drop of the diameter $d$ when $p$ grows, while the clustering coefficient $C$ changes very slow. The diameter decreases fast because even a small addition of short-cuts which takes place during the rewiring process, reduces significantly the average distance between any pair of nodes.
These two properties, namely high clustering and small-world effect, agree with findings for many real networks. However, the degree distribution is similar to that of ER random graphs. There is no natural possibility to produce a power-law (or generally fat-tailed) degree distribution in this model.

\begin{figure}
\center
\includegraphics*[width=14cm]{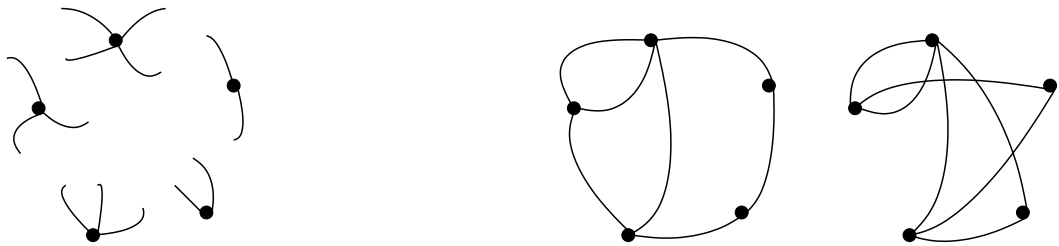}
\caption{\label{fig:mr}Example of the Molloy-Reed construction of pseudographs. We start from $N=5$ empty nodes with $k_i$ ``half-edges'' (left-hand side) connected to node $i$. The numbers $k_i$ are taken independently from some distribution $\Pi(k)$. On the right-hand side we show two possible configurations obtained by pairing half-edges.} 
\end{figure}

Now we shall describe the {\bf Molloy-Reed model} \cite{ref:mr,ref:mr2, ref:mr3} or configuration model, which allows for construction of pseudographs as well as simple graphs. In the Molloy-Reed model, to build a graph with $N$ nodes one generates a sequence of non-negative integers $\{k_1,k_2,\dots,k_N\}$, almost always as independent identically distributed numbers from a desired distribution $\Pi(k)$ and interprets $k_i$'s as node degrees. The only requirement is that the sum $k_1+k_2+\dots +k_N=2L$ is even. In the first step, each integer $k_i$ represents a hub consisting of node $i$ and $k_i$ outgoing ``half-edges''. 
In the second step these ``half-edges'' are paired randomly to form undirected links which now connect nodes (see figure \ref{fig:mr}). 
The number of links fluctuates around $N\langle k \rangle/2$ if no additional constraint is imposed.
In general, this procedure leads to pseudographs since sometimes an edge can be created between already connected nodes.
To restrict to simple graphs one has to stop the procedure every time when a multiple or self-connecting link is created, and to start it from the beginning. This can be very time-consuming, especially for degree distributions with heavy-tails, where it is unlikely to produce only a single link between nodes of high degree. Thus sometimes for practical purposes one does not discard the whole network but only the last move, and chooses another pair of half-edges. This introduces correlations to the network and an uncontrolled bias to the sampling. In other words, graphs are not sampled uniformly \cite{ref:uni}.

Using this model Molloy and Reed have shown that for networks with uncorrelated degrees the giant component emerges when the following condition is fulfilled:
\bq
	\sum_k k(k-2) \Pi(k) >0.
\eq
For $\Pi(k)$ being Poissonian one gets the well-known result for ER graphs: $\bar{k}_c=1$. The most important property of the model is that it allows for power-law degree distribution. Indeed, up to finite-size effects, the distribution $\Pi(k)$ is reproduced correctly.
The average path length $\bar{l}$ has also been calculated \cite{ref:mejn2}; it grows as $\ln N $ with the system size, so again one has a small-world behavior. The clustering coefficient is proportional to $\bar{k}/N$ so it vanishes as for random graphs, but the proportionality coefficient depends on $\Pi(k)$ and may be quite large for heavy-tailed distributions.

Finally, we shall mention the {\bf Maslov-Sneppen algorithm} \cite{ref:masl} used for obtaining a randomized version of any network.
The original motivation was to examine whether the appearance of degree-degree correlations and other non-trivial properties observed in some biological networks could be entirely attributed to the power-law degree distribution. The basic step in this algorithm involves rewiring of two edges. One selects two edges: $i\to j$ and $k\to l$, and then one rewires their endpoints to get $i\to l$ and $k\to j$. If this move leads to multiple- or self-connections, one rejects it and tries with another pair of edges. To obtain a randomized (``thermalized'') version of the given network, one repeats this move many times.
The algorithm preserves degrees of all nodes, so at the end of randomization the degree distribution is the same as for the original network.
However, thermalization breaks any correlations between nodes which might be present at the beginning. In a sense, one obtains a new network being maximally random for the given sequence of degrees $\{k_1,\dots,k_N\}$. In next sections we will see that this algorithm is also very helpful for generating graphs in a micro-canonical ensemble.

The four models presented above clearly belong to the class of equilibrated networks because every node on the network has statistically the same properties. Nodes have no individual attributes which would be correlated with nodes' labels, as one can see if one repeats the process of generation of networks many times. In the next subsection we shall explain in a more detailed way what it means and how to define an ensemble of equilibrated graphs. We shall see that graphs from the ensemble can be generated in a process of thermalization which homogenizes the network.

\subsection{Canonical ensemble for ER random graphs}
The basic concept in the statistical formulation is that of statistical ensemble. The statistical ensemble of networks is defined by ascribing a statistical weight to every graph in the given set. Physical quantities are measured as weighted averages over all graphs in the ensemble. The probability of occurrence of a graph during random sampling is proportional to its statistical weight, thus the choice of statistical weights affects the probability of occurrence and, in effect, also ``typical'' properties of random graphs in this ensemble. For convenience, the statistical weight can be split into two components: a fundamental weight and a functional weight. If the functional weight is independent of the graph, graphs are maximally random. The fundamental weight tells one how to probe the set of ``pure'' graphs uniformly, so that each graph in the ensemble is equiprobable. In other words, the fundamental weight defines an uniform measure on the given set of graphs and should be fixed. The functional weight is the parameter of the model.

What is the most natural candidate for the fundamental weight for graphs?
Consider simple graphs with a fixed number of nodes. We can choose the uniform measure by saying that in this case all unlabeled graphs are equiprobable, or alternatively that all labeled graphs are equiprobable. These two definitions give two different probability measures since the number of ways in which one can label graph's nodes depends on graph's topology. It turns out that the latter definition is in many respects better and we will stick to it. For instance, with this definition ER graphs have a uniform measure and thus are maximally random. 
There are also some practical reasons.
First, in the real world as well as in computer simulations node are labeled\footnote{In real-world networks one can always distinguish nodes for example by names of Web pages, people, scientific papers etc. On a computer, nodes are obviously labeled by their representation in computer memory.}. Second, it is not easy to determine whether two unlabeled graphs are identical or not. The problem of graph isomorphism has certainly NP-complexity but it is unknown if it is NP-complete \cite{ref:isomorph}.  

For pseudographs, the fundamental weight is most naturally defined by saying that fully labeled graphs, that is having nodes and edges' endpoints labeled, are equiprobable in the maximally random case. One can show that for this choice each unlabeled graph has the weight equal to the symmetry factor of Feynman diagrams generated in the Gaussian perturbation field theory \cite{ref:bjk1,ref:bck}. 

\begin{figure}
\center
\includegraphics*[width=13cm]{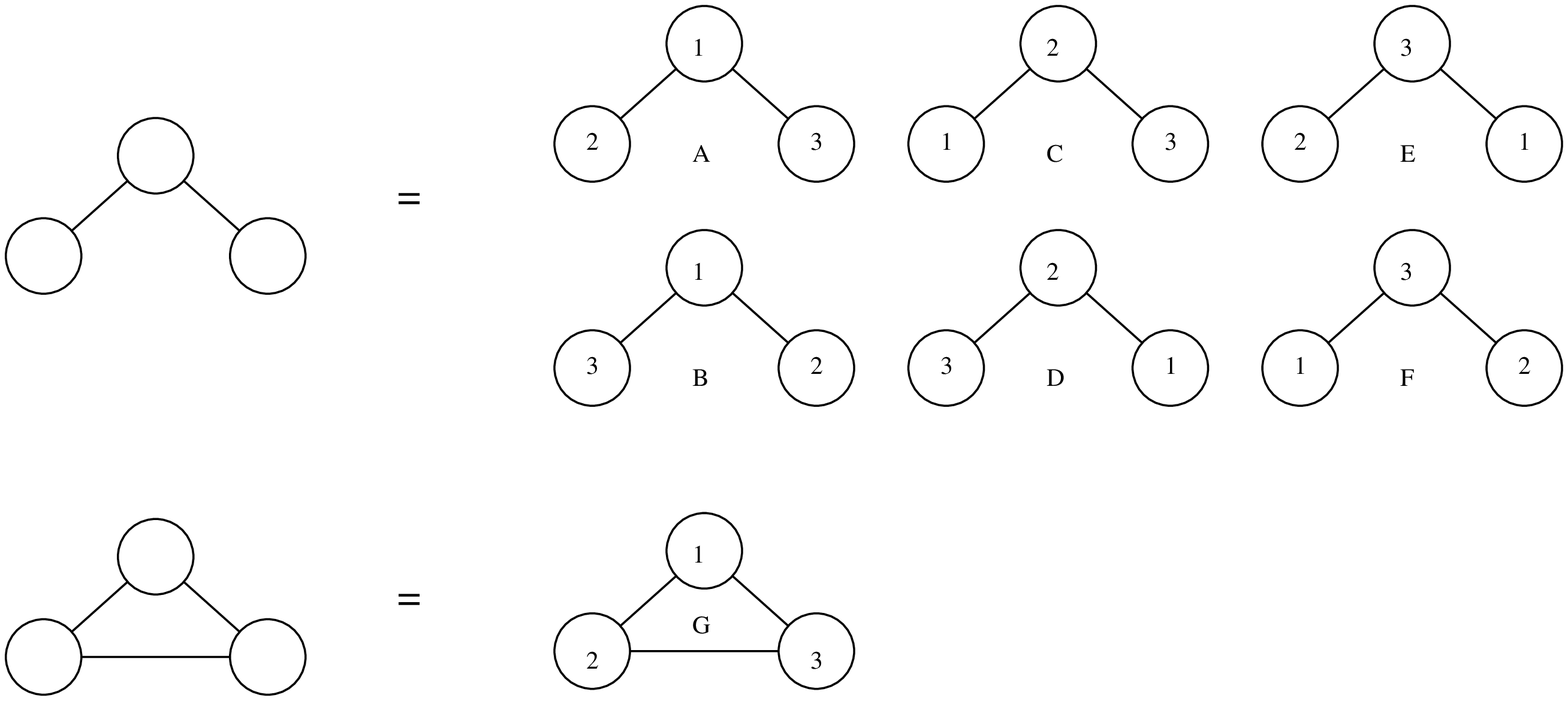}
\caption{\label{fig:shape}Top: the given unlabeled graph can be realized as three different labeled graphs. A is equivalent to B, C to D and E to F since they have the same adjacency matrix and hence are identical. Bottom: a triangle-shaped graph has only one realization as a labeled graph.}
\end{figure}

Let us concentrate on simple graphs. Consider again an ensemble of Erd\"os-R\'{e}nyi's graphs  with $N$ labeled nodes and an arbitrary number $L$ of (unlabeled) links. Since each ER graph from this ensemble can be in a one-to-one way represented as a symmetric $N\times N$ adjacency matrix we see that the uniform measure in this  ensemble is alternatively defined by saying that all such matrices are equiprobable. What about unlabeled graphs? Are they equiprobable in this ensemble? An unlabeled graph is obtained from a labeled one by removing labels. We immediately see that each unlabeled graph can be obtained from many different labeled graphs. Let us consider the unlabeled one shown on the left-hand side in the upper part of Fig.~\ref{fig:shape}. Since there are three nodes one can naively think that there are $3!$ labeled graphs corresponding to this shape as shown on the left-hand side of the figure. Actually, it turns out that there are only three distinct ones in the sense of having distinct adjacency matrix.
Graphs A, C, E are distinct, but B is identical to A, D to C, and F to E:
\bq
\A_{\mbox{\scriptsize A}} = \A_{\mbox{\scriptsize B}} = 
\left( \begin{array}{ccc}
	0 & 1 & 1 \\
	1 & 0 & 0 \\
	1 & 0 & 0
\end{array}
\right), \quad
\A_{\mbox{\scriptsize C}} = \A_{\mbox{\scriptsize D}} = 
\left( \begin{array}{ccc}
	0 & 1 & 0 \\
	1 & 0 & 1 \\
	0 & 1 & 0
\end{array}
\right), \quad
\A_{\mbox{\scriptsize E}} = \A_{\mbox{\scriptsize F}} = 
\left( \begin{array}{ccc}
	0 & 0 & 1 \\
	0 & 0 & 1 \\
	1 & 1 & 0
\end{array}
\right) \ .
\eq
In other words there are three labeled graphs having this shape. On the other hand, if one takes the shape in the lower line of Fig.~\ref{fig:shape} one can see that there is only one labeled graph corresponding to it, since all others have the same adjacency matrix.
In view of this we see that the probability of occurrence of the upper shape is three times larger than of the lower one since the upper is realized by three adjacency matrices while the lower has only one realization.

\begin{figure}
\center
\includegraphics[width=7cm]{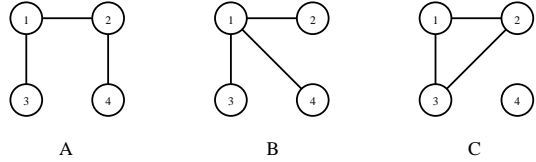}
\caption{\label{fig:shape2}Three possible graphs for $N=4,L=3$, for each of them one of possible labellings is shown. 
The total number of different labellings of these graphs is: $n_A=12$, $n_B=4$,
$n_C=4$.}
\end{figure}

Let us consider now an ensemble of Erd\"os-R\'{e}nyi graphs with $N=4$, $L=3$. The set consists of only three distinct unlabeled graphs A, B, C shown in Fig.~\ref{fig:shape2}. 
Each graph has a few possible realizations as a labeled graph. One can label four vertices of A in $4!=24$ ways corresponding to permutations of nodes $1-2-3-4$, but only $n_A=12$ of them give distinct labeled graphs. It is so because every permutation has its symmetric counterpart which gives exactly the same labeled graph, e.g. $1-2-3-4$ and $4-3-2-1$. Similarly, one can find that there are $n_B=4$ labeled graphs for $B$ and $n_C=4$ for $C$. One can check that indeed by dropping three links at random on four nodes one gets these numbers of labeled ER shapes. Altogether, there are $n_A+n_B+n_C=20$ labeled graphs in the given set. Because all labeled graphs are assumed to be equiprobable, the shapes A, B, C have the following probabilities of occurrence during the random sampling: 
\bq
	p_A=\frac{n_A}{n} = \frac{3}{5}, \quad
	p_B=\frac{n_B}{n} = \frac{1}{5}, \quad
	p_C=\frac{n_C}{n} = \frac{1}{5}. \label{pABC}
\eq
We see that (unlabeled) ER graphs are not equiprobable - the distribution is uniform only for labeled graphs. 
Let us denote the statistical weights for A, B, C by $w_A,w_B,w_C$. They are proportional to probabilities of configurations and hence $w_A:w_B:w_C = p_A:p_B:p_C$. There is a common proportionality constant in the weights, which we for convenience choose so that the weight of each labeled graph is $1/N!$. For this choice we have $w_A=1/2, \quad w_B=1/6, \quad w_C=1/6$. The larger is the symmetry of a graph topology, the smaller is the number of underlying labeled graphs and thus the smaller is the statistical weight. The choice $1/N!$ compensates the trivial factor of permutations of indices, and thus removes overcounting - however, for graphs with fixed number of nodes this particular choice does not influence on any physical properties.

We now apply the above ideas to define an ensemble of ER graphs with arbitrary $N,L$. The partition function $Z(N,L)$ for the Erd\"os-R\'enyi ensemble can be written in the form:
\bq
	Z(N,L) = \sum_{\alpha'\in lg(N,L)} \frac{1}{N!} \;\;= \sum_{\alpha\in g(N,L)} w(\alpha),	\label{eq2:er}
\eq 
where $lg(N,L)$ is the set of all labeled graphs with given $N,L$ and $g(N,L)$ is the corresponding set of (unlabeled) graphs. The weight $w(\alpha) = n(\alpha)/N!$, where $n(\alpha)$ is the number of labeled versions of graph $\alpha$. We are interested in physical quantities averaged over the ensemble. The word ``physical'' means here that the quantity depends only on graph's topology and not on how nodes' labels are assigned to it. It is a natural requirement. The average of a quantity $O$ over the ensemble is defined as
\bq
\left\langle O\right\rangle \equiv 
\frac{1}{Z(N,L)} \sum_{\alpha'\in lg(N,L)} O(\alpha') \frac{1}{N!} \;\;= 
\frac{1}{Z(N,L)} \sum_{\alpha\in g(N,L)} w(\alpha) O(\alpha)  .
\eq
We shall refer to the ensemble with fixed $N,L$ as to a {\bf canonical ensemble}. The word ``canonical'' is used here to emphasize that the number of links $L$ is conserved like the total number of particles in a container with ideal gas remaining in thermal balance with a source of heat. Although there is no temperature here, the analogy is close because, as we shall see later, these graphs can be indeed generated in a sort of thermalization process.

The partition function $Z(N,L)$ can be calculated by summing over all adjacency matrices $\A$ which are symmetric, have zeros on the diagonal and $L$ unities above the diagonal \cite{ref:homnasz}. The result is:
\bq
  Z(N,L) = \frac{1}{N!} \binom{\binom{N}{2}}{L} ,
\eq
which agrees with simple combinatorics: there are $\binom{\binom{N}{2}}{L}$ ways of choosing $L$ links among all possible $\binom{N}{2}$ edges. In a similar manner, summing over adjacency matrices, one can calculate averages of various quantities. As an example let us consider the node degree distribution $\Pi(k)$:
\bq
	\Pi(k) = \left\langle \frac{1}{N} \sum_i \delta_{k,k_i} \right\rangle ,  \label{piq}
\eq
where one can use an integral representation of the discrete delta to get \cite{ref:homnasz}
\bq
	\Pi(k) = \binom{\binom{N-1}{2}}{L-k} \binom{N-1}{k} / \binom{\binom{N}{2}}{L}  . \label{eq2:exact_p}
\eq
This is an exact result for ER random graphs. It reduces in the limit $\bar{k}=\mbox{const},\,N\rightarrow\infty$ to the Poissonian distribution (\ref{eq2:poiss}).

\subsection{Grand-canonical and micro-canonical ensemble of random graphs}
So far we have discussed the canonical ensemble of Erd\"os-R\'{e}nyi graphs with $N,L$ fixed. If we allow for fluctuations of the number of edges, we get the binomial model. The probability of obtaining a labeled graph with given $L$ is $P(L)=p^L (1-p)^{\binom{N}{2}-L}$.
Thus the partition function is
\bq
Z_0(N,\mu) = \sum_L \sum_{\alpha\in lg(N,L)} \frac{1}{N!} P(L(\alpha)) 
= (1-p)^{\binom{N}{2}} \sum_L \left(\frac{p}{1-p}\right)^L 
\sum_{\alpha\in lg(N,L)} \frac{1}{N!} .
\eq
The factor $(1-p)^{\binom{N}{2}}$ is inessential for fixed $N$ and can be skipped. The new partition function reads
\bq
 Z(N,\mu) =  \sum_L \exp (-\mu L) \; Z(N,L) ,\label{grand}
\eq
where $\frac{p}{1-p}\equiv \exp(-\mu)$ or equivalently $\mu = \ln \frac{1-p}{p}$. The weight of a labeled graph $\alpha$ is now $w(\alpha)= \exp(-\mu L(\alpha))/N!$, where $\mu$ is a constant which can be interpreted as a chemical potential for links in the {\bf grand-canonical} ensemble (\ref{grand}). Notice that the function (\ref{grand}) can be regarded as the generating function for $Z(N,L)$. One can calculate the average number of links or its variance as derivatives of the grand-canonical partition function with respect to $\mu$:
\ba 
	\langle L \rangle &=& -\partial_\mu \ln Z(N,\mu), \\
	\langle L^2 \rangle -\langle L \rangle^2 &=& \partial^2_\mu \ln Z(N,\mu).
\ea
Like for the canonical ensemble of ER graphs, the sum of states can be done exactly:
\bq
	Z(N,\mu) = \sum_{L=0}^{\binom{N}{2}} e^{-\mu L} \frac{1}{N!} \binom{\binom{N}{2}}{L} = 
  \frac{1}{N!} (1+e^{-\mu})^{\binom{N}{2}}  .
\eq
It is easy to see that for fixed chemical potential $\mu$ the average number of links behaves as
\bq
\langle L \rangle = p \frac{N(N-1)}{2} = 
\frac{1}{1 + e^\mu} \frac{N(N-1)}{2}   .
\label{Lp}
\eq
Thus for $N\rightarrow\infty$ the graphs become dense; $\bar{k}$ increases to infinity. We know that this pathology can be cured by an appropriate scaling of the probability $p$: $p \sim 1/N$. Since $\mu = \ln \frac{1-p}{p}$, this corresponds to $\mu \sim \ln N$. In this case $L$ is proportional to $N$. The corresponding graphs become sparse and the mean node degree is now finite. The situation in which $\mu$ scales as $\ln N$ is very different from the situation known from classical statistical physics, where such quantities like chemical potential 
$\mu$ are intensive and do not depend on system size $N$ in the thermodynamic limit $N\rightarrow\infty$. Moreover, the entropy $S= \ln Z(N,L)$ is not extensive - one can show that
\bq
	S = \frac{\bar{k} - 2}{2} N \ln N + \frac{2+\bar{k}-\bar{k}\ln \bar{k}}{2} N +O(\ln N)  ,
\eq
so the system is not ``normal'' in the thermodynamical sense for $\bar{k}\neq 2$. Only when $\bar{k} = 2$, that is if $N=L$, the entropy becomes extensive. This means that each graph from this set can be partitioned so that we get two sets of graphs A and B, with $N_A+N_B=N$ nodes, and the partition function for A+B being just the product of the partition functions for A and B. In other words, almost every graph in A+B can be constructed by taking two graphs: one from A and the second one from B, and joining two of their nodes by a link. In classical statistical physics this means that interactions between A and B take place only on the boundary which can be neglected in the thermodynamical limit. In the context of ER graphs, the case $N=L$ must therefore corresponds to the set of tree-like graphs - the number of loops must be small and they must be short (local).

As mentioned, the difference between canonical and grand-canonical ensembles gradually disappears in the large $N$ limit. It is easy to see why. In a canonical ensemble of sparse graphs the average degree $\bar{k} = 2L/N $ is kept constant when $N\to\infty$ while in a grand-canonical it fluctuates around $\langle k \rangle  = 2\langle L\rangle /N = \bar{k}$, if $\mu$ is properly chosen. However, the magnitude of
fluctuations around the average disappears in the large $N$ limit since
\bq
\langle L^2 \rangle - \langle L\rangle^2 = 
\binom{N}{2} \frac{e^{-\mu}}{(1+e^{-\mu})^2}  ,
\eq
and for $\mu\sim\ln N$ the relative width $\sqrt{\langle L^2 \rangle - \langle L\rangle^2} / \langle L \rangle \sim N^{-1/2} \rightarrow 0$, so effectively the system selects graphs with $\langle k \rangle  =\bar{k}$.

Apart from the canonical and grand-canonical ensembles, one can define a {\bf micro-canonical ensemble} of ER random graphs. By analogy with classical physics, we define it as a set of all equiprobable graphs with prescribed sequence of degrees $\{k_1,\dots,k_N\}$ which plays the role of the microstate. Then the canonical ensemble is constructed by summing over all sequences obeying the conservation law $k_1+\dots+k_N=2L$. It looks similar to the construction of Molloy and Reed, and indeed, it is its special case. We shall make use of the micro-canonical ensemble in Chapter 3 in the context of dynamics on graphs. 

\subsection{Weighted equilibrated graphs}
In the previous section we described ensembles for which all labeled graphs had the same statistical weight. They were just ER or binomial random graphs and thus had well known properties. In section 2.1.1 we pointed out however, that most of these properties do not correspond to those observed for real world networks. But the framework of statistical ensembles is very general and flexible and it allows one to model a wide class of random graphs and complex networks with non-trivial properties. Consider the same set of graphs as in the Erd\"os-R\'enyi model but now to each graph in this set, in addition to its fundamental weight $1/N!$, we ascribe a functional weight $W(\alpha)$ which may differ from graph to graph so that graphs are no longer uniformly distributed. By tuning the functional weight one can make that typical graphs in the ensemble will be scale-free or have more loops, etc. One has a freedom in choosing the functional weight. The only restriction on $W(\alpha)$ is that it should not depend on the labeling because graphs need to remain equilibrated. We stress that we still have the same set of graphs but now they may have distinct statistical weights. 

The partition function for a weighted canonical ensemble can be written as
\bq
	Z(N,L) = \sum_{\alpha'\in lg(N,L)} (1/N!)\, W(\alpha') \;\;= \sum_{\alpha\in g(N,L)} w(\alpha) W(\alpha)  , \label{eq2:can_g}
\eq
where as before $w(\alpha) = n(\alpha)/N!$ counts labeled graphs. For $W(\alpha)=1$ we recover the ensemble of ER graphs. The simplest non-trivial choice of $W(\alpha)$ is a family of product weights:
\bq
	W(\alpha) = \prod_{i=1}^N p(k_i)  ,\label{eq2:prod}
\eq
where $p(k)$ is a semi-positive function depending on degree $k_i$ of node $i$. This functional weight is local in the sense that it depends only on individual degrees which are a local property of the graph. It does not introduce explicitly correlations between nodes, so we will call random graphs generated in this ensemble {\bf uncorrelated networks}. One should, however, remember that the total weight does not entirely factorize because the fundamental weight $w(\alpha) = n(\alpha)/N!$ written as a function of node degrees $w(k_1,k_2,\dots,k_N)$ does not factorize since the number $n(\alpha)$ of labelings is not a product of any local property of the graph but is a global feature. There is also another factor which prevents the model from a full factorization and independence of node degrees, namely the constraint on the total number of links 
$2L = k_1+k_2+\dots + k_N$ which for given $L$ and $N$ introduces correlations between $k_i$'s. For example, if one of $k_i$'s is
large, say $\approx 2L$, then the remaining ones have to be small in order not to violate the constraint on the sum. The effect gradually disappears in the limit $L\to\infty$ for a wide class of weights $p(k)$ since then the canonical ensemble and the grand-canonical ensemble, for which $L$ does not need to be fixed, become equivalent \cite{ref:snd-stat-mech}.

The weight (\ref{eq2:prod}) is especially well-suited for studying ensembles with various degree distributions and no higher-order  correlations. To see how $p(k)$ is related to $\Pi(k)$, let us first discuss the analogous ensemble of weighted pseudographs. They can model networks where self-interactions of nodes are important, as for example ecological networks which describe predator-prey relations where cannibalism is often present. A pseudograph can be represented by a symmetric adjacency matrix $\A$ whose off diagonal entries $A_{ij}$ count the number of links between nodes $i$ and $j$, and the diagonal ones $A_{ii}$ count twice the number of self-connecting links attached to node $i$. Each adjacency matrix represents a certain labeled graph, but now, due to possibility of multiple links, we label also edges and call such a graph a {\bf fully labeled graph}. To each fully labeled graph we ascribe a configurational weight $1/N!(2L)!$. The weight of each labeled graph (where only nodes are labeled) having adjacency matrix $\A$ is then
\bq
	\frac{1}{N!} \left( \prod_{i} \frac{1}{2^{A_{ii}/2} \left( A_{ii}/2\right)!} \right) \prod_{i>j} \frac{1}{A_{ij}!}
	=  \frac{1}{N!} \prod_{i} \frac{1}{A_{ii}!!} \prod_{i>j} \frac{1}{A_{ij}!},\label{wp}
\eq
where the origin of all symmetry factors is the same as in case of Feynman diagrams and stems from possible ways of labeling links (see e.g. \cite{ref:homnasz}). The key points behind introducing pseudographs are: i) the set contains the subset of all simple graphs which we are interested in, and, ii) despite a complicated form of Eq.~(\ref{wp}), the canonical partition function can be easily evaluated. Let us rewrite the formula (\ref{eq2:can_g}) for $Z(N,L)$ for pseudographs with functional weight (\ref{eq2:prod}):
\bq
	\frac{1}{N!} \sum_{\vec{q}} \delta_{\sum_i q_i -2L} \prod_i p(q_i) \sum_{
		\begin{array}{c} \scriptstyle A_{ii}= \\  \scriptstyle 0,2,4,...\\  \scriptstyle i=1..N \end{array}
		} 
	\sum_{
	  \begin{array}{c} \scriptstyle A_{ij}= \\ \scriptstyle 0,1,2,... \\ \scriptstyle i>j \end{array}
		} \prod_i \frac{\delta_{\sum_{k<i} A_{ik} + A_{ii} +\sum_{k>i} A_{ki} -q_i}}{A_{ii}!!} \prod_{i>j} \frac{1}{A_{ij}!}.
\eq
Using the standard integral representation of the delta function we can rewrite all sums over $A_{ij}$ as
\bq
	\oint \prod_i \frac{\dd z_i}{2\pi i} \sum_{A_{ii}=0,2,4,\dots} \frac{z_i^{-1-q_i+A_{ii}}}{A_{ii}!!} 
	\sum_{A_{ij}=0,1,2,\dots, i>j} \prod_i z_i^{\sum_{k<i} A_{ik} +\sum_{k>i} A_{ki}} \prod_{i>j} \frac{1}{A_{ij}!}.
\eq
The sum over diagonal elements gives a product of factors $e^{z_i^2 /2}$. The sum over $A_{ij}$ is also easy to calculate and reads $\prod_{i>j} e^{z_i z_j}$. Putting the two results together we find the following factor: $e^{\sum_{i,j} z_i z_j/2}$. Therefore, the partition function is
\bq
	Z(N,L) = \frac{1}{N!} \sum_{\vec{q}} \delta_{\sum_i q_i -2L} \oint \prod_i \frac{\dd z_i}{2\pi i} p(q_i) z_i^{-1-q_i} e^{\frac{1}{2}\left(\sum_{i} z_i\right)^2}.	\label{eq2:znl00}
\eq
The last, quadratic term can be expanded by means of the Hubbard-Stratonovich identity:
\bq
	\exp\left(\frac{A^2}{2}\right) = \frac{1}{\sqrt{2\pi}} \int \dd x \exp\left( -\frac{x^2}{2}- A x\right). \label{eq2:hs}
\eq
The discrete delta giving conservation of links can be written as a contour integral, so we get
\bq
	Z(N,L) = \frac{1}{N!} \oint \frac{\dd y}{2\pi i} y^{-1-2L} \int \frac{\dd x}{\sqrt{2\pi}} e^{-x^2/2} \left[ \oint \frac{\dd z}{2\pi i} \sum_q p(q) \left(\frac{y}{z}\right)^q \frac{e^{xz}}{z} \right]^N.
\label{eq2:xxx}
\eq
The integral over $\dd z$ yields $(xy)^q/q!$. Changing variables: $y\to v=xy$ and changing the order of integration over $\dd x$ and $\dd v$ we immediately obtain
\bq
	Z(N,L) = \frac{1}{N!} \int \frac{\dd x}{\sqrt{2\pi}} e^{-x^2/2} x^{2L} \oint \frac{\dd v}{2\pi i} \left[ \sum_q p(q) \frac{v^q}{q!}\right]^N
	v^{-1-2L}
	= \frac{(2L-1)!!}{N!} \oint \frac{\dd v}{2\pi i} v^{-1-2L} F^N(v), \label{eq2:partps}
\eq
where we have defined the following generating function for weights $p(q)$:
\bq
	F(v) = \sum_q p(q) \frac{v^q}{q!}. \label{eq2:fv}
\eq
Up to now, these results are strict. However, the integral over $\dd v$ is often hard to calculate for finite $N,L$. Fortunately, the partition function (\ref{eq2:partps}) can be calculated in the thermodynamical limit. The saddle-point integration yields:
\bq
	\ln Z(N,L) \approx N \ln F(v_0) - (2L+1)\ln v_0 + \ln \frac{(2L-1)!!}{N!} + \dots, \label{eq2:logznl}
\eq
with $v_0$ being a solution to the equation:
\bq
	v_0 \frac{F'(v_0)}{F(v_0)} = \bar{k}.
\eq
We are now ready to calculate $\Pi(k)$. Since all nodes are equivalent in the equilibrated network with product weights (\ref{eq2:prod}), the degree distribution can be obtained by a simple differentiation of the partition function:
\bq
	\Pi(k) = \frac{p(k)}{N Z(N,L)} \frac{\partial Z(N,L)}{\partial p(k)} =  p(k) \frac{1}{N} \frac{\partial\ln Z(N,L)}{\partial p(k)},
	\label{eq2:pikdef}
\eq
and by applying Eq.~(\ref{eq2:logznl}) we finally arrive at
\bq
	\Pi(k) = \frac{p(k) v_0^k}{k! F(v_0)}. \label{eq2:pikv0}
\eq
This result has been derived in the thermodynamical limit for the canonical ensemble of pseudographs. If we try to do the same for simple graphs, the calculation of the partition function is more complicated, because if we exclude multiple and self-connections, the weight of each labeled graph is identical, and the entries $A_{ij}$ of the adjacency matrix assume now only two possible values $0$ and $1$. This leads to a change of the factor $e^{\frac{1}{2}\left(\sum_{i} z_i\right)^2}$ in Eq.~(\ref{eq2:znl00}) to 
\bq
	\prod_{i>j} (1+z_i z_j) = e^{\frac{1}{2} \sum_{i\neq j} \ln (1+z_i z_j)}.
\eq
The integrals over $\dd z_i$ cannot be done in a straightforward way. One can, however, use the following expansion:
\bq
	\sum_{i\neq j} \ln (1+z_i z_j) = \sum_{n=1}^\infty \frac{(-1)^n}{n} \sum_i z_i^{2n} - \sum_{n=1}^\infty \frac{(-1)^n}{n} \left(\sum_i z_i^n \right)^2, \label{eq2:pert}
\eq
and, in order to get the factorization of $z_i$'s, to apply the H-S identity (\ref{eq2:hs}) to each quadratic term in the second sum over $n$. This leads to the following, rather formal, integral:
\ba
	Z(N,L) = \frac{1}{N!} \oint \frac{\dd y}{2\pi i} y^{-1-2L} \int \dd x_1 \cdots \int \dd x_\infty \left(\prod_{n=1}^\infty \sqrt{\frac{-n(-1)^n}{2\pi}} e^{\frac{n(-1)^n}{2}x_n^2}\right) \nonumber \\
	\times \oint \frac{\dd z_1}{2\pi i} \cdots \oint \frac{\dd z_N}{2\pi i} \sum_{q_1,...,q_N} \left(\prod_i p(q_i) y^{q_i} z_i^{-1-q_i} \prod_{n=1}^\infty e^{\frac{(-1)^n}{2n}z_i^{2n} + x_n z_i^n } \right). \label{eq2:znlprodi}
\ea
If we look at Eq.~(\ref{eq2:pert}) as a perturbative expansion, the integral over $\dd x_n$ gives a ``product'' correction of $n$th order to $Z(N,L)$. Taking only first few terms in $n$ we get an approximation of $Z(N,L)$, but because we know that $Z(N,L)$ is finite, it is not necessary to take all of them. If we restrict ourselves only to the first order $n=1$ we get
\bq
	\frac{1}{N!} \oint \frac{\dd y}{2\pi i} y^{-1-2L} \frac{1}{\sqrt{2\pi}} \int \dd x_1 e^{-x_1^2/2} \left[ \sum_q p(q) \oint \frac{\dd z}{2\pi i} \left(\frac{y}{z}\right)^q \frac{e^{-z^2/2+x_1 z}}{z} \right]^N. \label{eq2:part1ord}
\eq
This is indeed a partition function for pseudographs but with single self-connections excluded. Multiple connections and double, triple, etc. self-connections are still present. Changing variables $y\to v=x_1 y$ and evaluating the integral over $\dd z$ we have
\bq
 \frac{1}{N!} \frac{1}{\sqrt{2\pi}} \int \dd x_1 e^{-x_1^2/2} x_1^{2L} \oint \frac{\dd v}{2\pi i} v^{-1-2L} \left[ \sum_q p(q) v^q \sum_{m=0}^\infty \left(-\frac{1}{2x_1^2}\right)^m \frac{1}{m! (q-2m)!} \right]^N,
\eq
and because the integral over $\dd x_1$ is dominated by the region $x_1\sim \sqrt{L} \propto \sqrt{N}$, only the first term in the sum over $m$ contributes in the limit of $N\to\infty$. We end up with a partition function like in Eq.~(\ref{eq2:partps}) for pseudographs. As a by-product we can also estimate the characteristic value of $z\approx (q+1)/x_1\sim 1/\sqrt{N}$ in the integral over $\dd z$ in Eq.~(\ref{eq2:part1ord}). Let us consider now the product of integrals in Eq.~(\ref{eq2:znlprodi}) and try to estimate the characteristic values of $x_1,\dots,x_\infty$ and $z_1,\dots,z_N$ in order to convince ourselves that integrals over $\dd x_2,\dd x_3,\dots$ can be neglected in the thermodynamical limit. Assuming that in the limit $N\to\infty$ the integral is dominated by a single saddle point, we must find the maximum of the function:
\bq
	\sum_{n=1}^\infty \left( \frac{n(-1)^n}{2} x_n^2 + \sum_i \frac{(-1)^n}{2n} z_i^{2n} + x_n z_i^n \right) - \sum_i (1+q_i) \ln z_i.
\eq
The differentiation with respect to $z_i$ and $x_n$ gives the following set of equations:
\ba
	\forall n=1,\dots,\infty:&& (-1)^n n x_n + \sum_i z_i^n = 0, \\
	\forall i=1,\dots,N:&& \sum_n  (-1)^n z_i^{2n} + n x_n z_i^n = q_i+1.
\ea
The integrals over $\dd z_i$ as well as the sums over $q_i$ factorize, thus we can skip indices $i$ because characteristic values of all $z_i$'s and all $q_i$'s are equal. This allows for solving these equations. We have
\ba
	|z| &\sim & \frac{1}{\sqrt{N}}, \\
	|x_n| &\sim & \frac{N^{1-n/2}}{n},
\ea
so $x_1\sim \sqrt{N}$ but $x_n$'s for higher $n$ tend to zero in the thermodynamic limit. This means that the only significant contribution to Eq.~(\ref{eq2:znlprodi}) is from the integral over $\dd x_1$. Therefore, Eq.~(\ref{eq2:part1ord}) is a good approximation. We notice that in the limit $N\to\infty$ this equation is identical to Eq.~(\ref{eq2:xxx}) which we had before for pseudographs. Thus the degree distribution $\Pi(k)$ is again given by Eq.~(\ref{eq2:pikv0}). Let us now discuss some consequences of that formula. First, for $p(q)=1$ the generating function $F(v)=e^v$ and $\Pi(k)$ is Poissonian as it should be for equally weighted ER graphs. Second, to get any desired degree distribution $\Pi(k)$ one should take $p(q)=q!\Pi(q)$ and tune the average degree $\bar{k}$ so that $v_0=1$:
\bq
	\bar{k} = \bar{k}_c \equiv F'(1)/F(1).
\eq
In other words, the number of links and nodes must be carefully balanced to obtain a desired distribution $\Pi(k)$: $2L/N=\bar{k}=\sum_k k \Pi(k)$ in the limit of large graphs. For instance, to get a power-law distribution one should take $p(q)\sim q! q^{-\gamma}$ and adjust $N,L$ carefully. A very important example is the distribution for Barab\'{a}si-Albert model \cite{ref:barab}:
\bq
	\Pi(k) = \frac{4}{k(k+1)(k+2)}	\label{eq2:BAtree}
\eq
for $k>0$ and $\Pi(0)=0$, which will be discussed in next section. In order to obtain the ensemble with $\Pi(k)$ given by the above formula, one has to choose $p(k) = k! \frac{4}{k(k+1)(k+2)}$ for $k=1,2,\dots$, and $p(0)=0$. The mean of the distribution (\ref{eq2:BAtree}) is $\bar{k}_c=2$ so we have to take $N=L$ to adjust $\bar{k}$ to this value. If $L$ is too small, the degree distribution  falls off exponentially for large degrees as one can see from Eq.~(\ref{eq2:pikv0}), because then the saddle point $v_0<1$. When one exceeds the critical degree $\bar{k}_c$, the saddle-point approximation is no longer valid\footnote{See the discussion of the condensation in balls-in-boxes model in Sec. 3.2.2.} and the behavior depends on whether we consider simple- or pseudographs. For simple graphs, the degree distribution has no longer a power-law tail, but has a more complicated form. We must remember that for simple graphs Eq.~(\ref{eq2:pikv0}) is only an approximation. A very interesting behavior is observed for pseudographs. It has been shown \cite{ref:bbw} that a surplus of links condenses on a single node, thus $\Pi(k)$ has the same power-law distribution as for the critical degree $\bar{k}_c$, but with an additional delta peak whose position moves linearly with the system size $N$. This is the same phenomenon as in the ``Backgammon condensation'' taking place in the balls-in-boxes model \cite{ref:bbj}. We shall devote one section of Chapter 3 to this problem, so now we will only mention that this is related to the divergence of the series (\ref{eq2:fv}) when $2L/N$ exceeds the threshold $\bar{k}_c$. In fact, we shall see in Chapter 3 that the partition function for the balls-in-boxes model is given by the same formula as Eq. (\ref{eq2:partps}) for pseudographs and therefore the model can be mapped onto the balls-in-boxes model.

There is also another problem which should be mentioned here. Equation (\ref{eq2:pikv0}) is valid only for infinite sparse graphs, that is for $N\to\infty$ and $\bar{k}$ fixed. For finite $N$, the node degree distribution $\Pi(k)$ deviates from the limiting shape due to finite-size corrections, which are particularly strong for fat-tailed distributions $\Pi(k) \sim k^{-\gamma}$. As a result of structural constraints, the maximal node degree cannot be $\sim N$ but often it scales as some power of $N$ smaller than one. Corrections to the scale-free degree distribution for finite networks will be extensively discussed in section 3.1.

Let us mention also a particularly important subset of weighted graphs, namely weighted trees \cite{ref:bbjk}. Because of their special structure (no cycles), many results can be obtained analytically. For instance, for trees with product weights, similarly as for pseudographs one can calculate the expression for $\Pi(k)$:
\bq
	\Pi(k) = \frac{p(k) v_0^{k-1}}{(k-1)! F(v_0)}, \label{eq2:pikv0tree}
\eq
where the generating function $F(v)$ is now given by
\bq
	F(v) = \sum_{q=1}^\infty p(q) \frac{v^{q-1}}{(q-1)!}. \label{eq2:fvtree}
\eq
Therefore to get a power-law degree distribution one has to take $p(k)\sim (k-1)! k^{-\gamma}$. Similarly, one can calculate correlations \cite{ref:pbialas}:
\bq
	\eps(k,q) = \frac{\Pi(k)\Pi(q)(k+q-2)}{2},
\eq
and hence the assortativity coefficient from Eq.~(\ref{eq1:ac}), which for trees with BA degree distribution reads
\bq
	\mathcal{A} = \frac{2(69-7\pi^2)}{21-2\pi^2} \approx -0.1384,
\eq
showing that this network is disassortative. Trees will be more throughly discussed in section 2.3 in the context of comparing the properties of equilibrated and causal networks.

At the end we shall mention that one can define more complicated weights than those given by Eq.~(\ref{eq2:prod}). A natural candidate for a weight to generate degree-degree correlations on the network is the following choice \cite{ref:bl1,ref:b}:
\bq
	W(\alpha) = \prod_{l=1}^L p(k_{a_l},k_{b_l})  ,
\eq
where the product runs over all edges of graph, and the weight $p(k_a,k_b)$ is a symmetric function of degrees of nodes $a,b$ at the endpoints of link. One can choose this function to favor assortative or disassortative behavior \cite{ref:bl1,ref:b,ref:n3,ref:bp,ref:d2}. Similarly, one can tune the weights to mimic some other functional properties of real networks, like for example higher clustering \cite{ref:n2,ref:bjk2,ref:bjk3,ref:d1,ref:pn2}.

\subsection{Monte Carlo generator of equilibrated networks}
Only for a few models of random graphs, closely related to ER graphs, one can calculate almost all quantities of interest analytically. This is not the general case for weighted networks like those presented in the previous section. In some cases it is useful to support the discussion with computer simulations. Various methods have been proposed for generating random graphs, but usually each of them works only for one particular model or its variations. In this section we will describe a very general Monte Carlo method which allows one to study a wide class of random weighted graphs. The idea standing behind this method is to sample the configuration space with probabilities given by their statistical weights. Unfortunately, there is no general and efficient procedure that picks up an element from a large set with the given probability. The most naive algorithm in which one picks up an element uniformly and then accepts it with the probability proportional to its statistical weight has a very low acceptance rate when the size of the set is large. Because the number of graphs grows exponentially or faster\footnote{For instance, for ER model it grows faster than exponentially which results in a non-extensive entropy of graphs, see Sec. 2.1.3; for introduction on counting graphs see also the reference \cite{ref:enum}.}, one clearly sees that another idea must be applied. In this section we will discuss such an idea which is derived from a general framework of dynamical Monte Carlo techniques.

The idea is to use a random walk process, which explores the set of graphs, visiting different configurations with probability proportional to their statistical weights. Such a process is realized as a Markov chain (process) which has a unique stationary state with the probability distribution proportional to $W(\alpha)$. The Markov chain is defined by specifying transition probabilities $P(\alpha\rightarrow \beta)$ to go in one elementary step from a configuration $\alpha$ to $\beta$. The elementary step is a kind of transformation which carries over the current graph into another one. A convenient way to store these probabilities is to introduce a matrix $\pp$, called a Markovian matrix, with entries $P_{\alpha\beta} \equiv P(\alpha\rightarrow \beta)$. For a stationary process, the transition matrix $\pp$ is constant during the random walk. The process is initiated from a certain graph $\alpha_0$ and then elementary steps are repeated producing a sequence (chain) of graphs $\alpha_0\rightarrow \alpha_1 \rightarrow  \alpha_2 \rightarrow \dots$.
The probability $p_\beta(t+1)$ that a graph $\beta$ is generated in the $(t+1)$th step is given by:
\bq
	p_\beta(t+1) = \sum_\alpha p_\alpha(t) P_{\alpha \beta} ,
\eq
which can be rewritten as a matrix equation:
\bq
	\p (t+1) = \pp^{\tr} \p (t)  ,\label{interp}
\eq
where $\pp^{\tr}$ denotes the transpose of ${\pp}$ and $\p$ is a vector of elements $p_\alpha$. From general theory of Markovian matrices \cite{ref:markov} we know that the stationary state, characterized by the equation: $\p(t+1) = \p(t)$, corresponds to the left eigenvector of $\pp$ to the eigenvalue $\lambda=1$. If the process is ergodic, which means that any configuration can be reached by a sequence of elementary steps starting from any initial graph, and if the transition matrix fulfills detailed balance condition:
\bq
	\forall \alpha,\beta: \quad  W(\alpha) P_{\alpha \beta} = W(\beta) P_{\beta \alpha}   ,\label{eq2:db}
\eq
then the stationary state approaches the desired distribution: $p_\alpha(t) \rightarrow W(\alpha)/Z$ for $t \rightarrow \infty$. In other words, in the limit of infinite Markov chain, the probability of occurrence of graphs becomes proportional to their statistical weights and is independent of the initial graph. However, one must be careful while generating relatively short chains. First of all, the probabilities can strongly depend on the initial state, and one has to wait some time before one starts measurements, to ``thermalize'' the system, i.e. to reach ``typical'' graphs in the ensemble. Second of all, consecutive graphs in the Markov chain may be correlated, especially when the elementary step is only a local update. Therefore one has to find a minimal number of steps for which one can treat measurements on such graphs as independent.

Among many possible choices for probabilities $P_{\alpha\beta}$, which lead to the same stationary distribution, we shall use here the well-known Metropolis algorithm \cite{ref:metrop}, based on the following transition probability:
\bq
	P_{\alpha\beta} = \min\left\{1,\frac{W(\beta)}{W(\alpha)}\right\}  .\label{metrop}
\eq
The algorithm works as follows. For the current configuration $\alpha$ one proposes to change it to a new configuration $\beta$ which differs slightly from $\alpha$ and then one accepts it with the Metropolis probability (\ref{metrop}). Repeating this many times one produces a chain of configurations. The proposed elementary modifications (steps) should not be too large because then one risks that the acceptance rate would be small. Therefore, all algorithms which we propose below attempt in a single step to introduce only a small change to the current graph, by rewiring only one or two links. 

\begin{figure}
\center
\psfrag{i}{$i$} \psfrag{j}{$j$} \psfrag{k}{$n$}
\includegraphics[width=13cm]{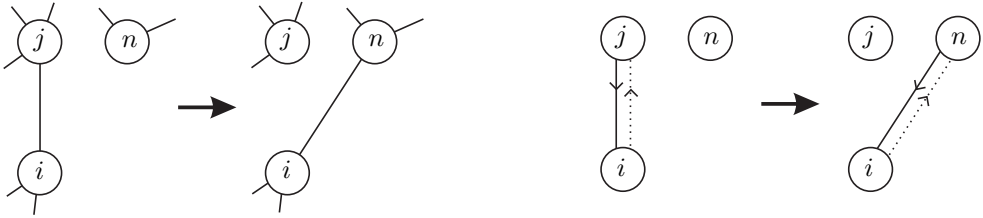}
\caption{\label{fig:tmove}The idea of ``T-move'': a random link (solid line) is rewired from vertex $j$ to a random vertex $n$ (left-hand side). Alternatively (right-hand side) a random, oriented link (dotted line) is rewired from vertex of its end $j$ to a random vertex $n$. The opposite link $j\rightarrow i$ is simultaneously rewired.}
\end{figure}

Let us try to apply these ideas to write a Monte Carlo algorithm for generating weighted graphs from the canonical ensemble. A good candidate for elementary transformation of a graph is rewiring of a link called ``T-move'' (see Fig.~\ref{fig:tmove}), because it does not change $N$ and $L$, fixed in the canonical ensemble. We choose a link $ij$ and a vertex $n$ at random, and rewire one of the endpoints of the link, say $j$, to $n$, forming a new link $in$ which replaces the old one $ij$. Sometimes it is easier to think about a rewiring of an oriented link $i\to j$ to $i\to n$, and simultaneously, $j\to i$ to $n\to i$. If the link $in$ is already present, or if $n=i$ we reject this move to prevent from forming a multiple- or self-connection. 
Then, each move is accepted with the Metropolis probability (\ref{metrop}). In a special case of the ensemble of Erd\"os-R\'enyi
graphs this probability is equal to one since every graph has the same weight. So ER graphs can be alternatively obtained by simple rewirings of any other graph - we call this process ``thermalization''.
The thermalization (homogenization) can be used to generate any ensemble of equilibrated networks. In fact, one could define equilibrated networks as graphs, which can be produced in a procedure like this, if weights do not depend on nodes' labels. This process destroys any correlations which might be present due to causal growth of the initial network.

One can show \cite{ref:homnasz} that, indeed, this algorithm produces labeled graphs with desired probabilities. We skip here the details. We would like, however, to point out two difficulties which can be encountered. First, it is not clear whether the ergodicity is not broken in the limit  $N\to\infty$ for models where the number of configurations grows  with $N$ faster than exponentially. Second, for some classes of (unphysical) weight functions, a local algorithm may not be ergodic. Consider for example weight functions $W(\alpha)$ of the form (\ref{eq2:prod}), with $p(k)$ being a function which is strictly positive on a support which has a gap in the middle -- an interval $k\in (k_1,k_2)$, where $p(k)=0$. In other words, there are no configurations in this ensemble which have a node with degree $k\in (k_1,k_2)$. A single rewiring can change degrees only by $\pm 1$, so it is not possible to change the value of $k$ from $k<k_1$ to $k>k_2$ since it would have to go through the forbidden region $(k_1,k_2)$. In this case, in order to avoid the difficulty one would have to invent an algorithm which is able to significantly change $k$ in a single move, to jump from one to another part of the support of the weight function $p(k)$.
We shall not, however, consider such unphysical weights $p(k)$. 
For weights, which are physically important, the support of the weight function is connected. In this case the acceptance probability reads
\bq
  P_a(\alpha\rightarrow\beta) = \min\left\{1,\frac{W(\beta)}{W(\alpha)}\right\} = \min\left\{1,\frac{p(k_j-1)p(k_n+1)}{p(k_j)p(k_n)}\right\}
  = \min\left\{1, \frac{w(k_n)}{w(k_j-1)}\right\},
\eq
where we have introduced an auxiliary function:
\bq
	w(k) = \frac{p(k+1)}{p(k)}. \label{eq2:wwdef}
\eq
The degrees $k_j,k_n$ are taken from the current graph $\alpha$. In the computer algorithm we prefer to use the weight function $w(k)$ instead of $p(k)$ to reduce computational cost and round-off errors. In fact, $w(k)$ can be exactly calculated for many important $p(k)$'s, which we are interested in. For example, to get the BA degree distribution in simple graphs, according to Eq.~(\ref{eq2:BAtree}), we have to choose $p(k)=k! \Pi_{BA}(k)$ and hence
\bq
	w(k) = \frac{k(k+1)}{k+3},
\eq
while for trees, because of the factor $(k-1)!$ in Eq.~(\ref{eq2:pikv0tree}),
\bq
	w(k) = \frac{k^2}{k+3}.
\eq

The rewiring procedure described above does not change $N$ and $L$. If we want to simulate weighted graphs from the grand-canonical ensemble, we have to choose another transformations which change the number of links $L$. Natural candidates for such transformations are two reciprocal transformations: adding and deleting a link. In order to add a link we have to choose two vertices to which the addition is attempted. To remove a link we pick up one link out of all $L$ existing in the graph.
These two transformations must be carefully balanced in order to get graphs with correct probabilities. If the frequency of the two transformations is the same, then the acceptance probabilities for each of them are given by \cite{ref:homnasz}
\bq
P_{\rm add}(\alpha\rightarrow\beta) = \min\left\{1, \exp(-\mu) \, \frac{N^2}{2(L_\alpha+1)} \, \frac{W(\beta)}{W(\alpha)} \right\} , \label{eq2:gcan_rown}
\eq
and
\bq
P_{\rm del}(\beta\rightarrow\alpha) = \min\left\{1, \exp(+\mu)  \, \frac{2L_\beta}{N^2} \, \frac{W(\alpha)}{W(\beta)}\right\} , \label{eq2:gcan_rown2}
\eq
respectively. Here $\mu$ is the chemical potential for links, defined in Eq.~(\ref{grand}) and  chosen to obtain a desired average number of links\footnote{For complicated weights, when analytical calculations of the correspondence $\left<L\right>\leftrightarrow \mu$ is impossible, one can tune $\mu$ during the simulation to obtain desired number of links.} $\left<L\right>$. As before, if we want to produce only simple graphs we must eliminate moves leading to self- or multiple connections.
One could modify the algorithm in many ways. For example, one could, instead of picking up a link as a candidate for removing, pick up two nodes at random and if there is a link between them, remove it. Then the fractions $N^2/2(L+1)$ and $2L/N^2$ would disappear from equations (\ref{eq2:gcan_rown}) and (\ref{eq2:gcan_rown2}). This would not change the probabilities of graph's occurrence, but it would affect the acceptance rate. For sparse networks such a modified algorithm is worse than the previous one because the chance that there is a link between two randomly chosen nodes is very small and for most of the time the algorithm would do nothing except looking for links that can be removed. On the other hand, the acceptance rate for the original algorithm is finite for $N\to\infty$ since then $\mu$ behaves as $\ln N$ and the factor $e^{\mu} 2L/N^2$ is of order $1$.

Let us consider now a version of this algorithm suitable for the product weights (\ref{eq2:prod}). The probability of acceptance of a new configuration by adding or removing a link between $ij$ reads:
\ba
\min \left\{ 1, \frac{N^2}{2(L+1)} 
e^{-\mu} w(k_i)w(k_j) \right\} & 
\mbox{for addition a link,} \nonumber \\
\min \left\{ 1, \frac{2L}{N^2} 
e^\mu \frac{1}{w(k_i-1)w(k_j-1)} \right\} &
\mbox{for deleting a link,} \nonumber
\ea
where $L$ and $k_i,k_j$ refer to the current configuration and $w(k)$ is given by Eq.~(\ref{eq2:wwdef}).

\begin{figure}
\center
\psfrag{i}{$i$} \psfrag{j}{$j$} \psfrag{k}{$l$} \psfrag{l}{$n$}
\includegraphics[width=5cm]{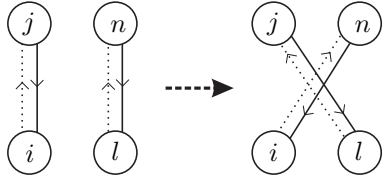}
\caption{\label{fig:xmove}The idea of ''X-move'': two oriented links (dotted lines) $ij$ and $ln$ chosen in a random way are rewired, exchanging their endpoints. Then the opposite links (solid lines) are also rewired. }
\end{figure}

Finally, let us say some words about the generation of graphs from the micro-canonical ensemble. Inspired by the Maslov-Sneppen algorithm preserving node degrees, as a local update we choose simultaneous rewirings shown in Fig.~\ref{fig:xmove}. We shall call this combination ``X-move''.
At each step of this algorithm one picks up two random links: $ij$ and $ln$, and rewires them to $in$ and $lj$. In case of the Maslov-Sneppen algorithm \cite{ref:masl}, the functional weight is $W_\alpha=1$ and hence all rewirings are accepted. In the general case, one can use this algorithm to generate graphs whose statistical weights depend for instance on the number of triangles, to get a high clustering coefficient, or to produce some higher-order correlations between nodes \cite{ref:pietsch}. The motivation is similar to that given by Maslov and Sneppen, namely if one tries to determine relations between the abundance of structural motifs and the functionality of network, it is very important to construct randomized networks which could serve as a benchmark.

We described the algorithms presented here and their implementation in \cite{ref:naszcpc}. Because we often have to do with sparse graphs, it is not needed to keep the whole adjacency matrix in computer memory. The data structure that we developed allows us to encode and simulate networks of size of order $10^6$.

\section{Causal (growing) complex networks}

In the previous section we discussed equilibrated networks, which can be constructed in a sort of thermalization or homogenization process. Clearly for such graphs, if they are labeled, a permutation of nodes' labels leads to the same set of graphs. In this section we shall discuss another kind of graphs, generated in a process of growth. A common feature of these graphs is that there is a natural labeling of nodes which corresponds to the order in which they were added to the graph. We call this labeling ``causal'', since it is always obvious which node is an ancestor of which. The corresponding graphs will be called {\bf causal networks}. The causality introduces a restriction on the number of ways in which the graph can be labeled. As we shall see, this restriction very strongly affects properties of typical networks in the ensemble.

In this section we shall discuss some famous models of growing networks. These models are more popular than equilibrated networks presented above and, indeed, they were first models reproducing many properties of real network. Although in some models the rules governing the process of growth may look somewhat artificial, it is instructive to study how methods of statistical physics can be applied to causal networks. Because of the growing nature of these networks, the rate equation approach is particularly well suited to study them. We will see, however, that also the concept of statistical ensemble can be very helpful in order to understand some features of these networks. 

\subsection{Models of growing networks}

As a first example of growing network we shall discuss the {\bf Barab\'{a}si-Albert (BA) model} introduced in a very seminal paper \cite{ref:bamodel}. This model triggered enormous activity in the field of complex networks. Similar models were proposed in different contexts and discussed several times in the past (for review see e.g. \cite{ref:willis}). The model has two basic ingredients: growth and preferential attachment. The latter means that new nodes added to the system prefer to attach to nodes with higher degrees. In effect, high degrees are further increased and become even higher\footnote{This is sometimes called St. Matthew's effect: ``For unto every one that hath shall be given, and he shall have abundance: but from him that hath not shall be taken away even that which he hath.'' (Matthew XXV:29, KJV).}. The model is defined as follows. Starting from a complete graph with $n_0$ nodes, at each step a new node is introduced and joined to $m$ previously existing nodes with the probability proportional to the degree of the node to which a new link is established. 
One can easily program this procedure on a computer, adding nodes one by one and attaching them according to the preferential attachment rule. There is also a slightly different version of the algorithm, a more tricky one, which instead of focusing on the nodes uses links as elementary objects. It is more effective, so let us shortly describe it. Each link $ij$ is viewed in this algorithm as a couple of directed links $i\to j$ and $j\to i$. In the algorithm one picks up  at random a directed link  and chooses the node which is at the endpoint of this link as a node to which a new link is going to be attached. The preferential attachment rule is in this way simply realized, because the number of links pointing onto a node with degree $k$ is equal to $k$. After $t$ steps of nodes' addition, the network consists of $n_0+t$ nodes and $mt+n_0(n_0+1)/2$ edges. For $m=1$, the graph generated by this procedure consists of trees planted on the initial graph. If the initial graph is a tree, so is the whole graph.

Later on we shall see that the degree distribution falls asymptotically as $k^{-3}$. In the limit $N\to\infty$ the distribution reads \cite{ref:dr2}
\bq
	\Pi(k) = \frac{2m(m+1)}{k(k+1)(k+2)} \Theta(k-m), \label{eq2:bafirst}
\eq
where $\Theta(x)$ is the step function: $\Theta(x)=1$ for $x\ge 0$, $\Theta(x)=0$ for $x<0$. By construction, nodes of degree smaller than $m$ are absent.
The degree distribution (\ref{eq2:bafirst}) is in accordance with distributions observed for some real networks like the citation network. The exponent $\gamma=3$ cannot be tuned in this version of the model. As we shall see below, a slight modification of the attachment rule will do the job. The next important property of the BA network is that the diameter grows as $\sim\ln N$, so it is a small-world. The clustering coefficient is rather small. For $m=1$, $C=0$ because the graph is essentially a tree. For $m>1$ many triangles appear\footnote{E.g. for $m=2$ for each new node one new triangle is also introduced.}, but their number is small in comparison to the number of connected triples in the limit $N\to\infty$. There are obvious correlations on BA networks between the age and the degree of nodes: the older node is, the higher degree it has. This is an effect of a pure growth in absence of any rewiring of links. In fact, this age-degree correlation is not observed in the WWW, for which the model was originally designed, because there are new web pages having sometimes more links than older ones.

As mentioned, many refinements have been introduced to the BA model to account for some of the experimentally observed facts. In particular, one can make the power-law exponent tunable by a simple modification of the attachment rule as proposed by Dorogovtsev, Mendes and Samukhin \cite{ref:dr2}. Here we shall refer to this model as to the {\bf DMS model}\footnote{There are also other models proposed by those authors called DMS models in the literature.} or as to the BA model with initial attractiveness. The algorithm is similar to that for the BA model. The only difference is that now a new node chooses the older one to which it creates a link, with probability $A_k$, called attachment kernel, proportional to its degree plus some constant:
\bq
	A_k = \frac{k+a_0}{\sum_i k_i+a_0}, \label{eq2:prefaa}
\eq
where $k$ is the degree of the old node and $a_0$ is called initial attractiveness. The model can be solved in the thermodynamic limit \cite{ref:dr2}. The degree distribution for $m=1$ reads
\bq
	\Pi(k) = \frac{(2+a_0)\Gamma(3+2a_0)}{\Gamma(1+a_0)}\frac{\Gamma(k+a_0)}{\Gamma(k+3+2a_0)}, \label{eq2:pikgena0}
\eq
that is $\Pi(k)\sim k^{-\gamma}$ with $\gamma=3+a_0$. The model can reproduce any power-law exponent larger than $2$ ($a_0>-1$), and therefore it can be adjusted to experimentally observed degree distributions for real-world networks. One can summarize this part of the discussion by saying that DMS model became very popular because of three important properties:  i) it yields scale-free networks with tunable exponent $\gamma$, ii) the networks are small-worlds, iii) the model is easy to handle in the numerical and analytical treatment.
Actually, DMS networks can be easily generated but not so easy as BA ones.
The innocently looking term $a_0$ in the attachment kernel changes the algorithm complexity, because one cannot apply the trick with picking up directed links at random instead of nodes. One has to work with nodes and choose them with a probability changing after each step, which increases the computational cost. Fortunately, it was shown in \cite{ref:kr} that the model with $m=1$ is equivalent to a model of {\bf growing network with re-direction (GNR)}.  The GNR network is constructed as follows. Starting from some small initial graph like in the BA model, at each time step one chooses a node $i$ with equal probability from the set of existing nodes. Then one introduces a new node which is attached with probability $1-r$ to $i$, and with probability $r$ to its ancestor\footnote{Because the network is growing (causal), one can always decide which node is older and fix the ancestor-descendant hierarchy.}. With the choice $r=1/(a_0+2)$ the GNR model is equivalent to the DMS tree model with initial attractiveness $a_0$. 

One can consider even more general attachment kernels than Eq.~(\ref{eq2:prefaa}). For instance, one can assume that $A_k$ behaves asymptotically as $k^\alpha$ for large $k$. When $\alpha<1$ that is for sub-linear kernels, the degree distribution is exponentially suppressed \cite{ref:kr}. When $\alpha>1$, links tend to condense on one or more nodes, depending on the value of $\alpha$: for instance for $\alpha>2$ almost all links condense on a single node. This is the ``winner takes all'' situation. We have mentioned a similar behavior in the previous section while discussing pseudographs, but there only a finite fraction of links condensed. The situation presented here is more similar to the condensation of balls on inhomogeneous networks which will be discussed in section 3.2.2.

\subsection{Rate equation approach}
\begin{figure}
\center
\psfrag{xx}{$k$} \psfrag{yy}{$\Pi(k)$}
\includegraphics*[width=15cm, bb=11 242 722 527]{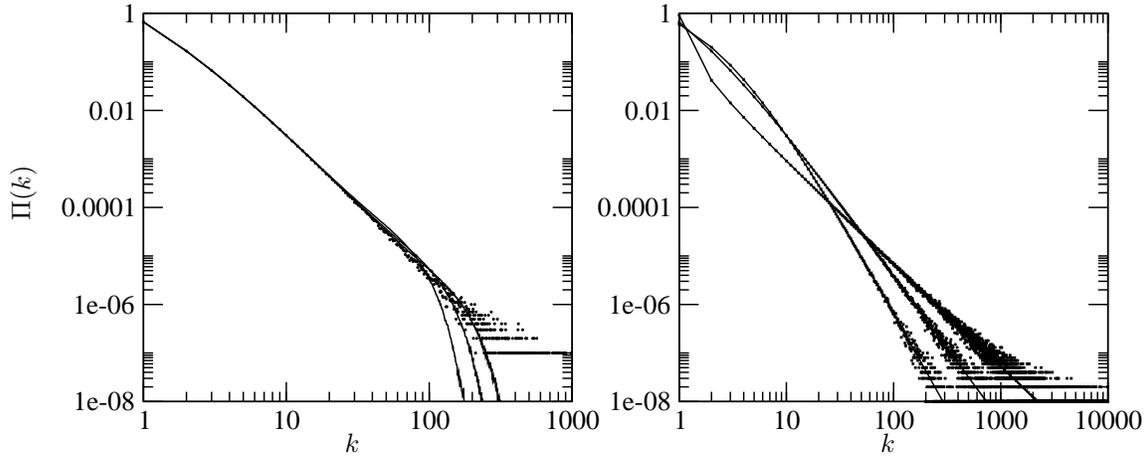}
\caption{\label{fig:bamodel}Left: Degree distribution for pure BA model with $m=1,n_0=2$ for networks of different sizes. Solid lines from left to right: $N=1000,2000,4000$, averaged over $10^6$ generated networks. Circles: $\Pi(k)$ for a single network with $N=10^7$. As $N$ grows, plots approach theoretical distribution $\Pi(k)\sim k^{-3}$. One also sees that averaging over the ensemble is (up to finite-size effects) equivalent to averaging over one large network (self-averaging). Right: plots of $\Pi(k)$ for DMS models with $N=10^6$ and various $a_0$, compared to the theoretical distributions (\ref{eq2:pikgena0}). The plots correspond to $a_0=2.1$ (the smallest slope), $a_0=3$ and $a_0=4$ (the largest slope). All results are averaged over 100 networks generated in the GNR model, equivalent to the DMS model.}
\end{figure}

In this subsection we shall discuss rate equations and show how to use them to calculate asymptotic degree distribution for a growing network. We shall follow the approach developed in \cite{ref:kr}. First, for simplicity we shall consider BA model with $m=1$, that is the ensemble of growing trees with linear attachment kernel.
The quantity of interest is $N_k(N)$, the number of nodes having degree $k$ when the total size of the network is $N$.
Assume that the initial graph consists of two nodes joined by an edge. This means that initially we have $N_k(2)=2\delta_{k,1}$. For $m=1$ the growth process does not introduce cycles, so the graph remains a tree. The assumption about the initial configuration is not crucial but it simplifies calculations. At each time step a new node is attached to an old node with probability equal to $k/\sum_q q N_q$, where $k$ is the degree of the old node. Because the sum of all degrees gives $2L$, this probability is simply $\frac{k}{2L}$. 
The process of growth is random, $N_k(N)$ may change by $0$ or $1$.
We can formally write:
\bq
	N_k(N+1) = N_k(N) + \xi(k,N), \label{eq2:xi}
\eq
where $\xi(k,N)$ is a random variable which may assume values $0,1$. Having the probability distribution of $\xi$ we could generate $\xi(k,N)$ at any time step $N$ and simulate the process of growth to get $N_k(N)$. But we are interested not in a particular distribution of degrees for one network, but in ``typical'' properties of all BA graphs. Therefore we should consider the average $\left<N_k(N)\right>$ rather than $N_k(N)$. The average is over an ensemble of all graphs of size $N$ which can be generated by the growth process. One can show \cite{ref:dr2} that this average exists in the limit $N\rightarrow \infty$ and that the system self-averages, which means that for $N\rightarrow \infty$ the averages over the ensemble are equal to the averages over one network picked up from this ensemble.
Taking the average of both sides of Eq.~(\ref{eq2:xi}) we get
\bq
	\left<N_k(N+1)\right> = \left<N_k(N)\right> + \left<\xi(k,N)\right>. \label{eq2:xi2}
\eq
The form of the average of the random variable $\xi$ can be deduced from the process of growth. Let us focus at some node $i$ having degree $k$. As a result of an addition of new node to the network, $i$ can get a new link with probability $k/2L$. Thus the average change of $\left<N_k(N)\right>$ will be $-\left<N_k(N)\right> k/2L$ because it happens only when the new node chooses one of $N_k(N)$ possible nodes with degree $k$. But $\left<N_k(N)\right>$ can also increase by $\left<N_{k-1}(N)\right> (k-1)/2L$ if the new node connects to any node with degree $k-1$. The last contribution to $\left<\xi\right>$ comes from addition of a new node with degree $k=1$ and is equal to $\delta_{k,1}$. Thus the full equation for the rate of change of $\langle N_k(N)\rangle$ reads:
\bq
	\left<N_k(N+1)\right> = \left<N_k(N)\right> + \delta_{k,1} +\frac{k-1}{2(N-1)} \left<N_{k-1}(N)\right> - \frac{k}{2(N-1)} \left<N_{k}(N)\right>,
	\label{eq2:rN}
\eq
where we take advantage of the fact that for trees $L=N-1$. The equation is exact for any $N$, not only in the thermodynamic limit, and could be solved for $\left<N_k(N)\right>$. Using this equation one can also calculate $\Pi(k)\equiv \left<N_k(N)\right>/N$, i.e. the degree distribution averaged over the ensemble of BA tree graphs. As we shall see it is not an easy task (see Chapter 3, Sec. 3.1.2). It can be simplified by neglecting finite-size corrections in the limit of large networks, in which case the degree distribution can be calculated by substituting $\left<N_k(N)\right>=N\Pi(k)$ and assuming that $\Pi(k)$ tends to a stationary state\footnote{One can show that $N_k$'s from Eq.~(\ref{eq2:rN}) grow as $\sim N$ \cite{ref:kr} for large $N$ and therefore $\Pi(k)$ has a stationary state in the thermodynamical limit.}. In this case one gets:
\bq
	\Pi(k) = \delta_{k,1} + \frac{k-1}{2} \Pi(k-1) - \frac{k}{2} \Pi(k) + O(1/N).
\eq
In the thermodynamical limit the term $O(1/N)$ can be neglected. Rearranging this equation:
\bq
	(k+2)\Pi(k) = (k-1)\Pi(k-1)+2\delta_{k,1} ,
\eq
one immediately obtains
\ba
	\Pi(1) &=& 2/3, \\
	\Pi(k) &=& \frac{k-1}{k+2}\Pi(k-1), \quad \forall k>1, \label{eq2:pikk}
\ea
and by iterating Eq.~(\ref{eq2:pikk}) 
one eventually arrives at the following degree distribution for $k\geq 1$:
\bq
	\Pi(k) = \frac{4}{k(k+1)(k+2)}.	\label{eq2:pureBA}
\eq 
Following \cite{ref:kr}, let us apply the same method for a general attachment kernel $A_k$. Now, the probability that a new node will be attached to the older one with degree $k$, is $A_k/A(N)$ where $A(N)$ is a normalization coefficient:
\bq
	A(N) = \sum_k A_k N_k(N).
\eq
In the limit $N\to\infty$, all $N_k\sim N$ and thus we can assume that $A(N)\approx N\eta$ where $\eta$ is some constant to be determined later. Proceeding exactly as above for pure BA model, we get the rate equation in the form:
\bq
	\Pi(k) = \delta_{k,1} + \frac{A_{k-1}}{\eta} \Pi(k-1) -\frac{A_{k}}{\eta} \Pi(k),
\eq
which can be solved with respect to $\Pi(k)$:
\bq
	\Pi(k) = \frac{\eta}{A_k} \prod_{j=1}^k \left(1+\frac{\eta}{A_j}\right)^{-1}.
\eq
The parameter $\eta$ can be obtained from the normalization of the degree distribution: $\sum_k \Pi(k) = 1$. If we now assume a shifted linear kernel like in the DMS model: $A_k = a_0+k$, we find
\bq
	\Pi(k) = \frac{\eta}{a_0+k} \frac{a_0+1}{a_0+1+\eta} \cdots \frac{a_0+k}{a_0+k+\eta} = \frac{\eta}{a_0+k} \frac{\Gamma(a_0+k+1)/a_0!}{\Gamma(a_0+k+\eta+1)/(a_0+\eta)!}, \label{eq2:pikdlug}
\eq
and $\eta=2+a_0$ as follows from $A(t)=\sum_k (k+a_0)N_k(N) = 2L+Na_0$. 
Inserting this into Eq.~(\ref{eq2:pikdlug}) we end up with Eq.~(\ref{eq2:pikgena0}). It reduces to the BA degree distribution (\ref{eq2:pureBA}) for $a_0=0$, that is for the purely linear attachment kernel. In figure \ref{fig:bamodel} we show plots of $\Pi(k)$ calculated analytically using Eq.~(\ref{eq2:pikdlug}) and measured in numerical simulations of networks generated by the GNR version of growing network algorithm for various $a_0$.

The same method allows one to determine $\Pi(k)$ for sub- and super-linear kernels, cited in the previous section, or to calculate degree distribution for non-tree growing networks \cite{ref:dr2,ref:rxb-thesis}. It can also be used to find degree-degree correlations \cite{ref:kr} by writing a rate equation for $N_{k,q}$, the number of nodes with degree $k$ attached to ancestor nodes of degree $q$. The exact result for BA is fairly complicated, but in the limit $k,q\to\infty$ with $y=q/k$ kept fixed, it simplifies to
\bq
	N_{k,q} \cong N k^{-4} \frac{4y(y+4)}{(1+y)^4}.
\eq
This function has a maximum at $y\approx 0.372$ which means that the ancestor node's degree is approximately $37\%$ of its descendant. The correlation function $\eps_{kq}$ defined in Sec. 1.3 and calculated from the formula:
\bq
	\eps(k,q) = \frac{N_{k,q}+N_{q,k}}{L},	\label{eq2:epsBA}
\eq
does not factorize: $\eps(k,q)\neq \eps_{\rm r}(k,q)$ which means that the network is correlated. A similar behavior is observed for shifted attachment kernels. The assortativity coefficient $\mathcal{A}$ defined in Eq.~(\ref{eq1:ac}) can be calculated for pure BA model. From Eq.~(41) in \cite{ref:kr}, and Eq.~(\ref{eq2:epsBA}) we obtain
\bq
	\eps(k,k) = \frac{2(5k^2-3k-2)}{k^2 (1+k)^2 (4k^2-1)} .
\eq
Using Eq.~(\ref{eq1:ac}) after some tedious but straightforward calculations we find:
\bq
	\mathcal{A} = \frac{33-24 \ln 4}{42-4\pi^2} \approx -0.1075,
\eq
which stands in a very good agreement with numerical simulations. This indicates that the BA growing tree network is disassortative.

Many improvements of BA growing network models have been proposed (for a review, see e.g. \cite{ref:physrep2006}). The growing BA network can be used as an initial configuration for the algorithms, like those described before, to generate scale-free networks with some features enhanced \cite{ref:xbs-assort}. In this way one can also extrapolate between causal and equilibrated networks.

\subsection{Statistical ensemble formulation of growing networks}
Although many properties of growing networks can be understood using rate equations, sometimes it is convenient to introduce the ensemble of causal graphs and calculate desired quantities from the partition function. As mentioned, such an ensemble cannot be thought as an ensemble with the Gibbs measure, in the usual statistical sense, but merely as an ensemble of networks which can be obtained in a growth process, if this process is terminated at some moment of time. In this subsection we shall define such an ensemble for trees with the product weight (\ref{eq2:prod}). As we shall see, the model is on the one hand solvable and on the other hand it exhibits non-trivial behavior. In particular, we shall be able to quantify the effects of causality.  

\begin{figure}
\center
\includegraphics*[width=10cm]{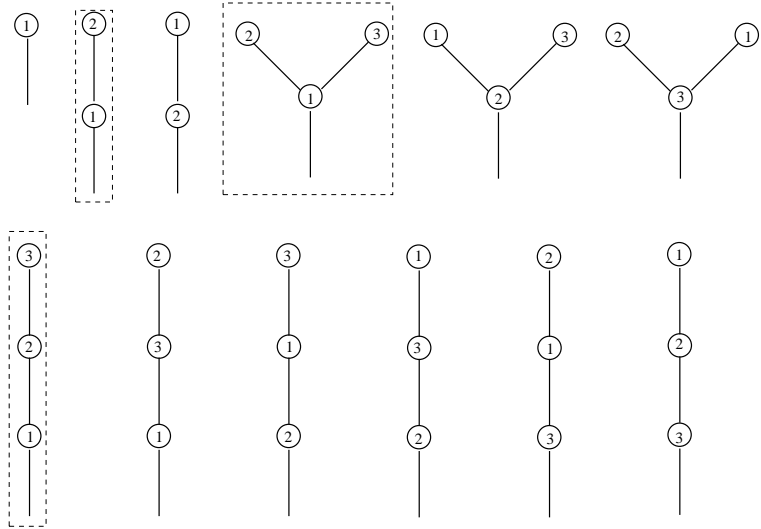}
\caption{\label{fig:causalrt}Planted rooted trees of size $N=1,2,3$. All causal labellings are surrounded by dashed rectangles.}
\end{figure}
\begin{figure}
\center
\psfrag{n1}{$N_1$}\psfrag{n2}{$N_2$}\psfrag{n3}{$N_3$}
\includegraphics*[width=10cm]{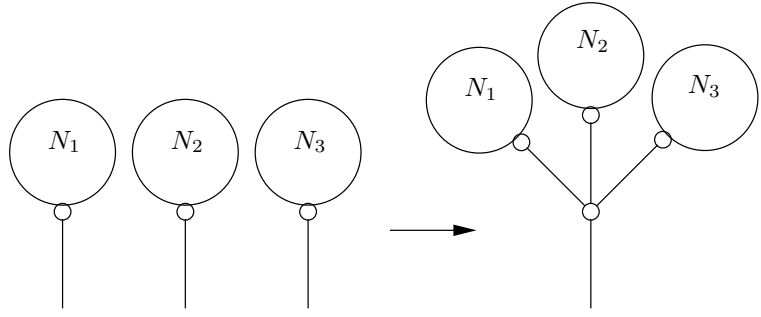}
\caption{\label{fig:treecomp} Construction of a new planted rooted tree from $q=3$ trees of size $N_1,N_2,N_3$. The large circle denotes the set of all trees of given size. One node (small circle) is distinguished by the stem joined to it. The new set of trees of size $N_1+N_2+N_3+1$ is obtained by joining the trees from the left-hand side to a new common rooted node (right-hand side). The new node has degree four since the root is counted as a link.}
\end{figure}
\begin{figure}
\center
\psfrag{x}{$x$} \psfrag{x0}{$x_0$} \psfrag{Z}{$Z$} \psfrag{Fx}{$F(x)$} \psfrag{F1X}{$\frac{1}{F(x)}$}
\psfrag{GZ}{$e^{-\mu(Z)}$} \psfrag{muZ}{$\mu(Z)$} \psfrag{mu0}{$\mu_0$} \psfrag{0,0}{$0,0$}
\includegraphics*[width=15cm]{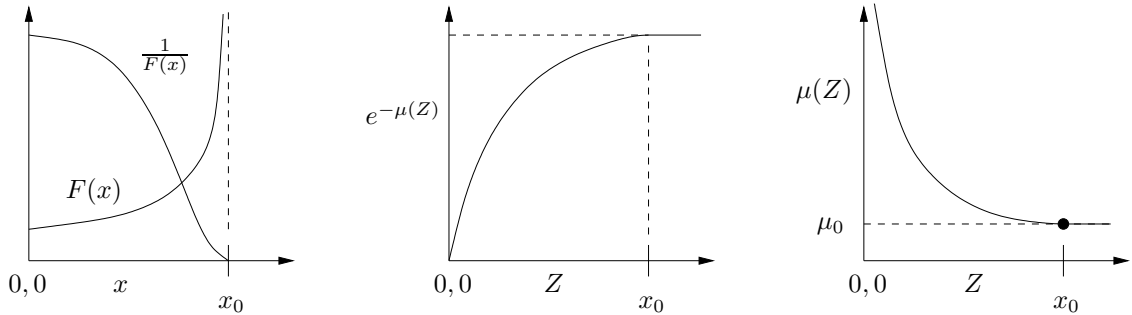}
\caption{\label{fig:gzmu}Typical behavior of functions discussed in section 2.2.3.}
\end{figure}

Because a tree with $N$ nodes has $L=N-1$ links, we shall define the canonical partition function which depends only on $N$:
\bq
	Z(N) = \frac{1}{N!} \sum_{\alpha\in lct(N)} n(\alpha) p(k_1) \cdots p(k_N),	\label{eq2:znct}
\eq
where the sum runs over all labeled causal tree graphs. The causal ordering of nodes' labels selects a relatively small fraction of all possible labeled trees. The calculation of Eq.~(\ref{eq2:znct}) is much simpler for planted rooted trees, i.e. trees with an additional link (a stem) attached to one of its nodes. The stem acts as an additional link which marks one node of the tree and increases its degree by one. Because only one node is marked, in the thermodynamic limit ensembles of trees and planted rooted trees have roughly the same properties.  In figure \ref{fig:causalrt} graphs with $N=1,2,3$ are sketched. Following \cite{ref:bbjk} we shall derive a recursion relation for $Z(N)$. First, we observe that every tree of size $N+1$ can be constructed from trees of sizes $N_1,\dots,N_q$ where $\sum_{i=1}^q N_i =N$, by attaching their stems to a common node (see Fig.~\ref{fig:treecomp}).
This new node is attached to a new common stem. Denoting by $n(N)$ the number of different labellings for the set of trees of size $N$ we have
\bq
	n(N+1) = \frac{N!}{N_1!\cdots N_q!} \frac{1}{q!} n(N_1) \cdots n(N_q).
\eq
The origin of factorials is the following. The whole tree has $N+1$ labels, but the smallest label must be attached to the root because of the causality. The remaining $N$ labels can be distributed arbitrarily. All $N_i!$ permutations of $N_i$ labels of a subtree are undistinguishable and thus they give the same graph. To avoid overcounting one divides by $N_i!$. This leads to the multinomial factor. In addition, $q$ subtrees can be permuted in $q!$ possible ways giving the same labeled graph, thus we have to divide by $q!$. 
The functional weight $W(N+1)$ for the set of compound trees also factorizes:
\bq
	W(N+1) = p(q+1) W(N_1) \cdots W(N_q).
\eq
Notice that the new node has degree $q+1$ because the root is counted as a link. The partition function (\ref{eq2:znct}) can be expressed as a sort of self-consistency equation for $Z(N)$: 
\ba
	Z(N+1) &=& \frac{1}{(N+1)!} \sum_{q=1}^\infty \sum_{N_1,\dots,N_q} \delta_{N,N_1+\cdots+N_q}
	\frac{N!}{N_1!\cdots N_q!} \frac{1}{q!} n(N_1)\cdots n(N_q)  \nonumber \\ &\times&  p(q+1) W(N_1) \cdots W(N_q) \nonumber \\
	&=& \frac{1}{N+1} \sum_{q=1}^\infty \frac{p(q+1)}{q!} \sum_{N_1,\dots,N_q} \delta_{N,N_1+\cdots+N_q} \prod_{i=1}^q Z(N_i), \label{eq2:zn+1}
\ea
where $Z(N)$ appears on both sides.
The sum goes over all subtrees $1,2,\dots, q$ of sizes $N_1,\dots,N_q$ with the only constraint given by the delta function. The constraint can be decoupled by introducing a grand-canonical partition function:
\bq
	Z(\mu) = \sum_{N=1}^\infty Z(N) e^{-N\mu}, \label{eq2:zmu}
\eq
which is just a generating function for the canonical partition functions $Z(N)$. Here $\mu$ is the chemical potential but contrary to the previous definition (\ref{grand}), it controls the average number of nodes, not links\footnote{Because $L=N-1$, the difference is in fact meaningless.}.
Multiplying both sides of Eq.~(\ref{eq2:zn+1}) by $(N+1)e^{-(N+1)\mu}$ and summing over $N=1,\dots,\infty$ we get:
\bq
	\sum_{N=2}^\infty N Z(N) e^{-N\mu} = e^{-\mu} \sum_{q=1}^\infty \frac{p(q+1)}{q!} \left( \sum_{N_i=1}^\infty e^{-\mu N_i} Z(N_i) \right)^q.
\eq	
If we add the term $Z(1)e^{-\mu}$ to both sides of this equation, the left-hand side  becomes just a derivative of $-Z(\mu)$ with respect to $\mu$, while the right-hand side is a sum over $q$ extended to the range $q=0,\dots,\infty$, which additionally includes the term for $q=0$. Thus we get
\bq
	Z'(\mu) = -e^{-\mu} F(Z(\mu)), \label{eq2:zprim}
\eq
where $F(x)$ is the generating function for the distribution $p(k)$ like in Eq.~(\ref{eq2:fvtree}):
\bq
  F(x) = \sum_{q=0}^\infty p(q+1) \frac{x^{q}}{q!}. \label{eq2:fvtree2}
\eq
This series may have a finite or infinite radius of convergence, $x_0$.
The equation (\ref{eq2:zprim}) can be integrated over $\dd\mu$. This yields
\bq
	e^{-\mu(Z)} = \int_0^Z \frac{\dd x}{F(x)},
\eq
and because\footnote{We can exclude the trivial case when all $p(q)$'s are zero.} $F(x)>0$ for $x>0$ and $F(x)\to\infty$ for $x\geq x_0$, the integral is bounded from above. Hence the chemical potential $\mu(Z)$ is bounded from below: $\mu\to\mu_0$ as $Z\to \infty$. This means that $Z$ as a function of $\mu$ has a singularity at $\mu_0$ (see Fig.~\ref{fig:gzmu}):
\bq
	\mu_0 = -\ln \int_0^{x_0} \frac{\dd x}{F(x)}.
\eq
From definition of the partition function (\ref{eq2:zmu}) we have that $Z(N)$ shall grow as $\sim e^{\mu_0 N}$ or faster. We assume now that the ensemble of trees is normal in the statistical-thermodynamical sense, that is $Z(N)\propto e^{\mu_0 N}$. As we have seen, this is not true for simple graphs, but the number of causal trees grows only as $(N-1)!$ \cite{ref:bbw} and not as $\sim (N^2)!$ for graphs. Therefore, many quantities as for instance degree distribution can be obtained from the critical value $\mu_0$. For example, according to Eq.~(\ref{eq2:pikdef}) the degree distribution reads
\bq
	\Pi(k) = p(k) \frac{\partial \mu_0}{\partial p(k)} = \frac{p(k)}{(k-1)!} \frac{\int_0^{x_0} \frac{{\rm d}x \, x^{k-1}}{F^2(x)}}{\int_0^{x_0} \frac{\dd x }{F(x)}}. \label{eq2:pikctree}
\eq
Thus, similarly as for simple graphs, by tuning $p(k)$ one can obtain any desired degree distribution. It is, however, not as trivial as in case of Eq.~(\ref{eq2:pikv0}) because the dependence on $F(x)$ is now more complicated. Some interesting distributions were investigated in \cite{ref:bbjk}. For instance, with the choice $p(k)=(k-1)!$, the generating functions reads $F(x)=(1-x)^{-1}$ and has the radius of convergence $x_0=1$. The integrals in (\ref{eq2:pikctree}) can be done analytically. The result is
\bq
	\Pi(k) = \frac{4}{k(k+1)(k+2)},
\eq
so one recovers the BA degree distribution. We can show that, indeed, causal trees with the product weight 
\bq
	W(\alpha)=(k_1-1)!\cdots (k_N-1)! \label{eq:kkkk}
\eq
form the same ensemble as BA growing trees. To this end, let us consider a set of all causal trees $\alpha$ which for a given degree sequence $k_1,\dots,k_N$ have the statistical weight $W(\alpha)$ given by Eq.~(\ref{eq:kkkk}). Imagine also that we have a Markov process which generates such trees. %We will show that it generates exactly the same trees as those generated by the growing process with the linear attachment kernel. 
First, we see that the number of possible causally labeled trees in this set is obviously the same as in the BA model. We have to check whether also the statistical weights are the same in both cases. Imagine that we take a graph $\alpha$ with $N$ nodes with degrees $k_1,\dots,k_N$ and attach a new node by linking it to a node $n$. We obtain a new configuration $\beta$, which has now $N+1$ nodes with degrees $k_1,\dots,k_n+1,\dots,k_N,1$.  The transition probability $\alpha\to\beta$ for a process which has a stationary distribution~(\ref{eq:kkkk}) is
\bq
	P(\alpha\to\beta) \propto \frac{W(\beta)}{W(\alpha)} = \frac{(k_1-1)!\cdots k_n! \cdots (k_N-1)! \cdot 1}{(k_1-1)!\cdots (k_N-1)!} = k_n  
\eq
and we see that it is identical to that for linear attachment kernel in the BA growth process.  In conclusion, this shows the equivalence of the two approaches.%, but as we will see below, the statistical ensemble approach has some advantages.

The formulation of the BA model of growing networks via statistical ensemble can be used to calculate degree-degree correlations or the average distance $\left<r\right>$ between any two nodes \cite{ref:bbjk}. For instance, it can be found that $\left<r\right> \cong (1/2)\, \ln N$, that is the BA network really displays the small-world phenomenon.

\section{Causal versus equilibrated networks}
So far we have discussed equilibrated and causal networks separately. We have shown that for both classes some properties of networks can be essentially the same, as for instance the power-law degree distribution. Now we shall compare these two ensembles and see that although $\Pi(k)$ can be identical in the thermodynamic limit, some other characteristics of the network topology are different for causal and equilibrated graphs. As shown in \cite{ref:bbw}, causal trees form only a small subset of all trees. The same is true for simple graphs. The fraction of causally labeled trees among all labeled trees is only $\sim N^{3/2} e^{-N}$. So the chance of picking up at random a causal tree from the set of all trees vanishes when $N$ grows. We shall show that geometrical properties of typical trees in this subset are quite different from those in the whole set.

\begin{figure}
\center
\psfrag{xx}{$r$} \psfrag{yy}{$G(r)$}
\includegraphics*[width=10cm,bb=37 146 519 423]{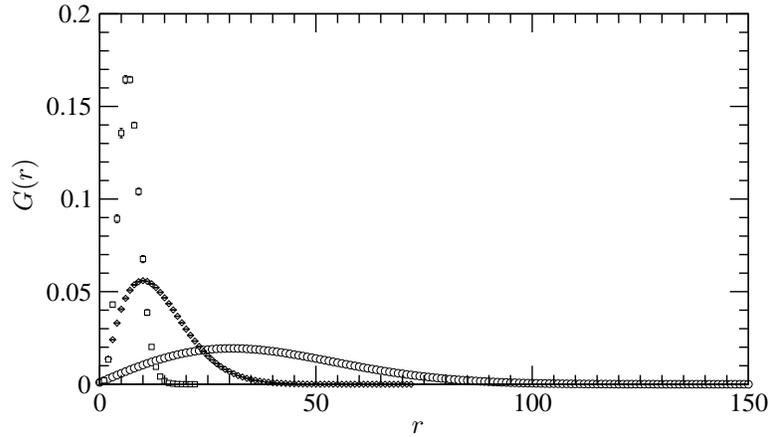}
\caption{\label{fig:hhct}The distance distribution $G(r)$ for unweighted equilibrated trees (circles), scale-free
equilibrated trees (diamonds) and scale-free causal trees (squares) of size $N=1000$. S-F causal trees are the shortest.}
\end{figure}

\begin{figure}[ht]
\center
\psfrag{xx}{$k$}
\psfrag{yy}{$\Pi(k)$}
\includegraphics*[width=10cm,bb=21 146 507 423]{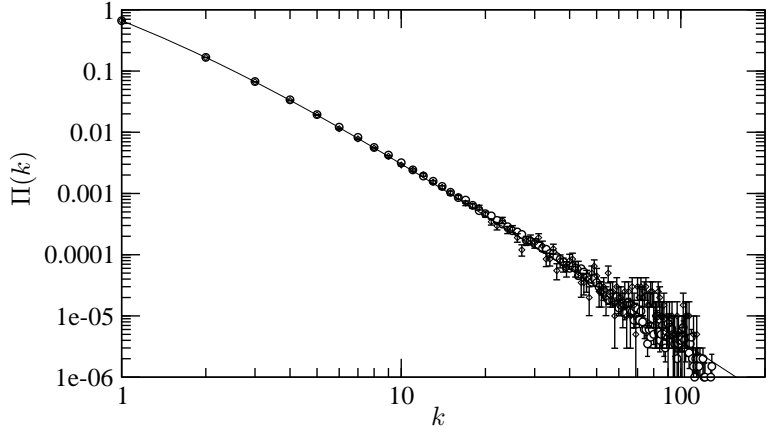}
\caption{\label{fig:chpi} The degree distribution for causal (diamonds) and equilibrated (circles) scale-free trees
measured in Monte Carlo runs for $N=1000$. The solid line stands for theoretical $\Pi(k) = \frac{4}{k(k+1)(k+2)}$.}
\end{figure}

Let us first consider the ensemble of unweighted equilibrated trees 
and the corresponding ensemble for causal trees. Here ``unweighted'' means that
all trees have the same functional weight equal to one. We can now calculate some geometrical quantities for
trees in the first and in the second ensemble. An example of such a quantity 
is the average distance $\left<r\right>$. In fact, one can calculate it analytically \cite{ref:bbjk,ref:kr,ref:adj,ref:b-correlations,ref:bbju}. For equilibrated trees it is 
\bq
	\left<r\right> \sim \sqrt{N},
\eq
which means that the fractal dimension of typical equilibrated trees is equal to 2. These trees are therefore rather elongated and certainly are not small-worlds so abundantly observed in nature. On the other hand, for causal trees,
\bq
	\left<r\right> \sim \ln N ,
\eq
hence the fractal dimension is infinite. This is because most of nodes concentrate around the oldest node. A similar observation was made for weighted trees with BA degree distribution \cite{ref:bbw}. An even better insight into geometrical properties of trees (or graphs) is provided by the distribution $G(r)$ of distances $r$ between all pairs of nodes:
\bq
	G(r) = \left< \frac{1}{N^2} \sum_{i,j} \delta_{r,r(ij)} \right>.
\eq
Here $r(ij)$ is the length of the shortest path between two nodes $i,j$. The average distance is the mean of this distribution: $\left<r\right> = \sum_r r G(r)$. 
In figure \ref{fig:hhct} we present a comparison of $G(r)$ for equilibrated and causal trees of the same size. 
Causal trees were generated using the BA model while for equilibrated trees we used the Monte Carlo algorithm described in section 2.1.5.
The weights $p(k)=4(k-1)!/(k(k+1)(k+2))$ were chosen according to Eq.~(\ref{eq2:pikv0tree}) to get the same degree distribution as in the BA model. In Fig.~\ref{fig:chpi} we see that indeed both types of trees have the same $\Pi(k)$, so one cannot easily distinguish to which ensemble the given tree belongs, by only measuring\footnote{We shall see in the next chapter that this statement is true only in the thermodynamic limit. For any finite $N$ there are finite-size corrections, which are different for both ensembles. To see a difference coming from the finite-size correction one has to have much better statistics than in figure \ref{fig:chpi}.} $\Pi(k)$. But one easily sees in Fig.~\ref{fig:hhct} that the causal trees are much shorter than the equilibrated ones.
\begin{figure}
\center
\psfrag{xx}{$x$}
\psfrag{yy}{$G_{\rm h}(x)$}
\includegraphics[width=7cm]{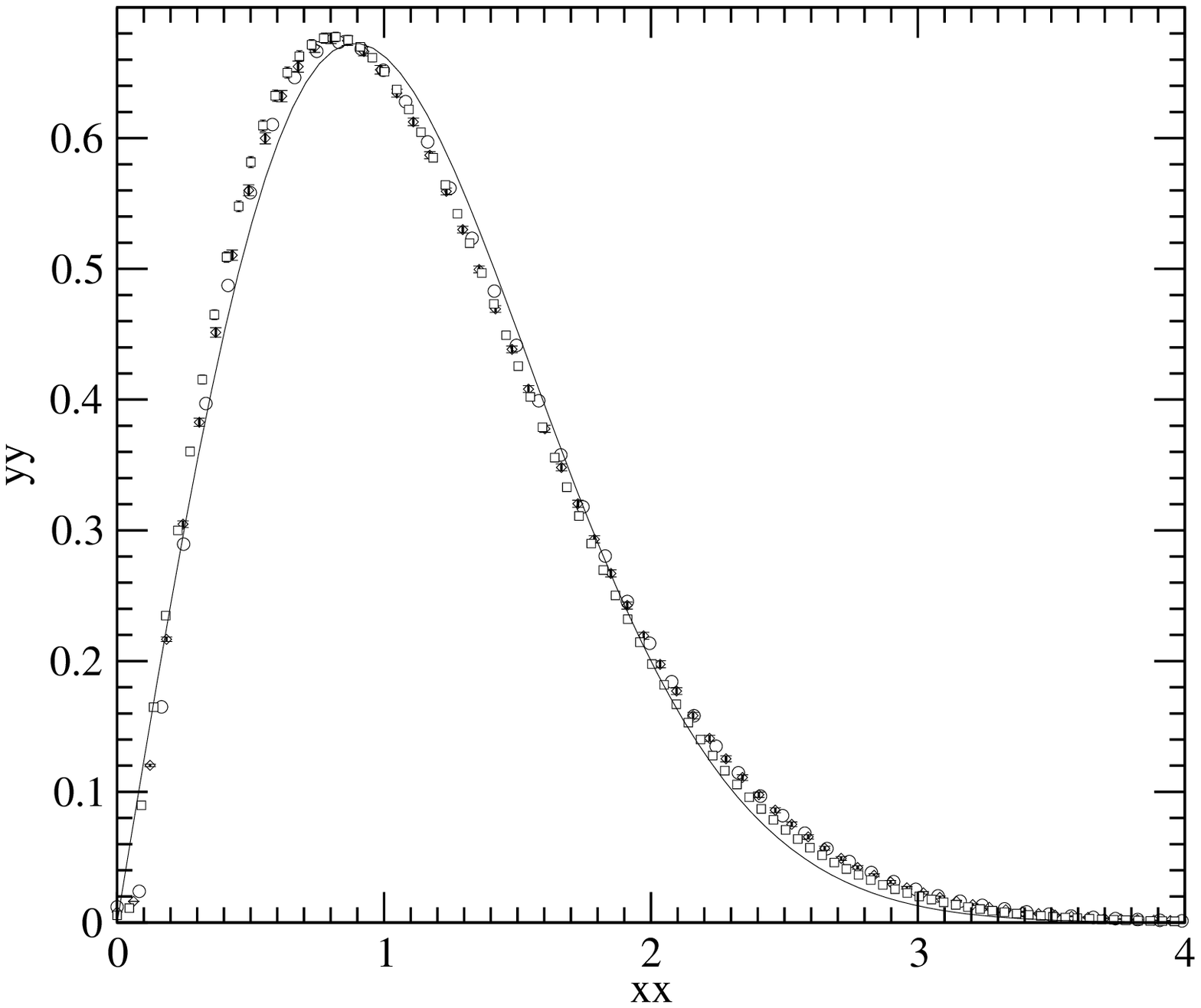}
\hspace{1mm}
\psfrag{yy}{$G_{\rm c}(x)$}
\includegraphics[width=7cm]{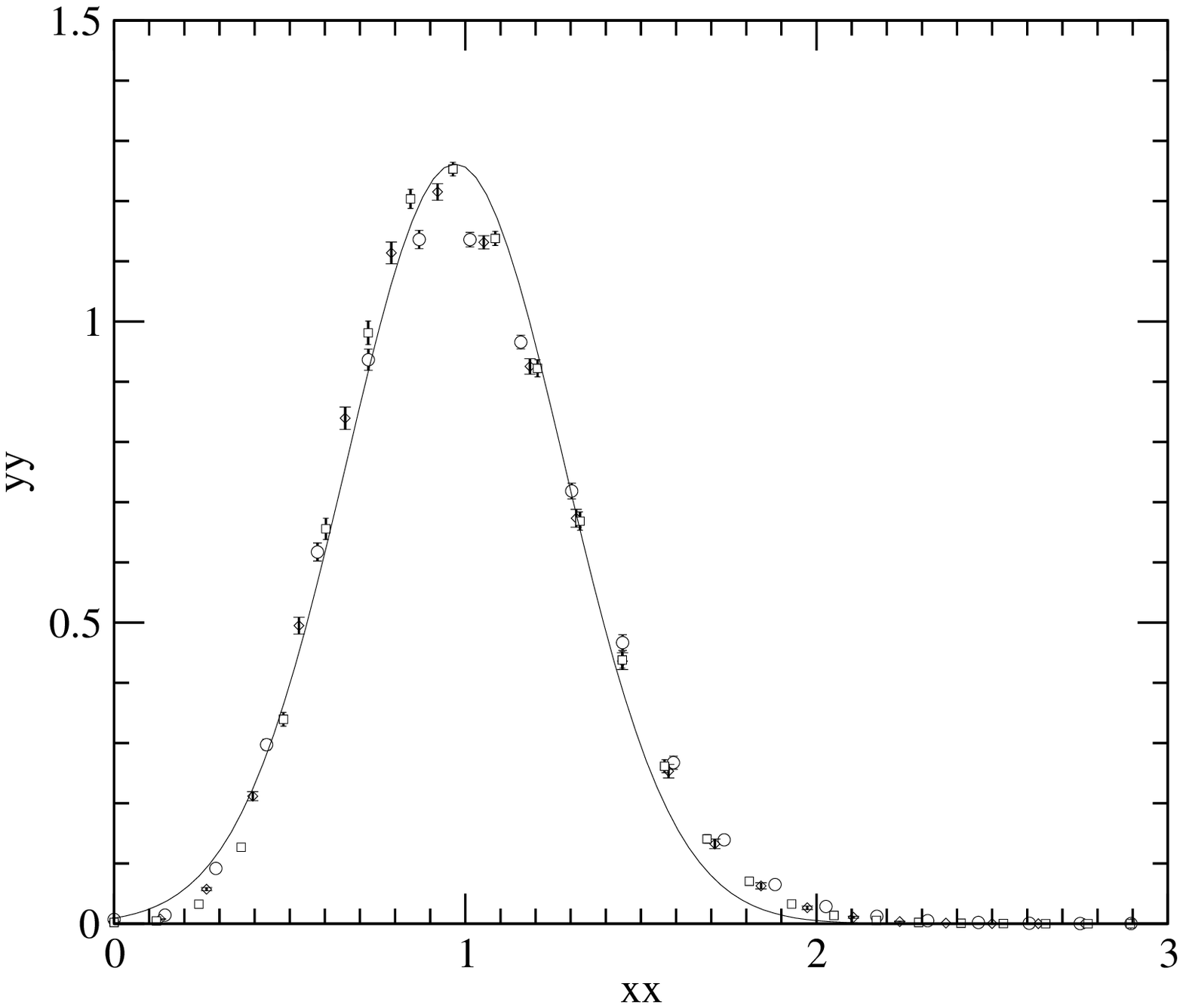}
\caption{\label{fig:collapsgr}Left: $G(r)$ for equilibrated trees plotted in the rescaled variable: $x=r/\sqrt{N/\ln N}$ for different 
sizes $N=500,1000,2000,4000$. Right: the same quantity for causal trees, for $x=r/\ln N$. Continuous lines are given by Eqs.~(\ref{eq2:ghx}) and (\ref{eq2:ghc}).}
\end{figure}
If we assume that the average distance scales for equilibrated trees like 
\bq
	\left<r\right>_{\rm h} \sim \sqrt{N/\ln N},
\eq
and for causal trees:
\bq
	\left<r\right>_{\rm c} \sim \ln N,
\eq
we can plot curves $G(r)$ for different $N$ in the rescaled variable $x\equiv r/\left<r\right>$ and observe that they collapse to some characteristic curves but different for each of the two ensembles (see Fig.~\ref{fig:collapsgr}). The function $G_{\rm h}(x)$ for equilibrated trees is well approximated by 
\bq
	G_{\rm h}(x) = Ax \exp(-Bx^2/2), \label{eq2:ghx}
\eq
while for causal trees by
\bq
	G_{\rm c}(x) = A' \exp(-(x-\bar{x})^2 /B'), \label{eq2:ghc}
\eq
with some parameters $A,A',B,B',\bar{x}$ fitted to data. So again, the average node-node distance is smaller for the causal trees than for the equilibrated ones with the same degree distribution. The effect is qualitatively the same when one considers simple graphs instead of trees. Thus the causality enhances the small-world effect by increasing the relative weight of graphs with clusters of nodes around the oldest vertices. 

\begin{figure}
\center
\psfrag{xx}{$k$} \psfrag{yy}{$\bar{k}_{\rm nn}(k)$}
\includegraphics*[width=10cm]{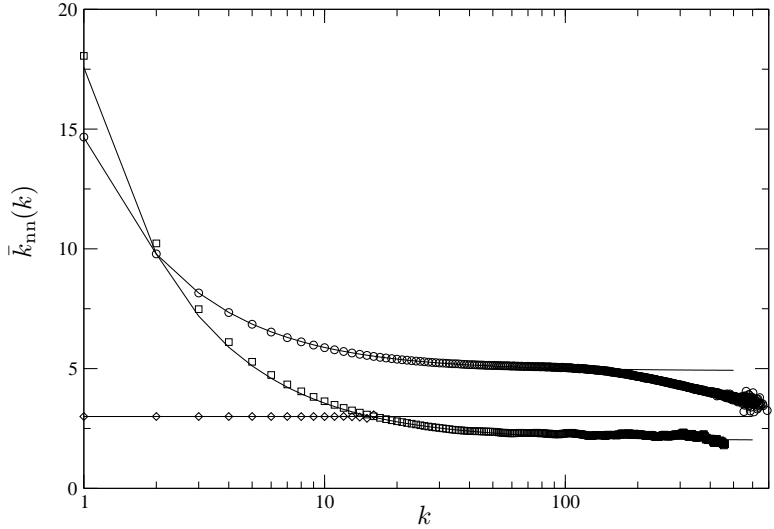}
\caption{\label{fig:kknn}Plots of $\bar{k}_{\rm nn}(k)$ for ER random graphs (diamonds), random trees with BA degree distribution (squares) and causal BA trees (circles). Both causal and random trees are disassortative, but they differ in approaching $k\to \infty$. The descent of experimental curves for large $k$ is caused by finite-size effects. Theoretical curves (solid lines) are calculated from Eqs.~(\ref{eq2:knnr}) and (\ref{eq2:knnc}).}
\end{figure}

There are many other differences between the causal and equilibrated networks.
We shall give one more example showing the difference in degree-degree 
correlations in both types of trees. A quantity which is commonly used to study these correlations is the average degree $\bar{k}_{\rm nn}(k)$ of the nearest neighbors of a node with degree $k$, defined in Sec. 1.3 and expressed through Eq.~(\ref{eq1:knnk}). For uncorrelated graphs it is $\eps_{\rm r}(k,q) = kq\Pi(k)\Pi(q) /\left<k\right>^2$ and thus
\bq
	\bar{k}_{\rm nn}(k) = \frac{\left<k^2\right>}{\left<k\right>},  
\eq
which gives a constant value $\bar{k}_{\rm nn}= 1+ \bar{k}$ for maximally random graphs in the ER model. In general case for equilibrated trees with an arbitrary degree distribution it can be shown \cite{ref:pbialas} that 
\bq
	\bar{k}_{\rm nn}(k) = 2+ \frac{1}{k}\left(\left<k^2\right> -4\right). \label{eq2:knnr}
\eq
This result differs from the corresponding one for causal trees. For instance, for causal trees with BA degree distribution \cite{ref:bwunpub},
\bq
	\bar{k}_{\rm nn}(k) = \frac{\left<k^2\right>}{2} \left( \frac{1}{2}+\frac{1}{k} \right). \label{eq2:knnc}
\eq
The second moment $\left<k^2\right>$ depends on the size $N$ and can be calculated for growing BA trees \cite{ref:bwis}:
\bq
	\left<k^2\right> = (2-2/N) H(N-1).
\eq
Here $H(n)=\sum_{i=1}^n 1/i$ is the harmonic number. The same formula performs well for random BA trees. In figure \ref{fig:kknn} we plot $\bar{k}_{\rm nn}(k)$ for BA causal and equilibrated graphs, and also for ER graphs. From Eq.~(\ref{eq2:knnr}) we get $\bar{k}_{\rm nn}(k)\to 2$ when $k$ approaches infinity, while for causal trees the limiting value is proportional to the second moment of the degree distribution, and thus diverges for $N\to\infty$. This means that the affinity of nodes with higher degrees and their tendency to cluster together grow with the size of tree. This is another argument supporting the conjecture that causal trees are more compact than the corresponding equilibrated ones.

\chapter{Applications to modeling complex networks}
In this chapter we present some further applications of the mathematical methods developed so far. First we shall quantify finite-size effects in networks. Usually, while discussing networks one calculates quantities of interest in the thermodynamic limit $N\to \infty$. As we will show such a procedure may lead to neglecting some important effects, which are seen for finite networks. In particular, the node-degree distribution exhibits for finite $N$ apparent deviations from the limiting distribution. We will find an explicit form of finite-size corrections to the scale-free behavior for growing networks and talk over corrections for homogeneous graphs.
A second problem which shall be discussed in this chapter concerns a very important class of phenomena which describe the dynamics of statistical processes on networks. On the example of a zero-range process we will show the usage of techniques developed in the previous chapter.

\section{Finite-size effects in networks}
In the preceding chapter we discussed some popular models of networks, for which we determined degree distribution, clustering coefficient, diameter etc., in the limit of infinite networks. The derivation of exact analytical result was possible because in this limit structural constraints, like for example that on the sum of degrees, become less important and some of them loose their virtue at all. For instance, we mentioned that in the thermodynamic limit the canonical ensemble for homogeneous graphs with fixed $L$ is  equivalent to the grand-canonical one where $L$ can in principle fluctuate\footnote{We have shown this explicitly for equally weighted random graphs. A more general situation is considered in \cite{ref:snd-stat-mech}.}. However, it is not the case for finite $N$ and one has to incorporate the effect of finite-size constraints into calculations.

One must be very careful while comparing models solved in the thermodynamic limit to real-world networks. For finite $N$, some local quantities like node degrees are bounded from above. There are also some effects resulting from network's features which are rare but can significantly change the picture for small graphs. For example, it is known that in many models, as for instance in the ER model, large graphs are  essentially trees, because the average number of cycles of finite length is constant and does not depend on the network size $N$. On the other hand, for smaller graphs, short loops play an important role. Their presence shapes the network and strongly affects its global properties. 

In next sections we shall discuss one type of finite-size effects, namely that which is related to the appearance of a cutoff in the degree distribution of finite networks. We will present our recent findings for various graphs and compare them to those from the literature.

\subsection{Cutoff in the degree distribution}
As we have pointed out in Sec. 1.3, for any finite network the power-law behavior of the degree distribution $\Pi(k)$ can hold only for values of $k$ significantly smaller than $N$. Both experimental data and theoretical models of scale-free networks indicate that the behavior of $\Pi(k)$ for $k \gg 1$ for a finite network exhibits two regimes: below some $k_{\rm max}$ it follows the power-law
behavior as in an infinite network while above $k_{\rm max}$ it displays a much faster decay. The characteristic degree $k_{\rm max}$ which separates these two regimes is called a cutoff. Intuitively, the cutoff comes about due to the fact that the overall number of links present in a finite, non-degenerated graph is restricted and so is also the degree of each node. Thus for any finite network the power-law behavior of the degree distribution is truncated. In consequence, many quantities calculated on finite networks significantly differ from their counterparts derived in the thermodynamic limit. One can see this effect for example when one calculates percolation thresholds for statistical systems on networks, like for instance those describing infection spreading for real diseases or computer viruses.

Many attempts were undertaken to estimate the position of the cutoff for different scale-free networks, most of them concentrated on sparse networks where the average degree $\bar{k}$ is fixed. This restricts the class of distributions $\Pi(k)$ to those which have a finite mean value, and the power-law tail exponent to the range $\gamma>2$, which is indeed observed for real networks.

\begin{table}[h]
%	\center	
	\begin{tabular}{|l|c|c|}
	\hline
	Article & $\alpha$ for $2<\gamma<3$ & $\alpha$ for $\gamma>3$ \\
	\hline
	(a) \cite{ref:dr3}, homog. simple graphs & $1/(5-\gamma)$ * & $1/2$ * \\
	\hline
	(b)  \cite{ref:zbak2}, homog. simple graphs & $1/2$ * & $1/(\gamma-1)$ * \\
	\hline
	(c) \cite{ref:extr2}, uncorrelated networks & $1/2$ & $1/(\gamma-1)$ \\
	\hline
	(d) Pseudographs & $1/(\gamma-1)$ & $1/2$ \\
	\hline
	(e) \cite{ref:kr2}, growing trees & \multicolumn{2}{|c|}{$1/(\gamma-1)$} \\
	\hline
	(f) \cite{ref:bck}, homog. trees & \multicolumn{2}{|c|}{$1/(\gamma-1)$} \\
	\hline
	\end{tabular}
	\ \
	\psfrag{xx}{$\gamma$}\psfrag{yy}{$k_{\rm max}$}\psfrag{aa}{$1/(\gamma-1)$}\psfrag{bb}{$1/(5-\gamma)$}\psfrag{cc}{$1/2$}
	\psfrag{zz}{$\alpha$}
	\hspace{5mm}\parbox{5.0cm}{\includegraphics*[width=5.0cm]{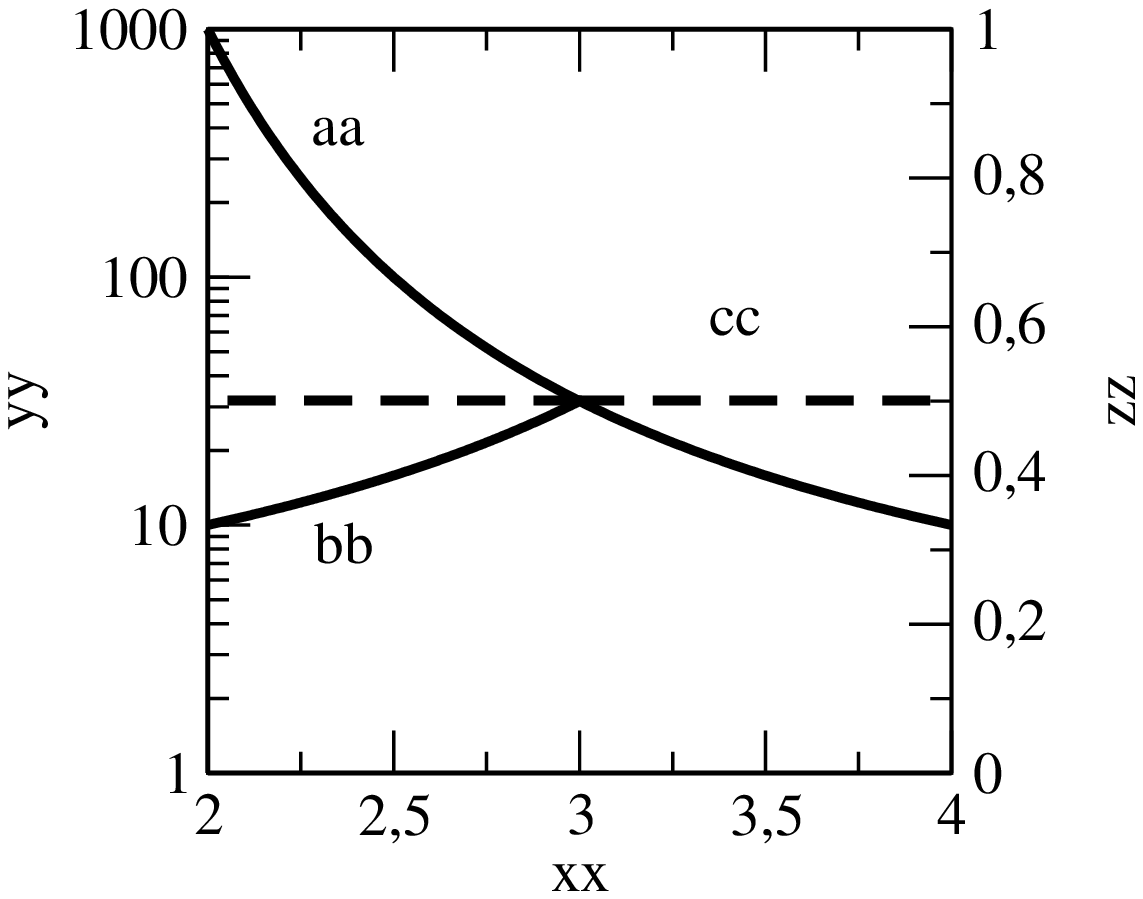}}
	\caption{\label{tab:cutoff}Some results for the exponent $\alpha$ in the cutoff $k_{\rm max}\sim N^\alpha$. A star (*) denotes two inconsistent predictions.	By pseudographs we understand graphs with self- and multiple-connections with the partition function (\ref{eq2:partps}), for which corrections to the degree distribution (\ref{eq2:pikv0}) can be found by observation that it is equivalent to so called balls-in-boxes model \cite{ref:bbj}, and calculated in the limit of large but finite $N$ \cite{ref:evans2}. In figure on the right-hand side we compare different exponents graphically. Left axis: $k_{\rm max}$ for the network of size $N=1000$, right axis: the value of $\alpha$.}
\end{table}

In general, the position of the cutoff $k_{\rm max}$ scales as $N^\alpha$ for large $N$ with an exponent $\alpha<1$. The value of the exponent depends on the type of network. In table \ref{tab:cutoff} we collect some values $\alpha$ calculated for S-F graphs of various type, together with references to the original papers where the quoted values were derived. The exponent $\alpha$ is calculated as a function of the exponent $\gamma$ in the power-law tail of the underlying degree distribution. In the discussion of S-F networks one should differentiate between equilibrated and growing networks, and simple graphs, pseudographs and trees. Note, however, that for $\gamma=3$, $\alpha=1/2$  in all cases. The value $\gamma=3$ is in a sense marginal because it separates the regime of anomalous fluctuations for $\gamma<3$ and of normal fluctuations for $\gamma>3$. In the former case $\langle k^2 \rangle$ is infinite in the limit $N\to\infty$ while in the latter one it is finite. 

Let us make some remarks on the results in the table. First, the result (a) has been found recently in \cite{ref:dr3} in the statistical ensemble approach. It is in disagreement with the results (b). The authors \cite{ref:dr3} claim that (b) gives only an upper bound on the value of the exponent $\alpha$ for $2<\gamma<3$.
Second, the scaling for trees seems to be the same for growing (e.g. the BA model) and equilibrated ones. Third, equilibrated pseudographs have different cutoffs than simple graphs or trees. Fourth, the result (c) applies only to hypothetic uncorrelated graphs with no correlations between degrees of nearest neighbors. Any finite network has certain correlations of such type, simply because of global constraints like that on the sum of degrees coming from the fixed number of links $L$. Therefore, as we have mentioned in Chapter 2, the two-point correlation function $\eps(k,q)\neq \eps_{\rm r}(k,q)$ even if we do not introduce correlations {\em explicite}. For instance, $\Pi(N-1)$ can be non-zero for simple graphs but then it is impossible to pick a link joining nodes both of degree $k=N-1$
as it would stem from Eq.~(\ref{eq1:uncorr}). The authors \cite{ref:extr2} are aware of this effect and conclude that for assortative networks the cutoff should be smaller than the one predicted in table \ref{tab:cutoff} while in case of  disassortativity it should be larger.

The results in the table were obtained with the help of different methods. 
For homogeneous networks many of them were based either on some simple probabilistic arguments or extreme values statistics.
Those methods allow one to determine the cutoff but not the shape of the function giving the finite-size correction to the degree distribution.
For growing networks, however, the shape of the corrections can be found. In \cite{ref:kr2} the BA model of growing tree network has been solved for finite $N$. The authors have calculated the mean number of nodes of a given degree for the network of size $N$ and deduced the form of the correction to the degree distribution for the pure BA model with $\gamma=3$. In the next section we shall present a more general method which also applies to other growing networks and we shall use it to determine the form of the cutoff function.
In the last section we shall present Monte Carlo simulations of networks which allow for the estimation of the cutoff function and the exponent $\alpha$, and we shall compare the results to those in table \ref{tab:cutoff}.

\subsection{Growing networks}
Here we would like to present the method of determining of the cutoff functions for growing networks. We shall explain it for the BA growing networks with initial attractiveness \cite{ref:dr3} described in Sec. 2.2.1. Some of results presented here were obtained in Ref. \cite{ref:kr2}. However, our approach is different and allows for solving more sophisticated variants of the model. Before we start, let us give some key points of the method here. We begin with the rate equation for the average number of nodes $N_k(N)$ of a given degree $k$. The average is taken over the canonical ensemble of growing trees as in Sec. 2.2.2. The solution of the rate equation in the limit of large $N$ gives, up to a normalization constant, the degree distribution $\Pi_\infty(k)$ for the infinite network.

For a finite network, we define $\Pi(k)$ as a product of the limiting degree distribution $\Pi_\infty(k)$ and a cutoff function $v(k,N)$ giving finite-size corrections. The recursion equation for this function can be obtained from that for $N_k(N)$. The next ingredient of the method is to consider moments of the cutoff function. One can derive recursive equations for the moments from the recursion relations for $v(k,N)$. The equations can be solved recursively and one can derive explicit asymptotic expressions for the moments for sufficiently large but finite $N$. The knowledge of all moments makes it possible to reconstruct the leading behavior of the cutoff function $v(k,N)$. This is the sketch of the method which we shall explain below in details. Although the idea is very simple, its implementation leads to quite complicated and lengthy calculations
which we omit here, referring the reader to the original paper \cite{ref:bwis}.

We start from the BA tree model with $m=1,a_0=0$ and thus $\gamma=3$. Like in Sec. 2.2.2, as an initial configuration we take the graph with $n_0=2$ nodes joined by a link (a dimer configuration), therefore $N_k(2)=2\delta_{k,1}$. At each time step a new node is added and connected to one of $N$ existing nodes in the system, with the probability proportional to the number of the preexisting links of the corresponding node, leading to a new network with $N+1$ nodes. According to Eq.~(\ref{eq2:rN}), the rate equation for the average number $N_k(N)$ has the form:
\bq
	N_k(N+1) = N_k(N) + \delta_{k,1} +\frac{k-1}{2(N-1)} N_{k-1}(N) - \frac{k}{2(N-1)} N_{k}(N),
	\label{rN}
\eq
where, for brevity, we have omitted the angle brackets denoting the average. The origin of all terms has been already explained in Sec. 2.2.2. This equation is exact for any $N$. In the limit of $N\to\infty$ it has a solution given by $N_k(N) \approx N\Pi_{BA}(k)$, where
\bq
	\Pi_{BA}(k) = \frac{4}{k(k+1)(k+2)} \label{piba}
\eq
is the degree distribution in the BA model. Here we are, however, interested in the general solution for $N_k(N)\equiv N \Pi(k,N)$, with $\Pi(k,N)$ being the degree distribution for a finite network. It is convenient to split $\Pi(k,N)$ into the product of the known function
$\Pi_{BA}(k)$ and an unknown function $v(k,N)$ giving finite-size corrections. With the substitution $N_k(N) = \Pi_{BA}(k) v(k,N)$, the equation (\ref{rN}) can be rewritten in terms of $v(k,N)$:
\bq
	v(k,N) = \frac{3}{2} \delta_{k,1} + \frac{2+k}{2(N-2)} v(k-1,N-1) - \frac{4-2N+k}{2(N-2)} v(k,N-1).
	\label{rv}
\eq
Multiplying now both sides of Eq.~(\ref{rv}) by $k^q$ and summing over $k=1,\dots,\infty$ we get
\bq
	m_q(N+1) =  \frac{1}{2N}\left( 3+ \sum_{i=0}^{q-1} c_{qi} m_i(N) + (2N+q+1)m_q(N) \right), 
        \label{rm}
\eq
where define moments $m_q(N)$ for the distribution $v(k,N)$ as follows:
\bq
	m_q(N) = \frac{1}{N-1} \sum_{k=1}^\infty k^q v(k,N).
\eq
The normalization constant $1/(N-1)$ has been chosen for the later convenience.
The initial condition reads
\bq
	m_q(2) = 3,	\label{ic1}
\eq
for all $q$ as can be found for the initial configuration. The coefficients $c_{qi}$ are given by:
\bq
c_{q0} = 3,\;\; \mbox{and}\; \;\, c_{qi} = 3\binom{q}{i}+\binom{q}{i-1}\;\;  \mbox{for}\;\; i>0.
\label{Cik}
\eq
The equation (\ref{rm}) can be solved recursively starting from the lowest moments $m_0, m_1, m_2, \dots$. From expressions for the first moments we can infer that the general solution has the form:
\bq
	m_q(N) = \frac{1}{\Gamma(N)} \sum_{i=0}^{q+1} \frac{B_{qi}}{\Gamma(2+i/2)} \Gamma(N+i/2), \label{gsm}
\eq
with some coefficients $B_{qi}$, yet unknown. The equation for coefficients $B_{qi}$ can be found by inserting (\ref{gsm}) into Eq.~(\ref{rm}). For large $N$, the leading behavior of $m_q(N)$ is controlled by the term proportional to $B_{q,q+1}$:
\bq
	m_q(N) \simeq B_{q,q+1} \frac{\Gamma\left[N+(q+1)/2\right]}{\Gamma(N)\Gamma(2+(q+1)/2)} 
        \simeq N^\frac{q+1}{2} A_q,
\eq
with $A_q \equiv B_{q,q+1}/\Gamma((5+q)/2)$. Each two consecutive moments $m_{q+1}$ and $m_{q}$ differ by 
a prefactor $N^{1/2}$, so clearly the cutoff function must have the form:
\bq
	v(k,N) \simeq N w(k/\sqrt{N}), \label{wq}
\eq
where $w(x)$ is a universal (independent of $N$) cutoff function having moments 
equal to $A_q$:
\bq
	A_q = \int_0^\infty \dd x\, w(x) x^q.	\label{akw}
\eq
Therefore, the leading correction to the degree distribution for a large but finite BA tree network is
\bq
	\Pi(k,N) = \Pi_{BA}(k) w\left(\frac{k}{\sqrt{N}}\right).
\eq
The exponent $\alpha=1/2$ stemming from this equation agrees with the result for trees presented in Table \ref{tab:cutoff}. The function $w(x)$ can be found in two ways. First, we can evaluate numerically Eq.~(\ref{rv}) for some large $N$ and then rescale variables according to Eq.~(\ref{wq}). Second, it can be obtained analytically by reconstructing it from the moments $A_q$, which express through the coefficients $B_{q,q+1}$. 
Without going into the details we quote the result for the moments $A_q$ \cite{ref:bwis}:
\bq
	A_q = \frac{(2+q)^2 q!}{\Gamma((3+q)/2)}.	\label{akfinal}
\eq
Using the asymptotic behavior of Eq.~(\ref{akfinal}) we can infer the form of the cutoff function $w(x)$ for large values of the argument:
\bq
	\ln A_q \approx \frac{1}{2} q \ln q.	\label{aklarge}
\eq
Let us now compare Eq.~(\ref{aklarge}) with the behavior of moments $I_q$ of the function $\exp\left[ -(x/\sigma)^\rho\right]$:
\bq
	I_q = \int_0^\infty x^q \exp\left[ -(x/\sigma)^\rho\right] \dd x = \frac{\sigma^{q+1}}{\rho} \Gamma
	\left( \frac{q+1}{\rho} \right).	\label{Ik}
\eq
For large $q$ the leading term of $\ln I_q \approx (q \ln q)/\rho$ with $\rho=2$ is the same as in Eq.~(\ref{aklarge}), i.~e. the tail of $w(x)$ defined by its higher moments falls like a Gaussian. The parameter $\sigma$ is found by comparing sub-leading terms in $I_q$ and $A_q$. The value $\sigma=2$ obtained in this way will be confirmed below by a direct calculation of $w(x)$.
To this end we define a generating function:
\bq
	M(z) = \sum_{q=0}^\infty A_q \frac{z^q}{q!}.
\eq
Comparing this definition with Eq.~(\ref{akw}) we see that $M(z) = \int_0^\infty \exp(zx) w(x) \dd x$ so that 
\bq
	M(-z) = \int_0^\infty \exp(-zx) w(x)\, \dd x
\eq
is the Laplace transform of $w(x)$. Therefore $w(x)$ is given by the inverse Laplace transform of $M(z)$ or, equivalently, by the Fourier transform of $M(-iz)$:
\bq
	w(x) = \frac{1}{2\pi i} \int_{-i\infty}^{i\infty} \dd z\, e^{zx} M(-z)	
	= \frac{1}{2\pi} \int_{-\infty}^{\infty} \dd z\, e^{izx} M(-iz).	\label{invlap}
\eq
Using the explicit form of coefficients $A_q$ we get
\bq
  M(z)= \sum_{q=0}^\infty \frac{(2+q)\Gamma(q+3)}{\Gamma(q+2)\Gamma((3+q)/2)} z^q.
	\label{Mzf}
\eq
This series has an infinite radius of convergence. The function $M(z)$ given by Eq.~(\ref{Mzf}) is a special case of a more general power series:
\bq
	M(z) = \mathcal{N} \sum_{q=0}^\infty \frac{(aq+b)\Gamma(q+\xi)}
        {\Gamma(q+\zeta)\Gamma(\chi q+\psi)} z^q,
	\label{MZgen}
\eq
belonging to the class of so called Fox-Wright $\Psi$ functions \cite{ref:moskowitz,ref:pagnini}. In \cite{ref:bwis} it has been shown that its inverse Fourier transform, that is $w(x)$, can be expressed through a combination of auxiliary functions $\tilde{f}_{\chi,\psi,\xi,\zeta}(x)$. In general, they are defined as follows:
\bq
	\tilde{f}_{\chi,\psi,\xi,\zeta}(x) =
	\sum_i \mbox{res}_{s_i} \left[ \frac{\Gamma(\xi-s)\Gamma(1-s)}{\Gamma(\zeta-s)\Gamma(\psi-\chi s)} x^{s-1}
	 \right]_{s=s_i}, \label{fftgen}
\eq
where the sum runs over all points $s_i$ at which either $\Gamma(1-s)$ or $\Gamma(\xi-s)$ has a pole. The above formula simplifies for $\xi,\zeta$ being positive integers $m,n$:
\bq
	\tilde{f}_{\chi,\psi,m,n}(x) = \sum_{q=0}^\infty (-x)^q \frac{(m-2-q)(m-3-q)\cdots (n-1-q)}
{\Gamma(\psi-\chi-\chi q)q!}. \label{fftsimpl}
\eq
The final formula for $w(x)$ for arbitrary $\chi,\psi,\xi,\zeta$ reads
\bq
	w(x) = \mathcal{N} \left( ax \tilde{f}_{\chi,\psi-\chi,\xi-1,\zeta-1}(x)+
	(b-a)\tilde{f}_{\chi,\psi,\xi,\zeta}(x)\right), \label{finalwxgen}
\eq
where $\tilde{f}$ is given either by Eq.~(\ref{fftsimpl}) or by more general Eq.~(\ref{fftgen}).
In our case, which corresponds to $\mathcal{N}=1,a=1,b=2,\chi=1/2,\psi=3/2,\xi=3,\zeta=2$, the function $w(x)$ is given by
\bq
	w(x) = x\tilde{f}_{1/2,1,2,1}(x)+\tilde{f}_{1/2,3/2,3,2}(x) = \sum_{q=0}^\infty \frac{(-x)^q}{q!}
	\left[ \frac{-qx}{\Gamma(1/2-q/2)}+\frac{1-q}{\Gamma(1-q/2)} \right].
\eq
After some algebraic manipulations we get the function $w(x)$ expressed as an infinite series:
\bq
	w(x) = 1-\frac{4}{\sqrt\pi} \sum_{q=1}^\infty x^{2q+1} \frac{(-1)^q q^2}{q!2^{2q}(2q+1)}.
	\label{finalwx}
\eq
One can check that it corresponds to a Taylor expansion of the result given in \cite{ref:kr2}:
\bq
	w(x) = \mbox{erfc}(x/2)+\frac{2x+x^3}{\sqrt{4\pi}} e^{-x^2/4},
\eq
where $\mbox{erfc}(z)$ is the complementary error function. The approximate result of Ref.~\cite{ref:dr1} is close to this exact formula. The series in Eq.~(\ref{finalwx}) is rapidly convergent, and if truncated at some $q_{\rm max}$, can be used in numerical calculations. One sees that the cutoff is indeed of a Gaussian type, with the variance $\sigma^2=4$ in agreement with the asymptotic behavior of $A_q$ discussed above.

\begin{figure}
\center
\psfrag{x}{$x$} \psfrag{wx}{$w(x)$}
\includegraphics[width=11cm]{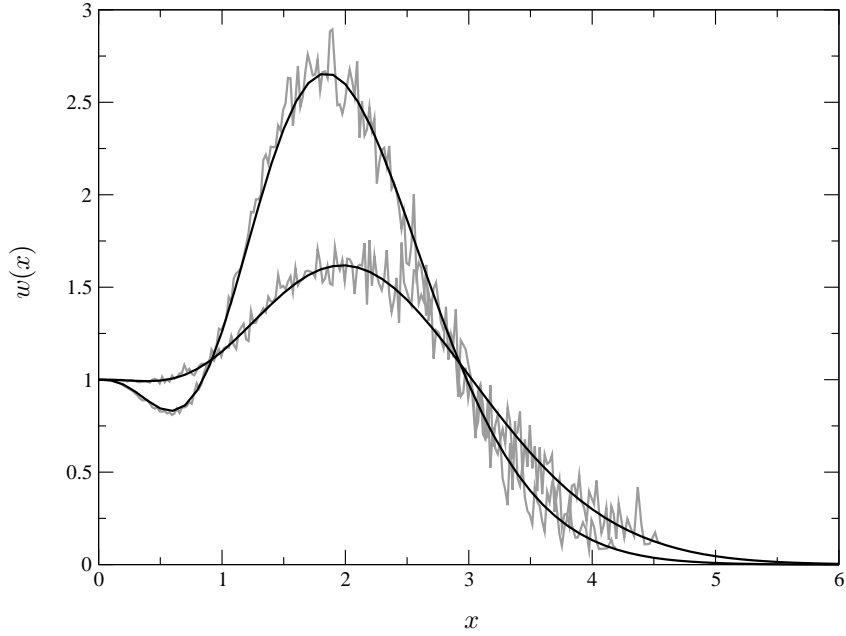}
\caption{\label{bagen2}Plots of the cutoff function $w(x)$ calculated from Eq.~(\ref{wxm0}) for the BA model without initial attractiveness, for the initial graph with $n_0=3$ (black lower line) and $5$ nodes (black upper line) agree very well with $w(x)$ obtained from averaged degree distribution for $2\times 10^4$ generated networks of size $N=10^4$ (gray lines).}
\end{figure}

These calculations can be easily extended to the case of an arbitrary initial graph. For example, if we assume that at the beginning we have a complete graph with $n_0$ nodes, after repeating all the steps of calculations, we will obtain the following formula for the moments:
 \bq
	A_q = \frac{\Gamma(1+n_0-\omega)}{\Gamma(n_0+3/2-\omega+q/2)} \left[
	\frac{\Gamma(4+q)}{2(q+1)} + \frac{m_0(n_0) \Gamma(2+n_0+q)}{\Gamma(n_0+2)} \right]
	\approx \exp\left(\frac{1}{2} q \ln q +O(q)\right),
	\label{akfinalm0}
\eq
with $\omega=n_0(3-n_0)/2$ and the zero-th moment $m_0(n_0)=(n_0+1)n_0/2$ being just the number of links in the initial graph. This allows us to infer the asymptotic behavior of $w(x)$ which is the same as for $n_0=2$. Therefore, the degree distribution for the BA tree model
without initial attractiveness has always a Gaussian cutoff whose position scales as $\sim N^{1/2}$.
The full expression for $w(x)$ takes the form:
\bq
	w(x) = \Gamma(1+n_0-\omega) \left[ \frac{1}{2} \tilde{f}_{\frac{1}{2},n_0+\frac{3}{2}-\omega,
	4,2}(x) + \frac{m_0(n_0)}{\Gamma(n_0+2)} \tilde{f}_{\frac{1}{2},n_0+\frac{3}{2}-\omega,
	2+n_0,1}(x) \right].	\label{wxm0}
\eq
In figure \ref{bagen2} we have plotted $w(x)$ calculated from Eq.~(\ref{wxm0}) together with the results of Monte Carlo simulations for finite-size networks. One readily sees that $w(x)$ strongly depends on the size of the seed graph $n_0$. This sensitivity to the initial conditions has just been reported in \cite{ref:kr2} as well as in Ref. \cite{ref:malarz} where another quantity has been measured.

Let us go now to the case of preferential attachment kernel $k+a_0$, that is to the model with initial attractiveness $a_0>-1$. From Eq.~(\ref{eq2:pikgena0}) we know that $\Pi_\infty(k) \sim k^{-\gamma}$ with the exponent $\gamma=3+a_0$. Like we said, the model is equivalent to the growing network with re-direction (GNR model), described in previous chapter, with the choice of the parameter $r=1/(a_0+2)$. In all numerical simulations showed in this section the GNR model is used. On the other hand, in analytical calculations we shall follow  the procedure, which we describe above for the pure BA model. Assuming that we start from the dimer configuration, using the recursion formula for $N_k(N)$ we get an equation for the moments $m_q(N)$ which can be solved in the form which involves some (yet) unknown coefficients $B_{q,q+1}$:
\bq
	m_q(N) = \frac{1}{\Gamma\left(N-1+\frac{a_0}{2+a_0}\right)(N-1)} \sum_{i=0}^{q+1} B_{qi}
	\frac{\Gamma\left(N+\frac{a_0+i}{2+a_0}\right)}{\Gamma\left(2+\frac{a_0+i}{2+a_0}\right)}.
\eq
For large but finite $N$ it simplifies to
\bq
	m_q(N) \simeq N^\frac{q+1}{2+a_0} A_q,
\eq
with $A_q=B_{q,q+1}/\Gamma\left[2+\frac{a_0+q+1}{2+a_0}\right]$. Therefore, the function $v(k,N)$ obeys the following scaling rule:
\bq
	v(k,N) \to N w\left(k/N^\frac{1}{2+a_0}\right), \label{wqgen}
\eq
where the function $w(x)$ has now moments $A_q$ depending on $a_0$. Equation (\ref{wqgen}) indicates that the cutoff scales as $N^{1/(\gamma-1)}$ where $\gamma=3+a_0$ is the exponent in the power law for $\Pi_\infty(k)$. This is in agreement with the result presented in table \ref{tab:cutoff}. For given $N$, the cutoff decreases when the exponent $\gamma$ increases. Practically, this implies that the power law in the degree distribution can hardly be seen for $\gamma>4$ because even for large networks with $N=10^6$ nodes the cutoff corresponds to the value of $k_{\rm max}\sim 100$ and the power-law extends maximally over $1-2$ decades in $k$. This partially explains the fact why S-F networks with $\gamma$ above $4$ are practically never encountered \cite{ref:cn2}.

\begin{figure}
\center
\psfrag{x}{$x$} \psfrag{wx}{$w(x)$}
\includegraphics[width=11cm]{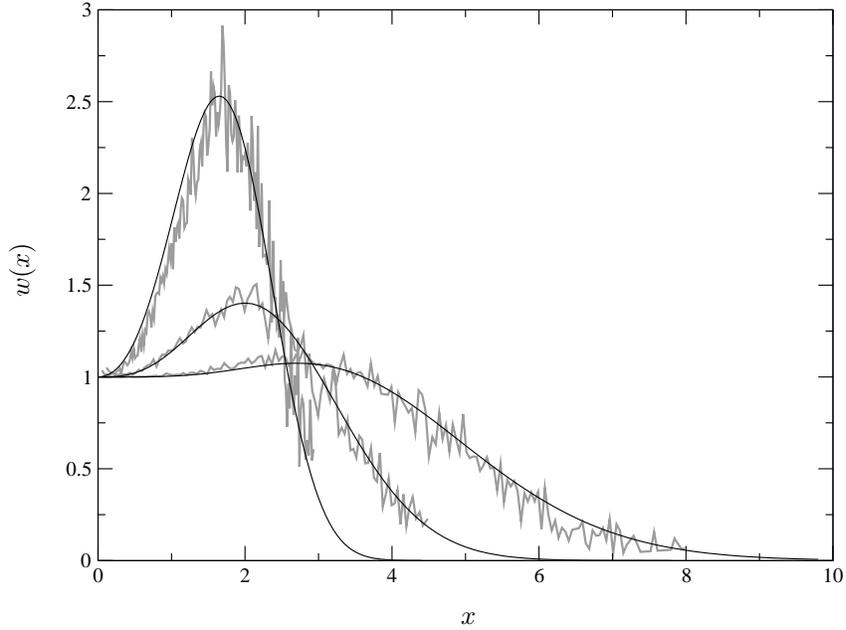}
\caption{\label{fig:bagen1}Plots of $w(x)$ calculated from Eq.~(\ref{finalwxgen}) for $a_0=-0.5$, $0$ and $1$ (solid lines from top to bottom) which correspond to $\gamma=2.5,3$ and $4$, respectively. The curves become flat with increasing $a_0$. The thick gray lines are $w(x)$ obtained from averaged degree distribution for $2\times 10^4$ generated networks of size $N=10^4$. The tails decay as $\exp(-x^\rho)$ with $\rho=3,2,3/2$, respectively, in agreement with numerical findings.}
\end{figure}

The moments $A_q$ can be found:
\bq
	A_q = 2\frac{\Gamma\left(1+\frac{a_0}{2+a_0}\right)}{\Gamma(5+2a_0)}
	 \frac{\left[ 6(q+2)+a_0(13+4a_0+3q)\right]
		\Gamma\left(4+2a_0+q\right)}{(1+q)\Gamma\left(\frac{5+3a_0+q}{2+a_0}\right)}.
		\label{akfinalgen}
\eq
The equation (\ref{akfinalgen}) is much more complicated than Eq.~(\ref{akfinal}) but it reduces to it for $a_0=0$. 
The leading term of $A_q$ is
\bq
	\ln A_q \approx \frac{1+a_0}{2+a_0} q \ln q.
\eq
Comparing this to Eq.~(\ref{Ik}) as it has been done before, one sees that for large $x$ the function $w(x)$ decays like $\exp\left[ -(x/\sigma)^\rho\right]$ 
with
\bq
	\rho=\frac{2+a_0}{1+a_0}=\frac{\gamma-1}{\gamma-2}.
\eq
This agrees very well with numerical results. The cutoff for $\gamma\neq 3$ is no longer Gaussian. For $2<\gamma<4$, as often found in real networks, $\rho$ is always greater than $1.5$ and therefore the finite-size cutoff cannot be approximated by a pure exponential decay, observed in some networks \cite{ref:cn2}. Exponential cutoffs found in such networks probably have different origin \cite{ref:stumpf}. The formula for $M(z)$ is still given by Eq.~(\ref{MZgen}) with the parameters $a,b,\dots$ expressed through $a_0$. The cutoff function $w(x)$ is given by Eq.~(\ref{finalwxgen}). For instance, for $a_0=-1/2$ which corresponds to $\gamma=2.5$ we get
\bq
	w(x) = \frac{\Gamma(2/3)}{3} \sum_{q=0}^\infty \frac{(-x)^q}{q!} \left[ x\frac{-9q}{2\Gamma(1-2q/3)}
	+ \frac{2-2q}{\Gamma(5/3-2q/3)} \right].
\eq
In Fig.~\ref{fig:bagen1} we plot $w(x)$ for $a_0=-1/2,\,0$ and $1$. For numerical calculations all series have been truncated. The truncation error is less than $10^{-4}$ in the plotted area. The results show that the curves become more flat when $a_0$ increases and agree well with $w(x)$ obtained in simulations of finite-size networks.

The initial graph has a great influence on the functional form of $w(x)$. We do not consider here the dependence on the size $n_0$ of the seed graph, but one can show that asymptotic properties of the cutoff function are insensitive to $n_0$ and therefore for $x$ being sufficiently large, $w(x)\sim \exp\left[ -(x/\sigma)^\rho\right]$ depends only on $a_0$, i.~e. only on the exponent $\gamma$ in the power-law $\Pi(k)\sim k^{-\gamma}$.

So far we have considered the model with $m=1$, restricting ourselves to the case when graphs are essentially trees and possible cycles can only come from the seed graph. The general case $m>1$ is much more complicated. Each of $m$ proper links of a newly introduced node has to be connected to one of $N$ preexisting nodes according to the preferential attachment rule. However, since multiple links are not allowed, the nodes to which links have been connected on this step have to be excluded from the set of nodes available for further linking. Thus, when a new proper link of a node is introduced, the probabilities of attaching it to one of the preexisting nodes are different depending on whether the link is the first, second, etc., of $m$. The rate equation for $N_k(N)$ can still be obtained in this case, but its structure is very involved. For example, for $m=2$ and $n_0=3$ (triangle as a seed graph) the full rate equation for $N_k(N)$ reads:
\ba
	N_k(N+1) = N_k(N) + \delta_{k,2} + \frac{1}{4N-6}\left[(k-1) \left(1-\frac{k-1}{4N-5-k}+S_N \right)N_{k-1}(N) \right. \nonumber \\
	\left. -k\left(1-\frac{k}{4N-6-k}+S_N \right)N_{k}(N) \right],
	\label{rNm=2}
\ea
where $S_N$ denotes the auxiliary quantity:
\bq
	S_N = \sum_{k=1}^N \frac{kN_k(N)}{4N-6-k}.
\label{EqS}
\eq
Due to the presence of the $S_N$ term, Eq.~(\ref{rNm=2}) is nonlinear in $N_k$, and contains $k$ in denominators. This makes impossible to  apply our method in a straightforward way to the exact equation. In this case the approximations are needed. The equations for $m>2$ are even more complicated because of increasing number of possible ways of distributing $m$ links at each time step. 

We can, however, make the following approximation. When the total number of links $L$ is large, the probability that at each step we pick up two or more links pointing onto the same node is small. When the size $n_0$ of the initial graph is much larger than $m$, this condition is fulfilled from the beginning and we expect that this approximation should work good. Within this approximation, the rate equation takes the form:
\bq
	N_k(N+1) = N_k(N) + \delta_{k,m} +\frac{k-1}{2(N-\omega)} N_{k-1}(N) - \frac{k}{2(N-\omega)} N_{k}(N).
	\label{rNm0}
\eq
The Kroenecker delta stands for the addition of one node with $m$ links at each time step. The remaining terms give the probability of preferential attachment like in Eq.~(\ref{rN}). The denominators must give the normalization $\sum_k k N_k(N)=2L$. Assuming that we start from a complete graph with $n_0$ nodes, we get the number of links $L=m(N-\omega)$ with 
\bq
	\omega=n_0(2m+1-n_0)/(2m).	\label{omm0}
\eq
The factor $m$ coming from $2L$ in the denominator cancels out with $m$ coming from $m$ possibilities of choosing links at each step. The choice of the same name ``$\omega$'' above, as in Eq.~(\ref{wxm0}) is not accidental. In fact, the cutoff function $w(x)$ is now given by a very similar formula:
\bq
	w(x) = \Gamma(1+n_0-\omega) \left[ \frac{1}{\Gamma(m+2)} \tilde{f}_{\frac{1}{2},n_0+\frac{3}{2}-\omega,
	3+m,2}(x) + \frac{m_0(n_0)}{\Gamma(n_0+2)} \tilde{f}_{\frac{1}{2},n_0+\frac{3}{2}-\omega,
	2+n_0,1}(x) \right],	\label{wxm2}
\eq
with $m_0(n_0)=n_0(n_0+1)/(m+1)$, which takes the form of Eq.~(\ref{wxm0}) for $m=1$. Thus the same scaling $k_{\rm max}\sim N^{1/2}$ holds also here.

Like before, we expect some dependence on the initial graph, but as far as the asymptotic properties of $w(x)$ are concerned the dependence should be negligible. Therefore, one should take the simplest possible initial configuration. The most natural choice is the complete graph with $n_0=m+1$ because then $\Pi(k)=0$ for all $k<m$ at each step of the growth process. However, one should remember that the approximation used here works well only for $m\ll n_0$ because Eq.~(\ref{rNm0}) with $\omega$ given by Eq.~(\ref{omm0}) approximates reasonably the full rate equation only if $m \ll N$ at each stage of the network growth. 

\begin{figure}
\center
\psfrag{x}{$x$} \psfrag{wx}{$w(x)$}
\includegraphics[width=11cm]{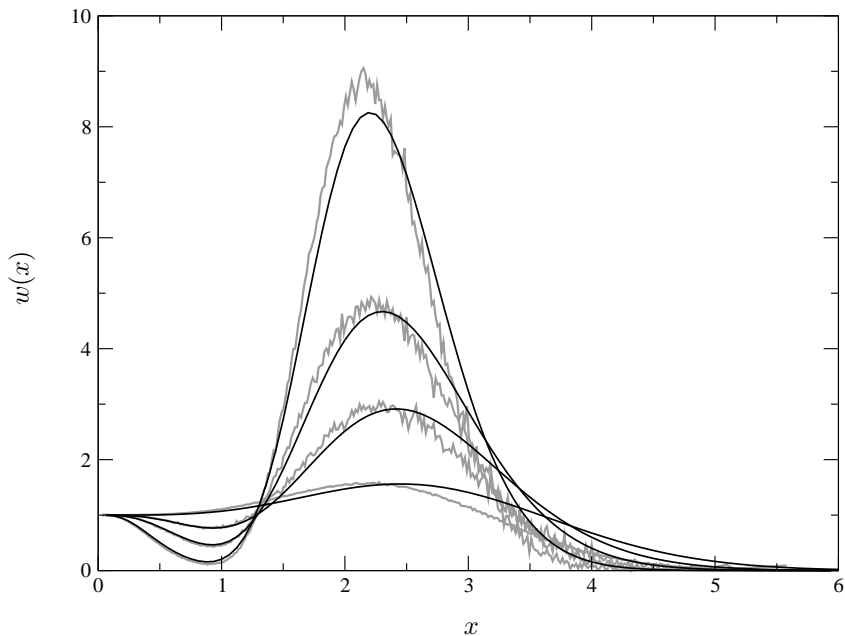}
\caption{\label{bagen3}The function $w(x)$ for $m=2$ and for $n_0=3,7,10$ and $15$ (curves from the flattest to the most peaked). The agreement between the analytical solution and the results of computer simulations is not as good as before due to approximate character of the solved equation. The experimental points are obtained for $N=4000$. The dip at about $x=1$ and the peak at about $x=2$ especially pronounced in the case of larger seed graphs mean that much more nodes with high degrees is present than it would be expected for the asymptotic power-law behavior of $\Pi_\infty(k)$.}
\end{figure}

In figure \ref{bagen3} we compare our approximate analytical solution with Monte Carlo simulations of BA networks initiated from complete graphs with different $n_0$. One can see a small deviation between the analytical and numerical curves. The largest difference is for $n_0=3$ and the smallest for $n_0=15$, confirming our earlier conclusion that the approximation is better for larger seed graphs. The asymptotic form of $w(x)\sim \exp(-x^2/4)$ is the same as for growing BA trees, regardless of how many new nodes $m$ we add per one time step. 

\subsection{Numerical simulations of equilibrated networks}
We have performed extensive numerical simulations of various networks 
to cross-check analytic results for the scaling of the cutoff function. 
The results for growing networks have been just presented in the previous section. As we saw, they were in a very good agreement with theoretical predictions, and also consistent with earlier results presented in table \ref{tab:cutoff}. Now we shall describe results of the numerical computation of the cutoff for equilibrated graphs. We used the Monte Carlo generator described in Section 2.1.5 and in Ref.~\cite{ref:naszcpc}. It performs a weighted random walk in the configuration space of the canonical ensemble. Each elementary step of the random walk is done using the T-rewiring and accepted with the Metropolis probability. We simulated three ensembles: equilibrated trees, equilibrated simple graphs, and equilibrated degenerated graphs. In all the cases we used basically the same algorithm except that in the first one we rejected rewirings violating the tree structure by introducing a cycle; in the second we rejected moves leading to multiple- or self-connections. This resulted in lowering the acceptance rate and algorithm efficiency, especially for tree graphs, in which case we had to extend the simulation time appropriately.

\begin{figure}
\center
\psfrag{xx}{$x$} \psfrag{yy}{$V_{\rm exp}(x)$} \psfrag{zz}{$N^{\alpha(\gamma-1)}P_{\rm exp}(x)$}
\psfrag{g=3}{$\gamma=3$}\psfrag{g=2.5}{$\gamma=2.5$}\psfrag{g=3.5}{$\gamma=3.5$}
\psfrag{a=0.5}{$\alpha=0.5$}\psfrag{a=0.55}{$\alpha=0.55$}\psfrag{a=0.4}{$\alpha=0.4$}\psfrag{a=0.667}{$\alpha=0.667$}
\includegraphics*[width=15cm,bb=13 28 792 596]{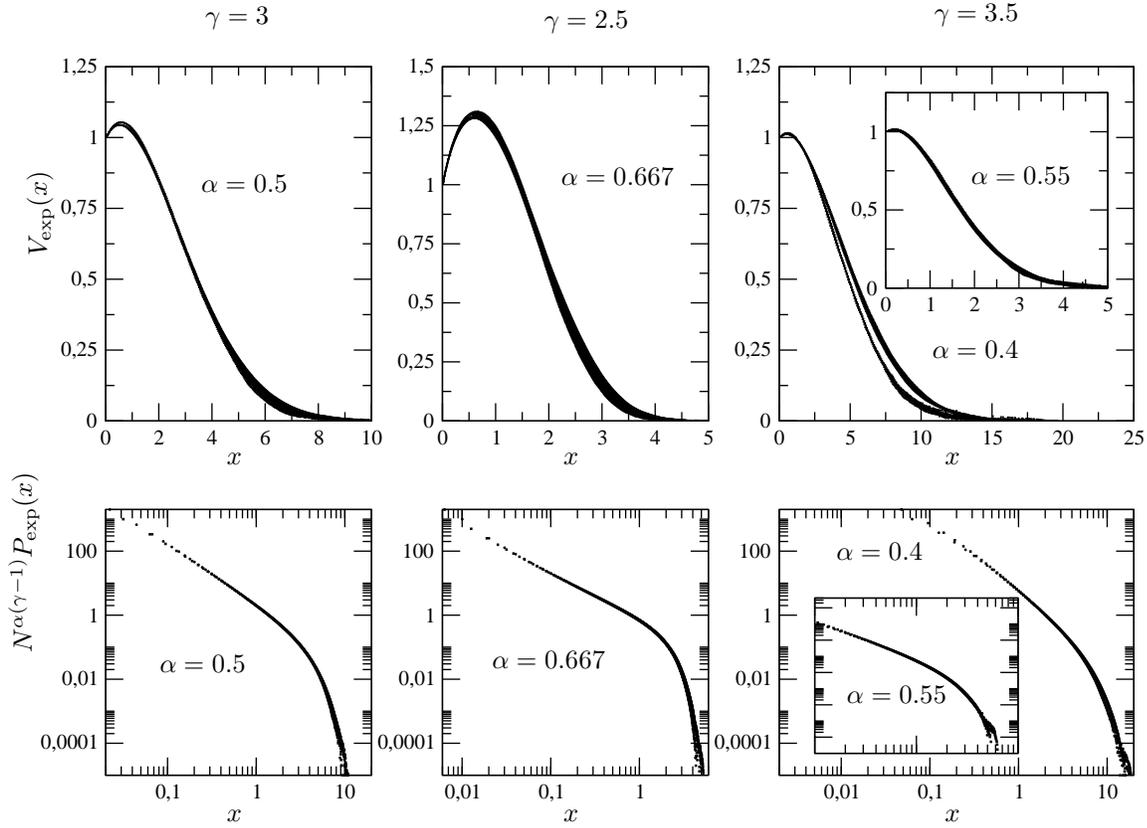}
\caption{\label{fig:cuttreeh}Top: plots of experimentally found cutoff function $V(k,N)$ for equilibrated trees, plotted in the rescaled variable $x=k/N^{\alpha}$. The curves for $N=1000$ and $N=2000$ collapse nicely for theoretical values of the cutoff exponent $\alpha$ for the first two plots ($\gamma=3$ and $2.5$) but significant difference is observed for $\gamma=3.5$. A better collapse has been found for $\alpha=0.55\pm 0.03$ (inset) than for theoretical value $0.4$. Bottom: plots of rescaled cumulative degree distributions $N^{\alpha(\gamma-1)}P(x)$ versus $x$.}
\end{figure}

In simulation of trees, as an initial configuration we chosen a GNR network with $a_0=\gamma-3$ adjusted to have the desired value of the exponent $\gamma$  in the tail of the node degree distribution. The asymptotic degree distribution, given by Eq.~(\ref{eq2:pikgena0}), has $\left<k\right> = 2$ for infinite GNR trees as it should\footnote{Since  $L=N-1$, the average degree is in fact $2-2/N$ but it converges fast to the asymptotic result.}. In order to
preserve this distribution in the process of homogenization we had to set the ratio-weight function $w(k)$ according to Eqs.~(\ref{eq2:pikv0tree}) and (\ref{eq2:wwdef}):
\bq
	w(k) = \frac{k(k+a_0)}{k+3+2a_0}.
\eq
This choice ensures that the mean value of the degree distribution stays at its ``critical'' value equal to $\bar{k}=2L/N \approx 2$ for trees. For densities of links below the critical one, one would observe an exponential suppression of the degree distribution for large $k$, and for the average degree above two, the scale-free behavior would be disturbed by the surplus of highly connected nodes, or even by a condensation of links at some singular node.
We simulated ensembles for three exponents $\gamma=2.5,\,3$ and $3.5$ for which the scaling exponent $\alpha$ should be $0.667$, $0.5$ and $0.4$, respectively. For each of them we took trees of two different sizes $N=1000$ and $N=2000$, and for each we made between four and six independent runs in order to estimate errors by means of the standard Jackknife method \cite{ref:jack}. We measured the degree distribution as well as a cumulative degree distribution (c.d.d.) defined as
\bq
	P(k) = \sum_{q=k}^\infty \Pi(q).
\eq
Because the degree distribution has the power-law tail $\Pi(k)\sim k^{-\gamma}$, the corresponding cumulative distribution behaves as $P(k)\sim k^{-\gamma+1}$. Any cutoff effects should be clearly visible also in $P(k)$. The advantage of using the cumulative distribution is that one does not need to make binning to reduce statistical errors. One makes a rank plot instead. From Eq.~(\ref{eq2:pikgena0}) we get the following formula for theoretical $P(k)$ for infinite graphs:
\bq
	P_\infty(k) = \frac{\Gamma(3+2a_0)\Gamma(k+a_0)}{\Gamma(1+a_0)\Gamma(2+2a_0+k)}.
\eq
For any large but finite network we expect, similarly as in section 3.1.2, some cutoff so that $P(k) \approx P_\infty(k)V(k/N^{\alpha})$. Here $V(x)$ would be some universal function. If it is so, we should observe a collapse of curves $P_{\rm exp}(x)/P_\infty(x)$ plotted in the rescaled variable $x=k/N^{\alpha}$. In figure \ref{fig:cuttreeh} we see that such a collapse indeed takes place for $\gamma=3$ and $2.5$. This means that theoretical values of $\alpha$ (see table \ref{tab:cutoff}) agree very well with the experiment. However, for $\gamma=3.5$ the collapse is much better for $\alpha=0.55\pm 0.03$ than for the theoretically predicted value $0.4$. This means that there are more nodes with high degrees than it is expected. In \cite{ref:extr2} it has been suggested, that in the case of disassortative networks like equilibrated trees presented here, the cutoff might be higher than $1/(\gamma-1)$. On the other hand, the assortativity coefficient $\mathcal{A}$ increases\footnote{It can be shown that for the scale-free degree distribution (\ref{eq2:pikgena0}), $\Tr \eps$ grows with increasing $\gamma$ and so grows the coefficient $\mathcal{A}$.} with $\gamma$ so the assortativity is bigger for $\gamma=3.5$ than for $2.5$ where we observe a perfect agreement. So it is not clear whether indeed the argument of Ref.~\cite{ref:extr2} is entirely correct. 

Let us now discuss equilibrated simple graphs. We will generate weighted graphs with the degree distribution (\ref{eq2:pikgena0}). In our Monte Carlo generator (see Sect. 2.1.5) we have to set the weight from Eq.~(\ref{eq2:wwdef}) to
\bq
	w(k) = \frac{(k+1)(k+a_0)}{k+3+2a_0}, \label{wqhg}
\eq
in order to get the stationary distribution given by Eq.~(\ref{eq2:pikgena0}) in the limit of $N\to \infty$. As before, we must keep the average degree equal to $2$, which is the mean value of the distribution (\ref{eq2:pikgena0}). As the initial graph we have chosen again a GNR tree, because from the very beginning it has the correct degree distribution equal to that produced by the graph rewiring process in the course of thermalization. The final results are the same when one begins with any random graph with $N=L$ but the convergence to the asymptotic distribution might be in this case much slower.

\begin{figure}
\center
\psfrag{xx}{$x$} \psfrag{yy}{$V_{\rm exp}(x)$} 
\psfrag{g=3}{$\gamma=3$}\psfrag{g=2.5}{$\gamma=2.5$}\psfrag{g=3.5}{$\gamma=3.5$}
\psfrag{a=0.5}{$\alpha=0.5$}\psfrag{a=0.38}{$\alpha=0.38\pm 0.02$}\psfrag{a=0.4}{$\alpha=0.40\pm 0.02$}\psfrag{a=0.35}{$\alpha=0.35\pm 0.03$}
\includegraphics*[width=15cm,bb=7 26 773 585]{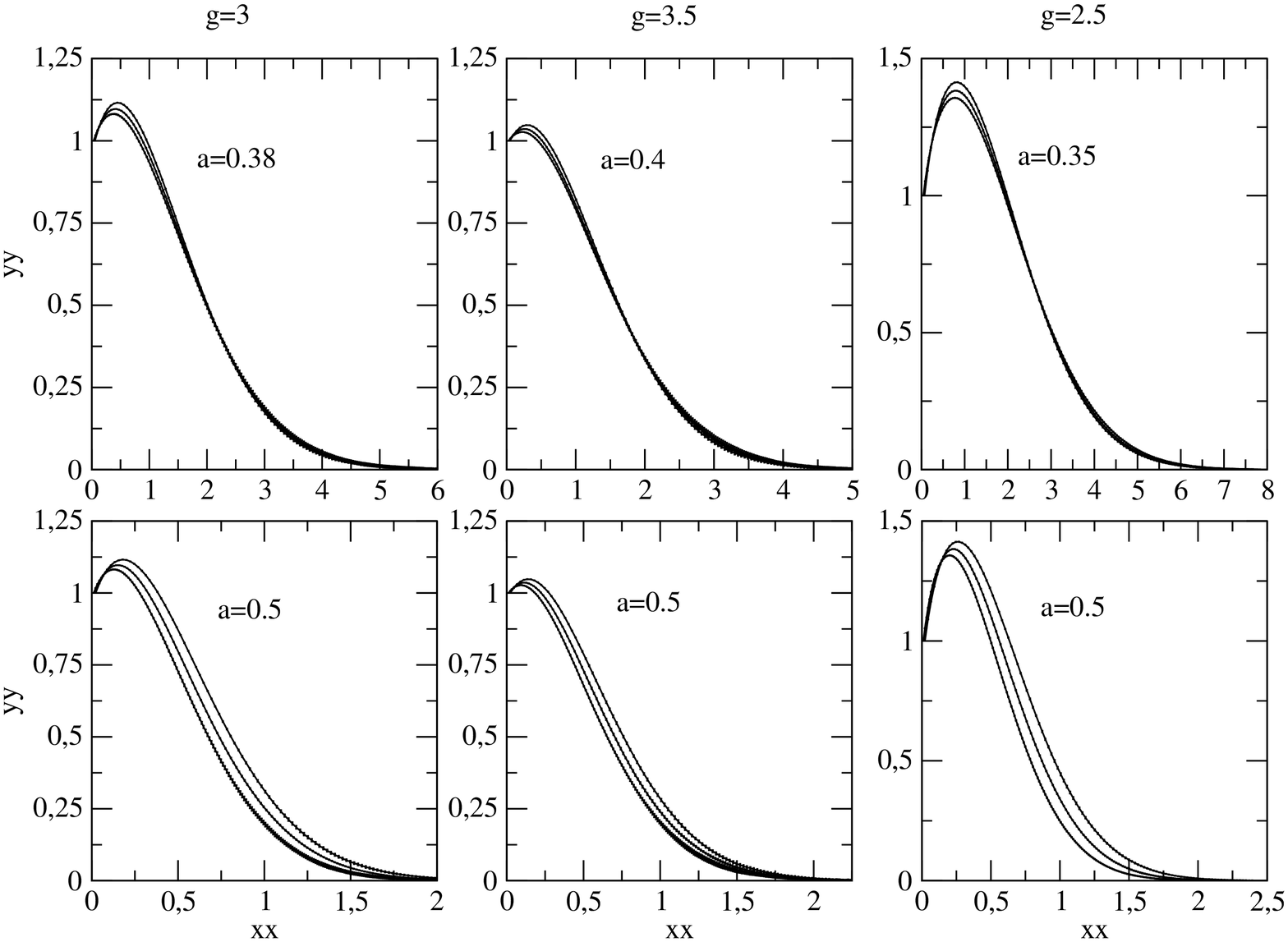}
\caption{\label{fig:cutgraph}Top: plots of experimentally found cutoff function $V(x)$ in the rescaled variable $x=k/N^{\alpha}$ for equilibrated graphs of different sizes, for $\gamma=2.5, \, 3$ and $3.5$. Top: plots for $\alpha$ chosen to obtain the best collapse of data (by-eye fit). Bottom: the same data but for $\alpha$ predicted in papers \cite{ref:dr3} and \cite{ref:zbak2} (the same result for both) for $\gamma=3$, in \cite{ref:dr3} for $\gamma=3.5$ and in \cite{ref:zbak2} for $\gamma=2.5$, in order to show the discrepancy with numerical simulations.}
\end{figure}

We simulated four ensembles: with $\gamma=3,\, 3.5,\, 2.5$ and $2.1$, each of them for three sizes $N=2000,4000$ and $8000$. The acceptance rate of the algorithm was better than in case of trees and thus we were able to examine larger systems. Before starting simulations we suspected that the data would collapse to a scaling function $V(x)$ better than for trees, because of less structural constraints.
Surprisingly, as we see in figure \ref{fig:cutgraph}, the collapse is worse and moreover, it takes place for different values of $\alpha$ than those give in table \ref{tab:cutoff} and predicted in either \cite{ref:dr3} or \cite{ref:zbak2}. In particular, for $\gamma=3$ where one expects $\alpha=1/2$, we measured $0.38\pm 0.02$.
For the case $\gamma=3.5$ we found that the measured value $0.40\pm 0.02$ agrees with \cite{ref:zbak2} which predicts $\alpha=0.4$, while for $\gamma=2.5$, $\alpha=0.35\pm 0.03$ is closer to the result of \cite{ref:dr3} which predicts $\alpha=0.4$.
For $\gamma=2.1$ (not shown in the picture) we found $\alpha=0.33\pm 0.01$, which also agrees quite well with \cite{ref:dr3}. To summarize, the numerical results presented in this section do not give a conclusive evidence which of the theoretical predictions, \cite{ref:dr3} or  \cite{ref:zbak2}, for the scaling exponent $\alpha$ of the cutoff is better. Actually, for $\gamma>3$ the numerical value is closer to that of \cite{ref:zbak2} while for $2<\gamma<3$ to that of \cite{ref:dr3}. One should, however, keep in mind that the numerical results are based on relatively small systems. For such networks, subleading finite-size corrections may be important and may interfere with the leading scaling behavior $k_{\rm max}\sim N^{\alpha}$. The question how $\alpha$ depends on $\gamma$ for equilibrated S-F networks remains open.

\begin{figure}
\center
\psfrag{xx}{$x$} \psfrag{yy}{$V_{\rm exp}(x)$} 
\psfrag{g=3}{$\gamma=3$}\psfrag{g=2.5}{$\gamma=2.5$}\psfrag{g=3.5}{$\gamma=3.5$}
\psfrag{a1}{$\alpha=0.55\pm 0.03$}\psfrag{a2}{$\alpha=0.52\pm 0.02$}\psfrag{a3}{$\alpha=0.66\pm 0.02$}
\includegraphics*[width=15cm,bb=24 274 727 596]{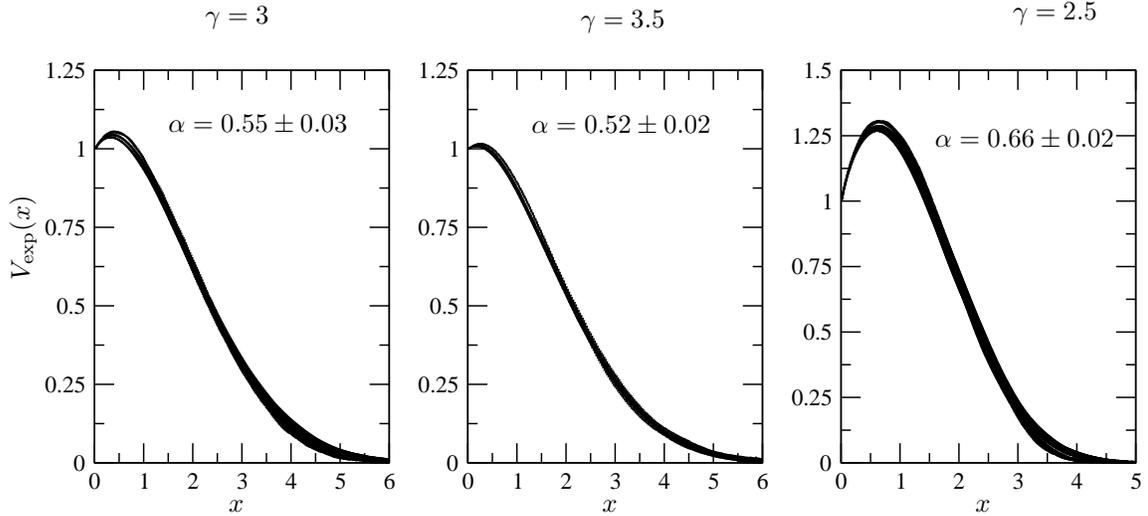}
\caption{\label{fig:pseudograph}Experimentally found $V(x)$ for equilibrated pseudographs of different sizes, for $\gamma=2.5, \, 3$ and $3.5$. The exponent $\alpha$ shown in figures gives the best collapse of curves.}
\end{figure}

Let us now discuss Monte Carlo results for pseudographs. As before we simulated ensembles for $\gamma=2.5,\, 3$ and $3.5$, for $N=1000,2000,4000$. The weight function $w(k)$ is the same as in Eq.~(\ref{wqhg}). \mbox{In figure \ref{fig:pseudograph}} we show the function $V(x)$ obtained from rescaled numerical data: $x=k/N^\alpha$. The exponent $\alpha$ has been chosen in order to obtain the best overlap of the data for different $N$. We see that for pseudographs the numerical results are consistent with theoretical predictions. In table \ref{tab:cutoff} we have $\alpha=1/2$ for $\gamma=3,\, 3.5$ and $2/3$ for $\gamma=2.5$, while the corresponding exponents determined experimentally are $0.55\pm 0.03,\, 0.51\pm 0.02$ and $0.66\pm 0.02$, respectively. For $\gamma=3$ the experimental value deviates a bit from the theoretical ones, while for the two remaining values of $\gamma$ the agreement is perfect. It is not surprising that experimental results for pseudographs exhibit the best agreement with theory, since they have almost no constraints on the structure of graph. In fact, they may be effectively reduced to the balls-in-boxes model \cite{ref:bbj}, which shall be discussed in next section.

\section{Dynamics on networks}
So far we have discussed networks as purely geometrical objects. However, many networks represent not only relationships between nodes but often they are viewed as a backbone of the complex system on which signals, matter, or other degrees of freedom can propagate. For example, networks are a convenient way of thinking about flows, transport, signal propagation, and information spreading between different objects. In other words, one is often interested in dynamics of some degrees of freedom which reside on the network and undergo some dynamical evolution transmitted by the structure of the network. A good example is the air transportation network with airports as nodes and flight connections as links. The dynamical variable in this case is the number of passengers or goods. The flow is proportional to the intensity of the air traffic between airports.
Another example is the voter model \cite{ref:voter,ref:voter2} which is used to mimic an opinion formation among different individuals.
The quantity which is this case ``transmitted'' on the network of acquaintance relations is the opinion. The distribution of opinions on the network is represented by discrete variables defined on nodes. In the simplest version, nodes change their state by copying the state of a randomly chosen neighbor. 
There are many other examples: the traffic of data packets in the Internet \cite{ref:int1}, epidemic spreading (``ill'' nodes infect other nodes) \cite{ref:ep1,ref:ep2,ref:ep3}, synchronization of coupled oscillators \cite{ref:osc1,ref:osc2,ref:osc3}, etc.

It is worth mentioning, that the observed dynamical properties in models defined on scale-free networks are often quite different from those on regular lattices. It is so because S-F networks have two important properties mentioned at the beginning of this thesis: the distance between every pair of nodes is relatively small, and the power law in the degree distribution leads to the emergence of nodes with a high number of connections. This inhomogeneity in the degree distribution seems to play a very important role. In this chapter we shall examine the influence of the node degree inhomogeneity on the dynamics of systems defined on networks. We shall use a very simple model, called zero-range process. This model is easy to handle analytically. For example, one can show that its steady state has a simple factorized form which makes it possible to solve the statics of the model analytically. Despite its simplicity, the model exhibits a very interesting behavior, e.g. it has a phase transition between condensed and uncondensed state.

In the first subsection we shall define a zero-range process and review its basic properties. In particular, we shall discuss the criterion for the condensation on homogeneous and inhomogeneous networks. Next, we shall discuss in details a derivation of most important results in the statistical ensemble approach.
In subsection 3 we shall study the dynamics of the condensate. We shall describe how the condensate is formed from a diluted state and how it behaves once it is formed. For example, we shall ask how much time the condensate needs to disappear from the node it occupies. In the last subsection we shall show how to obtain scale-free fluctuations in the inhomogeneous system. 

\subsection{Zero-range process}
The zero-range process (ZRP) is a particularly simple diffusive system which describes dynamics of balls (particles) on a given network. The balls hop from site to site on the network and the hopping rate depends only on the number of balls at that site from which the ball hops.
Despite its simplicity the model exhibits many interesting properties like phase separation, phase transition, long-range fluctuations and spontaneous symmetry breaking, observed in more complicated systems with mass transport.
Therefore, it has attracted a great attention recently \cite{ref:cc1,ref:cc2,ref:cc3,ref:cc4,ref:cc5,ref:cc6}.
In comparison to more realistic models, it has one advantage: the steady state of the system is known exactly and it assumes a very simple, factorized form. It is worth to mention here that static properties of the model are the same as in the balls-in-boxes model \cite{ref:bbj,ref:grav} developed earlier and successfully applied to explain such phenomena as for instance wealth condensation \cite{ref:wealth}, emergence of the Hagedorn fireball in hadron physics \cite{ref:haged1,ref:haged2} or a collapse of random geometry in the quantum gravity \cite{ref:grav,ref:grav2}. Although the statics of the zero-range process is relatively well known, its dynamics, in case when it takes place in an inhomogeneous system, has not been yet so thoroughly studied.

We shall consider a zero-range process on a connected simple graph with $N$ nodes and $L$ links. Each node $i$ of the graph is occupied by $m_i\geq 0$ identical balls and the total number of balls is fixed and equal to $M$.
The system undergoes the following dynamics: balls hop from non-empty nodes with rate $u(m)$, which depends only on the node occupation number $m$, to one of the nearest neighbors, chosen with equal probability. The function $u(m)$ is any non-negative function defined for $m=0,1,2,\dots\,$. For a node which has $k$ neighbors, the hopping rate per link emerging from this node is equal to $u(m)/k$ since each link is chosen with equal probability $1/k$.

It is easy to implement this type of dynamics on a computer. At each time step we pick $N$ nodes at random. From each of these nodes, occupied by at least one ball, a ball is moved to a node chosen with equal probability from its $k_i$ nearest neighbors. The move is accepted with probability proportional to $u(m_i)$, otherwise it is discarded. The jumping rate $u(m_i)$ must be properly normalized to be less than $1$. With such a definition, one unit of time corresponds to one sweep of the system that is to $N$ attempts of moving a ball.

The ZRP has a steady state. Following Ref.~\cite{ref:evans}, we shall present here a short derivation for an arbitrary network having adjacency matrix $A_{ij}$.
We are interested in the probability $P(m_1,\dots,m_N)$ of finding the system in a particular state with given number of balls on each site. In the stationary state, this probability must be constant, as a result of balance between hopping into and out of the given configuration:
\bq
	u(m_i)P(m_1,\dots,m_N) = \left[ \sum_{j\neq i} A_{ij} \frac{1}{k_j} u(m_j+1)P(\dots,m_j+1,\dots,m_i-1,\dots) \right] ,
	\label{eq3:pmm}
\eq
for each node $i$. The sum over $j$ includes only neighbors of node $i$, each of them gives the contribution $\propto 1/k_j$ since it has $k_j-1$ other neighbors than $i$. Assume now that $P(m_1,\dots,m_N)$ factorizes into some functions $\tilde{f}_i(m_i)$:
\bq
	P(m_1,\dots,m_N) = \frac{1}{Z(N,M)} \prod_{i=1}^N	\tilde{f}_i(m_i),
\eq
where $Z(N,M)$ is an appropriate normalization. Inserting this formula into Eq.~(\ref{eq3:pmm}) we have:
\bq
	\sum_{j\neq i}  A_{ij} \left[ \frac{1}{k_j} u(m_j+1) \tilde{f}_j(m_j+1)\tilde{f}_i(m_i-1) - \frac{1}{k_i} u(m_i)\tilde{f}_i(m_i)\tilde{f}_j(m_j) \right] = 0.
\eq
This equation is fulfilled only if each term of the sum over $j$ vanishes separately:
\bq
	\frac{1}{k_j} u(m_j+1) \frac{\tilde{f}_j(m_j+1)}{\tilde{f}_j(m_j)} = 
	\frac{1}{k_i} u(m_i) \frac{\tilde{f}_i(m_i)}{\tilde{f}_i(m_i-1)}.
\eq
The left-hand side depends on $m_j$ while the right-hand side on $m_i$. To be equal for any $m_i$ and $m_j$ they have to be a constant function, independent of $m$. Without loss of generality we can set it equal to one.
We get:
\bq
	\tilde{f}_i(m_i) = \frac{k_i}{u(m_i)} \tilde{f}_i(m_i-1).	\label{eq3:ppi-1}
\eq
Iterating this equation we come to the formula for $\tilde{f}_i(m)$:
\bq
	\tilde{f}_i(m) = k_i^{m_i} f(m_i),
\eq
where we have introduced the function $f(m)$ which is independent of $k_i$ and reads:
\bq
	f(m)= \prod_{k=1}^m \frac{1}{u(k)}, \;\;\; f(0)=1.	\label{pbyu}
\eq
The splitting into a site-dependent and a site-independent part is convenient when one considers regular graphs having all degrees $k_i$ the same.
The partition function $Z(N,M,\{k_i\})$ being a sum over all states $m_1,\dots,m_N$ has the form:
\bq
	Z(N,M,\{k_i\}) = \sum_{m_1=0}^M \cdots \sum_{m_N=0}^M \delta_{\sum_{i=1}^N m_i, M}	\prod_{i=1}^N f(m_i) k_i^{m_i}.	\label{part}
\eq
Here the Kronecker delta gives the conservation law of the total number of balls. For convenience we shall denote $Z(N,M,\{k_i\})$ of the original graph in short by $Z(N,M)$, skipping  the dependence on the sequence of degrees $\{k_i\}$. The partition function (\ref{part}) encodes the whole information about static properties of the system. It depends only on the node degrees and a detailed structure of the graph has no meaning. As we shall see, also dynamical quantities, like the typical life-time of the condensate, are mainly characterized only by the degree sequence $\{k_i\}$, if the graph has a small diameter.

In order to study static and dynamic behavior of the ZRP it is convenient to define an effective distribution of balls $\pi_i(m)$, that is the probability that site $i$ is occupied by $m$ balls, averaged over all configurations: $\pi_i(m) = \left<\delta_{m,m_i}\right>$. It can be calculated as follows:
\bq
	\pi_i(m_i) = \sum_{m_1} \cdots \sum_{m_{i-1}} \sum_{m_{i+1}} \cdots \sum_{m_N} P(m_1,\dots,m_N)
	\;\delta_{\sum_{i=1}^N m_i, M} = \frac{Z_i(N-1,M-m_i)}{Z(N,M)} k_i^{m_i} f(m_i), \label{pigeneral}
\eq
where $Z_i(N-1,M-m_i)$ denotes the partition function for $M-m_i$ balls occupying a graph consisting of $N-1$ nodes with degrees $\{k_1,\dots,k_{i-1},k_{i+1},\dots,k_N\}$. It is important not to think about $Z_i$ as of partition function
of the original graph with the $i$th node removed, but rather as of a new graph built of the old sequence of degrees without $k_i$.
We define also the mean occupation probability as the average over all nodes: 
\bq
	\pi (m) = (1/N) \sum_i \pi_i(m).
\eq
It is worth mentioning that for graph with $k_1=\dots=k_N\equiv k={\rm const}$, that is for a $k$-regular graph, the above formulas reduce to that known from the balls-in-boxes model and the distribution $\pi_i(m)=\pi(m)$ is the same for all nodes.
We will see below that sometimes $\pi(m)$ is indeed equal to $f(m)$. Therefore we shall call $f(m)$ ``bare'' occupation probability. 

The partition function can be calculated recursively:
\ba
	Z(N,M,\{k_1,\dots,k_N\}) = \sum_{m_N=0}^{M} \tilde{f}_N(m_N)
	\sum_{m_1,\dots,m_{N-1}} \delta_{\sum_{i=1}^{N-1} m_i, M-m_N}	\prod_{i=1}^{N-1} \tilde{f}_i(m_i) \nonumber \\
	 = \sum_{m_N=0}^{M} \tilde{f}_N(m_N) Z_N(N-1,M-m_N) = 
	 \sum_{m=0}^{M} \tilde{f}_N(m) Z(N-1,M-m,\{k_1,\dots,k_{N-1}\}).	\label{znmrec}
\ea
For $N=1$ the partition function reads simply $Z(1,M,k_1) = \tilde{f}_1(M)$. The formula (\ref{znmrec}) can serve for numerical calculations of the partition function for systems of order few hundreds nodes or more. Using it together with Eq.~(\ref{pigeneral}) we can compute the distribution of balls in a more efficient way that by Monte Carlo simulations. 

The knowledge of the partition function allows one to calculate correlation functions of higher order. For small systems we can calculate $Z(N,M)$ exactly from Eq.~(\ref{znmrec}), for large systems it is better to use the definition (\ref{pigeneral}) which allows for some approximations. In the thermodynamical limit of $N\to\infty$ it is therefore convenient to introduce the density of balls $\rho\equiv M/N$ and to study the limit of fixed $\rho$ while increasing $N$. As we shall see below, for large systems, i.e. for $N,M$ large, $\rho$ becomes a relevant parameter of the model.

The dynamical and static properties of the ZRP depend strongly on the function $u(m)$ and the degree sequence $\{k_i\}$. For the model described here we can distinguish two classes of systems. From now on, by a {\bf homogeneous system}\footnote{We mentioned before, that equilibrated networks were sometimes called ``homogeneous networks''. In this paper, however, we shall always use the term ``homogeneous'' while speaking about networks with equal degrees.} we shall understand the network with all $k_i$'s being equal. It is true for instance for a complete graph, one-dimensional closed chain or a $k$-regular random graph. In contrast, an {\bf inhomogeneous system} has a non-trivial degree sequence, with at least one degree different from others. This is the case for random graphs, star graphs and scale-free networks. 

\subsection{Condensation in the ZRP}
The reason why zero-range processes are so interesting is that under some conditions one observes a phenomenon of ``condensation'' in the steady state. In this phenomenon, a single node takes a finite fraction of all balls present in the system. The effect does not disappear in the thermodynamic limit. 
The condensation can be observed in the occupation distribution $\pi(m)$ as a separated peak, whose position moves almost linearly with the system size. 
Unlike the Bose-Einstein condensation which takes place in the momentum space, the above effect appears in the real space. Therefore it mimics such processes like the mass transport \cite{ref:cc1}, condensation of links in complex networks \cite{ref:bbw,ref:cc2} or phase separation \cite{ref:ph1,ref:ph2}.

The condition for the emergence of the condensation in homogeneous systems is well known \cite{ref:evans2,ref:evans}. On the other hand, until now only a few inhomogeneous systems have been examined \cite{ref:jdn,ref:kulki2,ref:kulki}.
In this section we summarize some results and discuss methods of derivations for the existence of condensate.
We begin with homogeneous systems and then we present our recent results for graphs with inhomogeneous degrees.

{\bf Homogeneous systems}.
For $k$-regular connected simple graphs, which we shall consider in this section, the static properties of the steady state depend only of the hopping rate $u(m)$, the number of nodes $N$ and the number of balls $M$, and are independent of the details of graph topology. The partition function assumes the form:
\bq
  Z(N,M) = \sum_{m_1=0}^M \cdots \sum_{m_N=0}^M \delta_{\sum_{i=1}^N m_i, M}
	\prod_{i=1}^N f(m_i).	\label{partsimple}
\eq
The factor $k^M$ has been dropped since it is constant for given $k$ and $M$. Similarly, for the distribution of balls we have
\bq
	\pi(m)= \frac{Z(N-1,M-m)}{Z(N,M)}  f(m). \label{pisimple}
\eq
From the definition (\ref{partsimple}) of $Z(N,M)$ we can obtain another formula for the distribution of balls:
\bq
	\pi(m) = N^{-1} f(m) \frac{\partial\ln Z(N,M)}{\partial f(m)},
\eq
and hence $\pi(m)$ is proportional to the derivative of the ``free energy'' $\ln Z$ and the bare occupation probability $f(m)$. Notice a similarity between this formula and that of Eq.~(\ref{eq2:pikdef}) for the degree distribution of equilibrated graphs. As we will see, indeed, there is a close relation of the ZRP and equilibrated pseudographs.

Now we shall study, how the behavior of $u(m)$ influences on the emergence of condensation. From Eq.~(\ref{eq3:ppi-1}) we see that for homogeneous system there is a correspondence between the hop rate $u(m)$ and the function $f(m)$:
\bq
	u(m) = f(m-1)/f(m) \quad \Longleftrightarrow \quad f(m)=f(m-1)/u(m),
\eq
and in many cases $f(m)$ is more convenient, so we will stick to it below.
Using the integral representation of Kronecker's delta we can rewrite the partition function as
\bq
	Z(N,M) = \oint \frac{\dd z}{2\pi i} z^{-M-1} \left[F(z)\right]^N,	\label{zbyint}
\eq
where $F(z)$ is an infinite series with coefficients given by $f(m)$:
\bq
	F(z) = \sum_{m=0}^\infty f(m) z^m. \label{eq3:deff}
\eq
Denote the radius of convergence of this series by $r$ (finite or infinite). The partition function (\ref{zbyint}) has the same form as the p.f.  (\ref{eq2:partps}) for pseudographs, up to a factor depending only on $N,L$. Therefore, equilibrated pseudographs can be mapped onto a homogeneous system of balls in boxes with $M=2L$. The degree distribution $\Pi(k)$ is then equivalent to the ball distribution $\pi(m)$. Therefore, many results presented below apply also to pseudographs from Sec. 2.1.4.
In the thermodynamical limit, the integral (\ref{zbyint}) can be rewritten as
\bq
	Z(N,M) \approx \oint \frac{\dd z}{2\pi i} \exp\left[ -N\left(\rho \ln z-\ln F(z)\right)\right],
\eq
and can be done using the saddle-point method:
\bq
	Z(N,M) \approx  \frac{1}{\sqrt{2\pi N G''(z_0)}}\frac{\left[F(z_0)\right]^N}{z_0^{M+1}}, \label{eq3:zbyintfin}
\eq
where $G(z)=-\rho \ln z+\ln F(z)$ and the saddle point $z_0$ is determined by the condition $G'(z_0)=0$, analogously to the case of pseudographs:
\bq
	\rho = z_0 \frac{F'(z_0)}{F(z_0)}. \label{eq3:sp-p}
\eq
The saddle point solution (\ref{eq3:zbyintfin}) holds as long as the Eq.~(\ref{eq3:sp-p}) has a real solution for $z_0$. If not, the formula (\ref{eq3:zbyintfin}) cannot be trusted and in fact, as we will see,  it breaks down in the condensed state. First, let us consider a situation when Eq.~(\ref{eq3:zbyintfin}) has a real solution for $z_0$. In this case the leading term in the free energy $\ln Z$ is given by $G(z_0)$. Differentiating this with respect to $f(m)$ we obtain the distribution of balls:
\bq
	\pi(m) = f(m) \frac{z_0^m}{F(z_0)}. \label{eq3:pibysp}
\eq
Hence, if $z_0=1$, $\pi(m)\propto f(m)$ which explains the name ``bare occupation function'' for $f(m)$.
One can directly see from the definition that $F(z)$ is an increasing function of $z$. Similarly one can see that the right-hand side of Eq.~(\ref{eq3:sp-p}) increases monotonically with $z_0$ as long as $z_0$ is smaller than the radius of convergence $r$. It means that $z_0$ increases when the density $\rho$ increases.
If the series (\ref{eq3:deff}) is convergent on the whole complex plane ($r\to\infty$), the saddle point solution for $\pi(m)$ is valid for any density $\rho$. This happens only when $f(m+1)/f(m)\to 0$ for $m\to\infty$, which corresponds to $u(m)\to\infty$. 
In turn, this means that there exists an effective repulsive force between balls preventing them from occupying a single site. So in this case balls tend to distribute uniformly on the whole graph, regardless of the density of balls $\rho$.
Consider now $u(m)\sim m^\delta$ for an arbitrary $\delta>0$. Then $f(m)$ is given by
\bq
	f(m) \propto \frac{1}{(m!)^\delta},
\eq
and it is clearly seen from Eq.~(\ref{eq3:pibysp}) that the distribution of balls falls faster than exponentially.

On the other hand, it is possible to choose $f(m)$ so that $F(z)$ has a finite radius of convergence. This situation happens when $u(m)$ tends to a constant for $m\to\infty$. Because multiplying $u(m)$ by a constant factor simply corresponds to rescaling the time axis, without loss of generality one can set $u(m) \to 1$ in the limit of large $m$.
To be more specific, let us consider the case $u(m) = 1+ \frac{b}{m}$, for which we find the following formula for $f(m)$ \cite{ref:god}:
\bq
	f(m) = \frac{\Gamma(b+1)m!}{\Gamma(b+m+1)} \cong \frac{\Gamma(b+1)}{m^{b}},
\eq
which falls like a power of $m$ for large $m$. In this case the series $F(z)$ has a finite radius $r=1$.
We must consider now two cases: $b\leq 2$ and $b>2$. For $b\leq 2$ the derivative $F'(z)$ goes to infinity when $z$ approaches one from below. But the ratio $F'(z)/F(z)$ is finite for $z<1$ and thus the density from Eq.~(\ref{eq3:sp-p}) can be arbitrarily large. The saddle-point approximation works well:
\bq
	\pi(m) \sim \frac{z_0^m}{m^b},
\eq
for all values of $\rho$. We see that in this case the distribution of balls falls off exponentially for large $m$. The case $b>2$ is different because the ratio $F'(z)/F(z)$ cannot grow above some critical $\rho_c$:
\bq
	\rho_c = \frac{F'(1)}{F(1)} = \frac{\sum_m m \, f(m)}{\sum_m f(m)} = \frac{1}{b-2}< \infty, 
\eq
at which Eq.~(\ref{eq3:pibysp}) cease to hold for real values of $z_0$. But of course we can put as many balls in the system as we want, so we can increase $\rho$ above $\rho_c$. What happens then? At the critical point $\rho=\rho_c$, the distribution of balls is given by
\bq
	\pi_c(m) = \frac{f(m)}{F(1)} \sim \frac{\Gamma(b+1)}{m^{b}}, \label{ref:pic}
\eq
in the thermodynamical limit. It has a finite-size cutoff for $N<\infty$ which scales as \cite{ref:evans2}
\ba
	&\sim & N^{1/(b-1)} \;\;\; \mbox{for} \; 2<b<3, \\
	&\sim & N^{1/2} \;\;\; \mbox{for} \; b>3,
\ea
exactly like for the cutoff of the degree distribution for degenerated graphs. For $\rho>\rho_c$ the saddle-point equation is no longer valid. To understand what happens then, recall the Bose-Einstein condensation. Below the critical temperature $T_c$, the fraction of particles occupying all energy levels above the ground state is equal to $(T/T_c)^{3/2}$ and it is less than one. The only way to keep the average number of particles fixed is to let them go into the lowest energy level which does not contribute to the partition function, which in the thermodynamic limit is given by the integral over energy \cite{ref:czyz}. The situation is a bit similar to the condensation of balls. Above the critical density some nodes take the surplus of balls, while the rest follows the critical distribution (\ref{ref:pic}).
In \cite{ref:grosk} one can find a complete proof. Here we only recall the main arguments \cite{ref:evans} standing behind it.
Assume that the system is deeply in the condensed phase, so that $M\gg N\rho_c$. Denote the excess of particles by $\Delta=M-\rho_c N$. The canonical weight of each configuration is
\bq
	P(m_1,\dots,m_N) = f(m_1)\dots f(m_N) \sim \left(\prod_{i=1}^N m_i\right)^{-b}. \label{eq3:pmcond}
\eq
Let us estimate the contribution to the partition function from states where the surplus of balls occupy one, or two nodes. The contribution to Eq.~(\ref{eq3:pmcond}) from a single-site condensate is $N\Delta^{-b}$.
From two-node condensation we however have $N(N-1)/2 \,\times (\Delta/2)^{-2b}$. Because $\Delta\approx M=\rho N$,
we get for these two states:
\bq
	N^{1-b} \rho^{-b} \;\;\; \mbox{and} \;\;\; (N-1)N^{1-2b} 2^{2b-1} \rho^{-2b},
\eq
respectively. The second expression disappears faster in the thermodynamical limit, so we infer that the condensate emerges on a single node taking $\Delta$ balls on the average. One can consider also higher-nodes states but they disappear even faster when $N\to\infty$. The condensate is seen as a peak $\pi_{\rm cond}(m-\Delta)$ in the distribution of balls. Since it occupies only one node, the area under the peak is equal to $1/N$.
Because in the remaining part of the system there are only $\rho_c N$ balls, the background of the distribution $\pi(m)$ is perfectly described by the saddle-point solution (\ref{ref:pic}). The complete expression for $\pi(m)$ including the condensate reads
\bq
	\pi(m) \approx \frac{\Gamma(b+1)}{m^{b}} + \pi_{\rm cond}\left(m-(M-\rho_c N)\right). %\frac{1}{N} \delta\left[m-(M-\rho_c N)\right]
\eq
The form of $\pi_{\rm cond}(m)$ has been investigated in \cite{ref:evans2} in the model with continuous masses $m_i$.
In the model considered here one can take a quasi-continuous limit by letting $M\to\infty$ and rescaling $u(m)$ properly. For $2<b<3$, the peak is approximated by
\bq
	\pi_{\rm cond}(x) \cong N^{-b/(b-1)} V_b \left( \frac{x}{N^{1/(b-1)}} \right),
\eq
with $V_b(z)$ given by the integral:
\bq
	V_b(z) = \int_{-i\infty}^{i\infty} \frac{\dd s}{2\pi i} e^{-zs+bs^{b-1}},
\eq
which falls as $|z|^{-b}$ for $z\to -\infty$ and faster than a Gaussian function for $z\to\infty$.
On the other hand, for $b>3$ in the vicinity of $m=\Delta$, the peak falls like a Gaussian:
\bq
	\pi_{\rm cond}(x) \cong \frac{1}{\sqrt{2\pi \sigma^2 N^3}} \exp(-x^2/2\sigma^2N),
\eq
with $\sigma^2 = \langle m^2\rangle - \langle m\rangle^2$ being the variance of $f(m)$. In both cases, the area under $\pi_{\rm cond}(m)$ is equal to $1/N$.

The picture we see now for $b>2$ is the following. Below the critical density the distribution of balls is given by a power law suppressed additionally by an exponential prefactor. At the critical point this prefactor vanishes and we observe a pure power law, disturbed only by finite-size effects. Above the critical point, the condensate emerges at one node chosen at random (spontaneous symmetry breaking) from all nodes. The condensate does not need to occupy this particular site for the whole time. In fact, we will see that it performs a kind of random walk through the system, but the process of melting and rebuilding the condensate is fast in comparison to the mean occupation time.
The condensate contains $M-\rho_c N$ balls on average.

At the end, let us mention the case $u(m)\to 0$ for large $m$. In this case $f(m)$ increases fast with $m$ and the series $F(z)$ has a zero radius of convergence. The critical density $\rho_c$ is zero and therefore the condensation takes place at any density $\rho>0$. Balls attract so strong that almost all of them fall into a single node chosen at random.

\begin{figure}
\center
\psfrag{xx}{$m$} \psfrag{yy}{$\pi(m)$}
\includegraphics*[width=15cm,bb=-1 242 810 527]{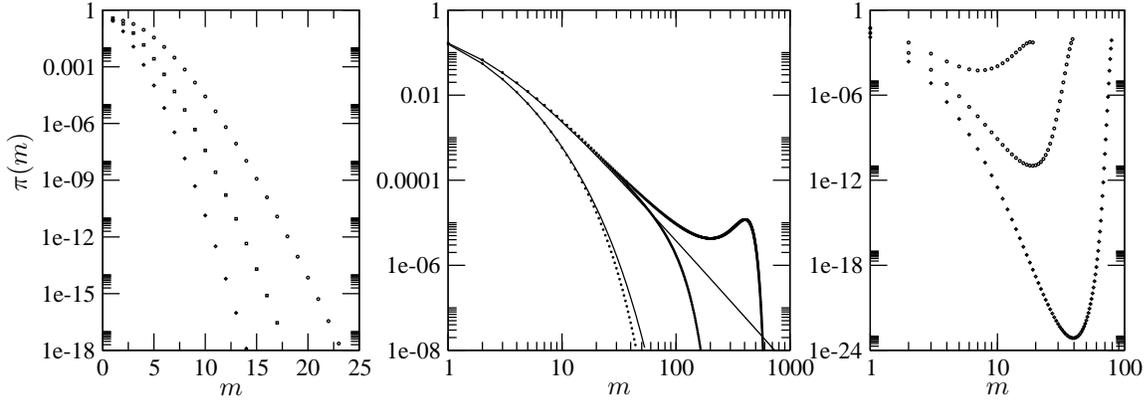}
\caption{\label{fig:kulkiex}Plots of the distribution of balls $\pi(m)$ for a homogeneous system with various $f(m),N$, and $\rho$. Left: $f(m)\sim 1/m!, \; N=40$, for $\rho=0.5$ (diamonds), $1$ (squares) and $2$ (circles), no condensation. Middle: $f(m)\sim m^{-3}$, $\rho_c=1$ and $N=400$. For $\rho=0.5$ (the left-most line) there is an exponential cutoff. At the critical density $\rho=\rho_c$ the power-law behavior is seen. Above $\rho_c$, the surplus of balls forms the condensate (the right-most line). The curves plotted with symbols are obtained by recurrential calculation of $Z(N,M)$ from Eq.~(\ref{znmrec}). Solid lines represent the saddle-point solution (\ref{eq3:pibysp}). Right: $f(m)\sim m!, \; N=40$, for $\rho=0.5, 1, 2$ (from left to right). The condensation is always present.}
\end{figure}

In Fig.~\ref{fig:kulkiex} we show $\pi(m)$ for the three different types of the hop functions $f(m)$ discussed above, for various densities $\rho$.
The data are obtained by means of the recursion formula (\ref{znmrec}) for the partition function and, if possible, compared with the saddle-point solution (\ref{eq3:pibysp}).

{\bf Inhomogeneous systems}. Now we focus on inhomogeneous networks, whose degrees vary from node to node. It turns out \cite{ref:evans,ref:jdn} that the effect of inhomogeneity is so strong that it completely dominates over the dependence on the hop rate $u(m)$ as long as the latter does not change too fast with $m$. For simplicity we can assume $u(m)=1$ and concentrate only on the effects coming from the inhomogeneity of degrees. Then, the zero-range process describes a gas of $M$ indistinguishable and non-interacting balls randomly walking on the given network.

The most interesting case of graph with inhomogeneous degrees is of course the S-F network. It was studied in \cite{ref:jdn,ref:jnd2} but because of its complicated structure, only very simple calculations of the static properties were possible. Here we decided to focus on the effect coming from the node with largest degree, say $k_1$. To further simply considerations, we just imagine that the effect can be well simulated by assuming the identity of the remaining degrees. In effect we are led to consider an almost $k$-regular graph which has one node of degree $k_1$ bigger than the others which are of degree $k_2=\dots=k_N=k$ \cite{ref:kulki}. We shall call it a {\bf single-inhomogeneity graph}. %The node $1$ has degree $k_1>k$ and $N-1$ remaining nodes have degrees . 

\begin{figure}
\center
\psfrag{a}{a)}\psfrag{b}{b)}
\psfrag{i}{$i$} \psfrag{j}{$j$}
\includegraphics[width=10cm]{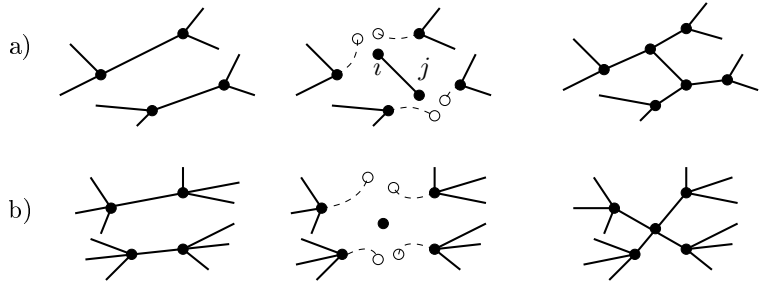}
\caption{\label{fig:kreggen}The illustration of the method for generating $k$-regular graphs, based on addition of new nodes: a) for odd $k$ (here $k=3$), b) for even $k$ (here $k=4$). }
\end{figure}

To construct $k$-regular graphs and the single-inhomogeneity graph one can use various methods. For instance, one can start from a random graph with given number of vertices and links and rewire it until all nodes will have desired degrees. Another method of building a $k$-regular graph is to start from  a complete graph with $k+1$ nodes and build the desired graph successively adding nodes and links. The procedure depends on the parity of $k$. If $k$ is odd, then the number of nodes has to be even because the number of links $L=Nk/2$ must be integer, otherwise the graph cannot be built. At each step we pick up $k-1$ existing links and split them so that nodes being formerly joined by these links, have now ``halves'' of them. Then we introduce two new nodes $i,j$ joined by a link.
Finally, one half of ``free ends'' of previously split links is joined to the newly added node $i$, and the other half to $j$ (Fig.~\ref{fig:kreggen}a). In this way every node has now $k$ neighbors. We repeat this process until the total number of nodes is equal to $N$.
In case of even $k$, the algorithm is similar, but we add only one new node per time step, and split $k/2$ existing links (Fig.~\ref{fig:kreggen}b). Sometimes, as a result of nodes' addition, multiple connections might arise. To prevent them, we discard such moves. For $k$ much smaller than $N$, they are rare and the acceptance of the algorithm is almost 100\%. We use the structure described in \cite{ref:naszcpc} to code the graphs.
%The statistical ensemble of $k$-regular graphs defined in this way is the same as the one obtained by thermalization with $X$-moves. It is just the ensemble of maximally random labeled graphs with all degrees equal to $k$.

A single-inhomogeneity graph with one node having degree $k_1>k$ can be then obtained from a $k$-regular graph with $N-1$ nodes by adding to it a new node and joining to it $k_1$ links coming from splitting $k_1/2$ randomly chosen links existing previously. If we get multiple connections, we discard this move and try again.

Before we calculate the partition function for a single-inhomogeneity graph it is convenient to calculate it for a $k$-regular graph.  Because we set $u(m)=1$ in this section, as mentioned before, hence also $f(m)=1$. For a $k$-regular graph the partition function $Z(N,M)$ from Eq.~(\ref{part}) is 
\ba
	Z_{\rm reg}(N,M) = \sum_{m_1=0}^M \cdots \sum_{m_N=0}^M \delta_{m_1+\dots+m_N, M} \; k^{\sum_i m_i} = k^M \frac{1}{2\pi i} \oint \dd z \, z^{-M-1} \left[F(z)\right]^N,
\ea
where now $F(z)$ reads
\bq
	F(z) = \sum_{m=0}^M z^m = \frac{1}{1-z}.
\eq
Using the expansion:
\bq
	\left(1-z\right)^{-N} = \sum_{m=0}^\infty \binom{-N}{m} (-z)^m  = \sum_{m=0}^\infty \binom{N+m-1}{m} z^m,
\eq
and Cauchy's theorem which selects only the term with $m=M$, we finally get an exact expression for the partition function of a $k$-regular graph:
\bq
	Z_{\rm reg}(N,M) = k^M \binom{N+M-1}{M}. \label{zreg}
\eq
The partition function for a graph with one irregular degree $k_1>k$ has the form:
\bq
	Z_{\rm inh}(N,M) = \sum_{m_1=0}^M (k_1)^{m_1} \sum_{m_2=0}^M\cdots \sum_{m_N=0}^M \delta_{M,m_1+\dots+m_N} k^{m_2+\dots+m_N}.
	\label{zsh1}
\eq
The sum over $m_2,\dots,m_N$ is just the partition function $Z_{\rm reg}(N-1,M-m_1)$ from Eq.~(\ref{zreg}).
After changing the variable from $m_1$ to $j=M-m_1$, the formula (\ref{zsh1}) can be rewritten as
\bq
	Z_{\rm inh}(N,M) = k_1^M \sum_{j=0}^M \alpha^j \binom{N+j-2}{j} ,
	\label{zinh}
\eq
where $\alpha=k/k_1$ describes the level of ``inhomogeneity''. Introducing an auxiliary function
\bq
	S(\alpha) = \sum_{j=0}^\infty (-\alpha)^j \binom{-(N-1)}{j} = \frac{1}{(1-\alpha)^{N-1}}, \label{auxs}
\eq
we obtain the following expression:
\bq
	Z_{\rm inh}(N,M) = k_1^M \left[ (1-\alpha)^{1-N} - c(M) \right],	\label{zinh2}
\eq
where $c(M)$ gives a correction for finite $M$. It tends to zero for $M\to\infty$:
\bq
	c(M) = \sum_{j=M+1}^\infty \alpha^j \binom{N+j-2}{j}. %\approx \frac{1}{(N-2)!} \int_M^\infty e^{F(m)} dm
\eq
This correction can be however quite large for $k_1\approx k$ because then $\alpha\approx 1$ falls slowly and the binomial term increases with $j$. We can estimate the correction by the saddle-point method, replacing the sum by the integral:
\bq
	c(M) \approx \frac{1}{(N-2)!} \int_M^\infty e^{F(j)} \dd j, \label{intc}
\eq
where we define a new function $F(j)$:
\bq
	F(j) = j \ln\alpha + \ln((N+j-2)!) - \ln(j!),
\eq
and use Stirling's formula for factorials. We get:
\bq
	\int_M^\infty e^{F(j)} \dd j \approx e^{F(j_*)} \int_M^\infty e^{\frac{1}{2}F''(j_*)(j-j_*)^2} \dd j = 
	e^{F(j_*)} \sqrt{\frac{-\pi}{2F''(j_*)}} \mbox{erfc}\left( (M-j_*)\sqrt{-F''(j_*)} \right),
\eq
where $\mbox{erfc}(x)$ denotes the complementary error function:
\bq
	\mbox{erfc}(x) = \frac{2}{\sqrt{\pi}} \int_x^\infty e^{-y^2} \dd y,
\eq
and $j_*$ is the point being a solution of the saddle-point equation $F'(j_*)=0$:
\bq
	j_* \approx \frac{\alpha (N-2)}{1-\alpha}. \label{eq3:miu}
\eq
Only leading terms in $F(j)$ were taken into account. Collecting all together one finds
\bq
	c(M) = \frac{\alpha^{\frac{\alpha(N-2)}{1-\alpha}}}{1-\alpha} \frac{((N-2)/(1-\alpha))!}{(\alpha(N-2)/(1-\alpha))!}
	\sqrt{\frac{\pi \alpha (N-2)}{2}} \frac{1}{(N-2)!} \mbox{erfc}\left( \frac{M(1-\alpha)-\alpha(N-2)}{\sqrt{\alpha(N-2)}}\right).	\label{cmfin}
\eq
The complete partition function $Z_{\rm inh}(N,M)$ is given by Eq.~(\ref{zinh2}) with $c(M)$ calculated by means of Eq.~(\ref{cmfin}). We can now calculate $\pi_1(m)$ that is the distribution of balls at the node with degree $k_1$:
\bq
	\pi_1(m) = \frac{Z_{\rm reg}(N-1,M-m)}{Z_{\rm inh}(N,M)} k_1^m,
\eq
where $Z_{\rm reg}(N,M)$ is the partition function for a $k$-regular graph. Making use of the formulas (\ref{zreg}) and (\ref{zinh2}) we get
\bq
	\pi_1(m) = \alpha^{M-m} \binom{M+N-m-2}{M-m} \left[ (1-\alpha)^{1-N} - c(M) \right]^{-1}. \label{pi1inh}
\eq
\begin{figure}
\psfrag{x}{$m$}
\psfrag{y}{$\pi(m)$}
\includegraphics*[width=8cm,bb=50 50 410 302]{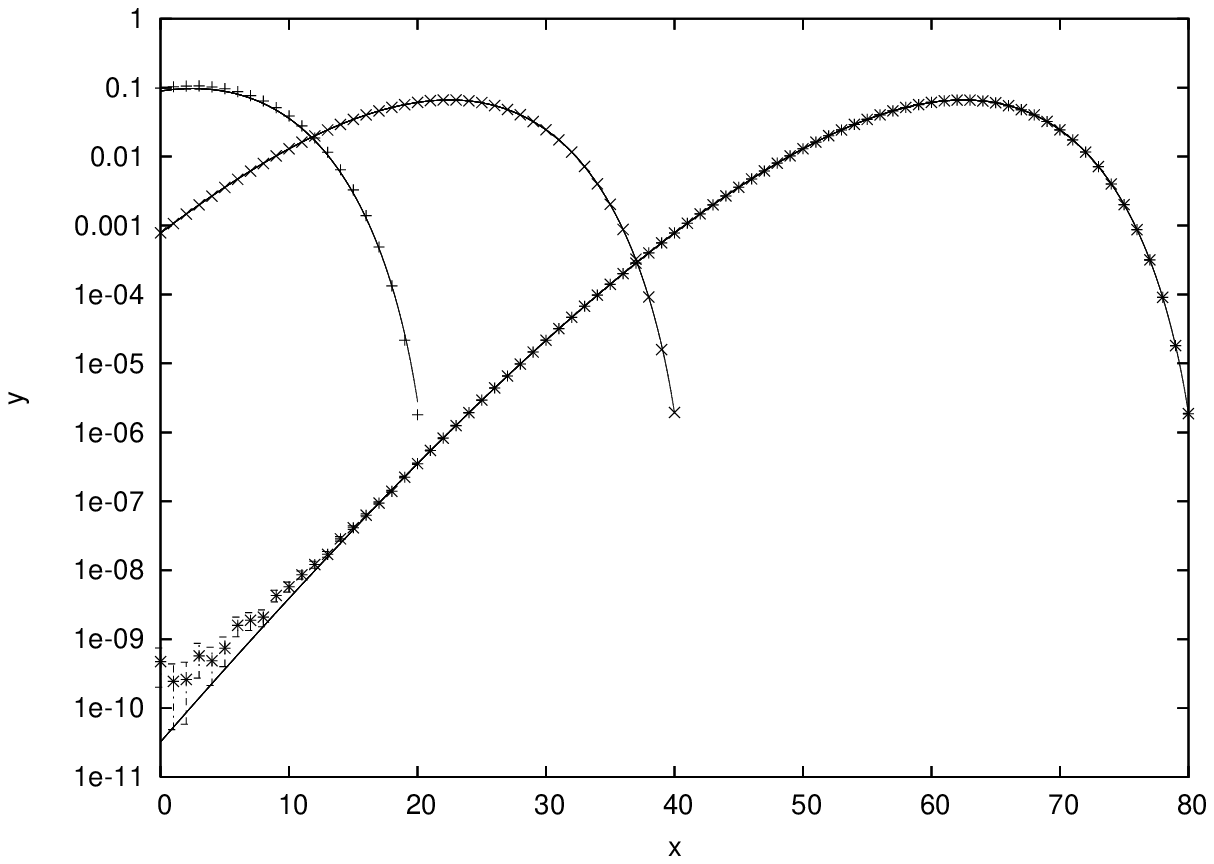}
\includegraphics*[width=8cm,bb=50 50 410 302]{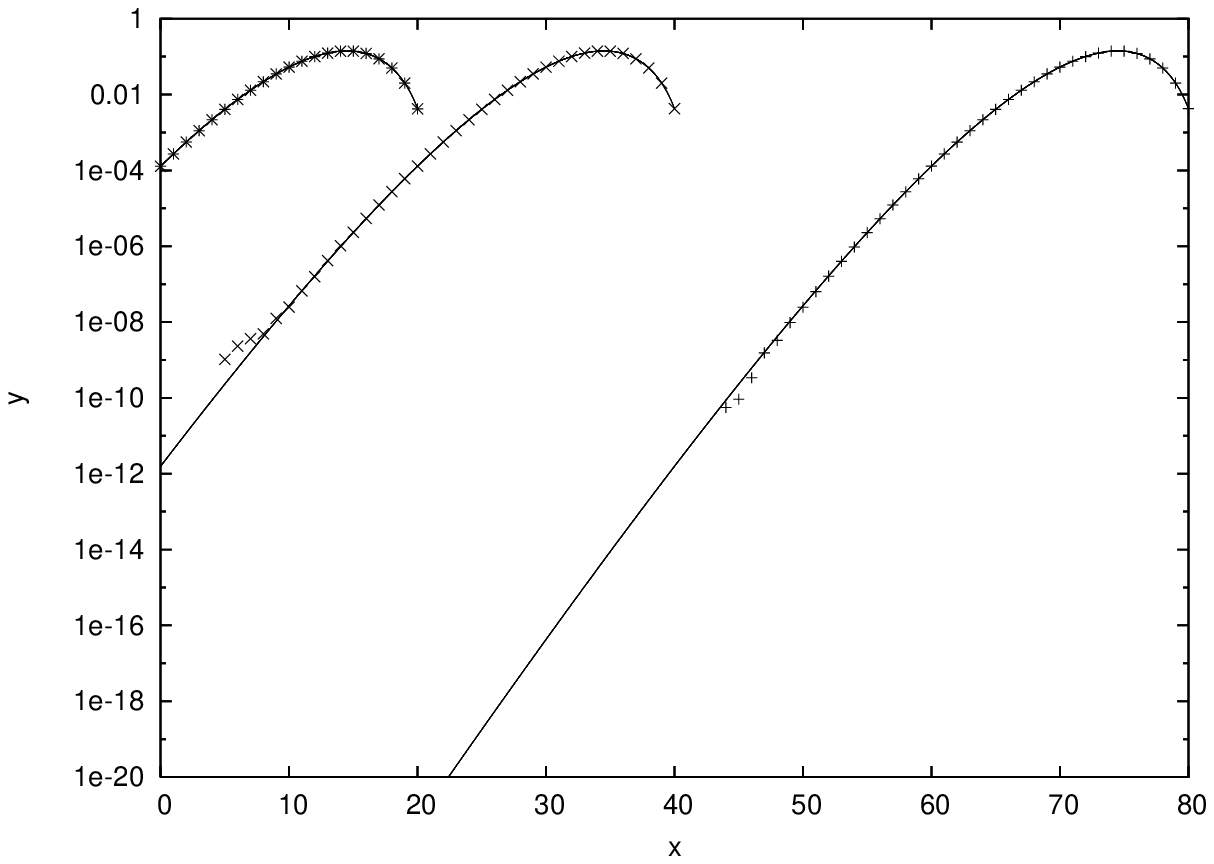}
\caption{\label{fig:f3}The distribution of balls at the singular node for single-inhomogeneity graphs with $k=4$, $N=20$, $k_1=8$ (left) and $k_1=16$ (right). The total number of balls is $M=20,40$ and $80$ from left to right, respectively. Points represent numerical data while solid lines the solution from Eq.~(\ref{pi1inh}).}
\end{figure}
\hspace{-2mm}
In figure \ref{fig:f3} we show a comparison between the analytic solution (\ref{pi1inh}) and the one obtained from Monte Carlo simulations. Neglecting inessential normalization, the equation (\ref{pi1inh}) has the following asymptotic behavior:
\bq
	\pi_1(m) \sim \exp(G(m)),
\eq
where 
\bq
	G(m) = -m \ln\alpha + \left(M+N-m-\frac{3}{2}\right)\ln(M+N-m-2)-\left(M-m+\frac{1}{2}\right) \ln(M-m).
	\label{gfunc}
\eq
Condensation takes place when $G(m)$ has a maximum for positive $m=m_*$. Taking the derivative of Eq.~(\ref{gfunc}) and neglecting terms of order $1/M^2$ in the corresponding equation $G'(m_*)=0$ we find
\bq
	m_* = M - \frac{\alpha}{1-\alpha} (N-2).
\eq
Let us calculate the mean number of balls at the first node:
\bq
	\left<m_1\right> = \sum_{m=0}^M \pi_1(m) m = M - \sum_{j=0}^M \pi_1(M-j) j .
\eq
In the large $M$ limit, the sum over $j$ can be calculated exactly:
\bq
	M - \frac{\sum_{j=0}^\infty j (-\alpha)^j \binom{-(N-1)}{j}}{\sum_{j=0}^\infty (-\alpha)^j \binom{-(N-1)}{j}} = M - \alpha \frac{\dd \ln S(\alpha)}{\dd\alpha}  = M - \frac{\alpha}{1-\alpha} (N-1) \approx m_*.	\label{m1inh}
\eq
The condensation occurs when an extensive number of balls is on the singular node. This happens  when $m_*>0$ or equivalently when $\left<m_1\right>>0$. Therefore, the critical density in the limit $N,M\to \infty$ with fixed density $\rho=M/N$ is
\bq
	\rho_c = \frac{\alpha}{1-\alpha}.
\eq
The condensation is possible only for the density $\rho>\rho_c$, exactly like in the Single Defect Site model \cite{ref:evans}. The critical density decreases with decreasing ratio $\alpha=k/k_1$ or, equivalently, with increasing inhomogeneity $k_1/k$. The site, which contains the condensate, has $N(\rho-\rho_c)+\rho_c$ balls on average, as follows from Eq.~(\ref{m1inh}). It is also easy to find that the distribution of balls $\pi_i(m)$ at any regular node with degree $k$ falls exponentially
\bq
	\pi_i(m) \propto \left(\frac{k}{k_1}\right)^m = \alpha^m,
\eq
with $\alpha<1$. Thus the condensation never appears on a regular node. When the system is in the condensed phase, the mean occupation of such a node is equal to $\rho_c$ independently on the total number of balls in the system.

Let us consider also a special example of a single-inhomogeneity graph called a star graph, which has one node of degree $k_1=N-1$ and $N-1$ nodes of degree $k=1$. Because $k_1$ increases when the system grows, the parameter $\alpha$ goes to zero as $1/N$. The critical density $\rho_c\to 0$ in the thermodynamical limit. Thus on a star graph the condensation appears for any finite density $\rho>0$.
We can calculate the variance of $m_1$ which is a measure of fluctuations. Introducing $\mu\equiv -\ln \alpha$ we have
\bq
	\left<(m_1-\left<m_1\right>)^2\right> = - \frac{\dd^2 \ln S(e^{-\mu})}{\dd\mu^2}, \label{eq3:pivar}
\eq
and inserting $\mu=\ln(N-1)$ for the star graph we get $\left(\frac{N-1}{N-2}\right)^2$
which tends to one when $N\to\infty$. Therefore, for almost all time the condensate has all balls but one, the mean value $\left<m_1\right>\approx M-1$ as it follows from Eq.~(\ref{m1inh}), and fluctuations are small. The occupation of other sites must be thus close to zero.

\subsection{Dynamics of the condensate}
One can address two natural questions while studying the dynamics of the condensate: i) how is it produced from a uniform distribution of balls, and, ii) how much time does it take to melt the condensate and rebuilt it at another site?
The answer to these questions is different for homogeneous and inhomogeneous systems. Moreover, in both cases the dynamics depends on the structure of network, not only on degrees. For instance, one can imagine that there is a bottleneck, e.g. a single link joining two larger parts of the network. The transport of balls on such graph will be different from the case when these two parts are strongly interconnected. As we will see, however, the structure of the network is not so important as one could think, and characteristic time scales are determined mainly by the size of the system and its (in)homogeneity.

\begin{figure}
\center
\psfrag{xx}{$N$} \psfrag{yy}{$\left<\tau\right>$}
\includegraphics*[width=8cm]{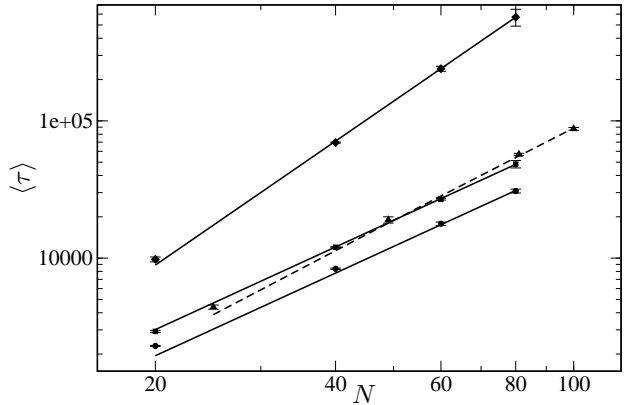}
\caption{\label{fig:coars}Time scales for coarsening dynamics of the condensate on various homogeneous networks. In numerical simulations we measure the average time $\left<\tau\right>$ after which the maximal number of balls exceeds $\Delta=M-\rho_c N$ at some node. The system starts from a uniform distribution of balls and the simulation is stopped when $m_{n_{\rm max}}\geq \Delta$. The time $\tau$ is collected and the procedure is repeated. At the end we calculate $\left<\tau\right>$. Repeating this a few times we estimate errors.
Circles: complete graph ($D>2$), squares: $4$-regular random graph ($D>2$), diamonds: 2d periodic lattice ($D=2$), triangles: one-dimensional closed chain ($D=1$). Lines are asymptotic solutions from Eq.~(\ref{eq3:coars}) with proportionality coefficient fitted to data.
In all cases $M=4N$ and $b=4$, hence the density $\rho\gg \rho_c=1/2$.}
\end{figure}

The emergence of the condensate from a state where all nodes have approximately equal occupation numbers has been investigated for homogeneous systems \cite{ref:evans}, \cite{ref:god}, \cite{ref:grosk}. 
The process can be divided into two stages. First, the surplus of balls $\Delta$ accumulates at a finite number of nodes. When this happens, small condensates exchange particles through the nearly-uniform background. This results in coarsening of many condensates which eventually form a single one with a larger number of balls. This process is very slow.
Assuming the jumping rate in the form $u(m)=1+b/m$ and that we are in the condensed phase $\rho>\rho_c$, the mean condensate size grows as $\Delta (t/\tau)^\delta$, where the characteristic time scale $\tau$ for coarsening dynamics has been estimated as
\bq
	\tau \sim \left\{ \begin{tabular}{ccc} $N^3$ & \mbox{for} & $D=1$, \\
																				 $N^2\ln N$ & \mbox{for} & $D=2$, \\
																				 $N^2$ & \mbox{for} & $D>2$, \end{tabular} \right. \label{eq3:coars}
\eq
and the exponent $\delta$ is inversely proportional to the power of $N$ in the expressions above .
Here $D$ is the dimension of the network, e.~g. $D=1$ for a closed chain, $D=2$ for a two dimensional lattice and $D=\infty$ for a complete graph.
In figure \ref{fig:coars} we present the comparison of these asymptotic formulas to computer simulations.

\begin{figure}
\center
\psfrag{yy}{$n_{\rm max}$}
\psfrag{xx}{$t$}
\includegraphics*[width=4.5cm]{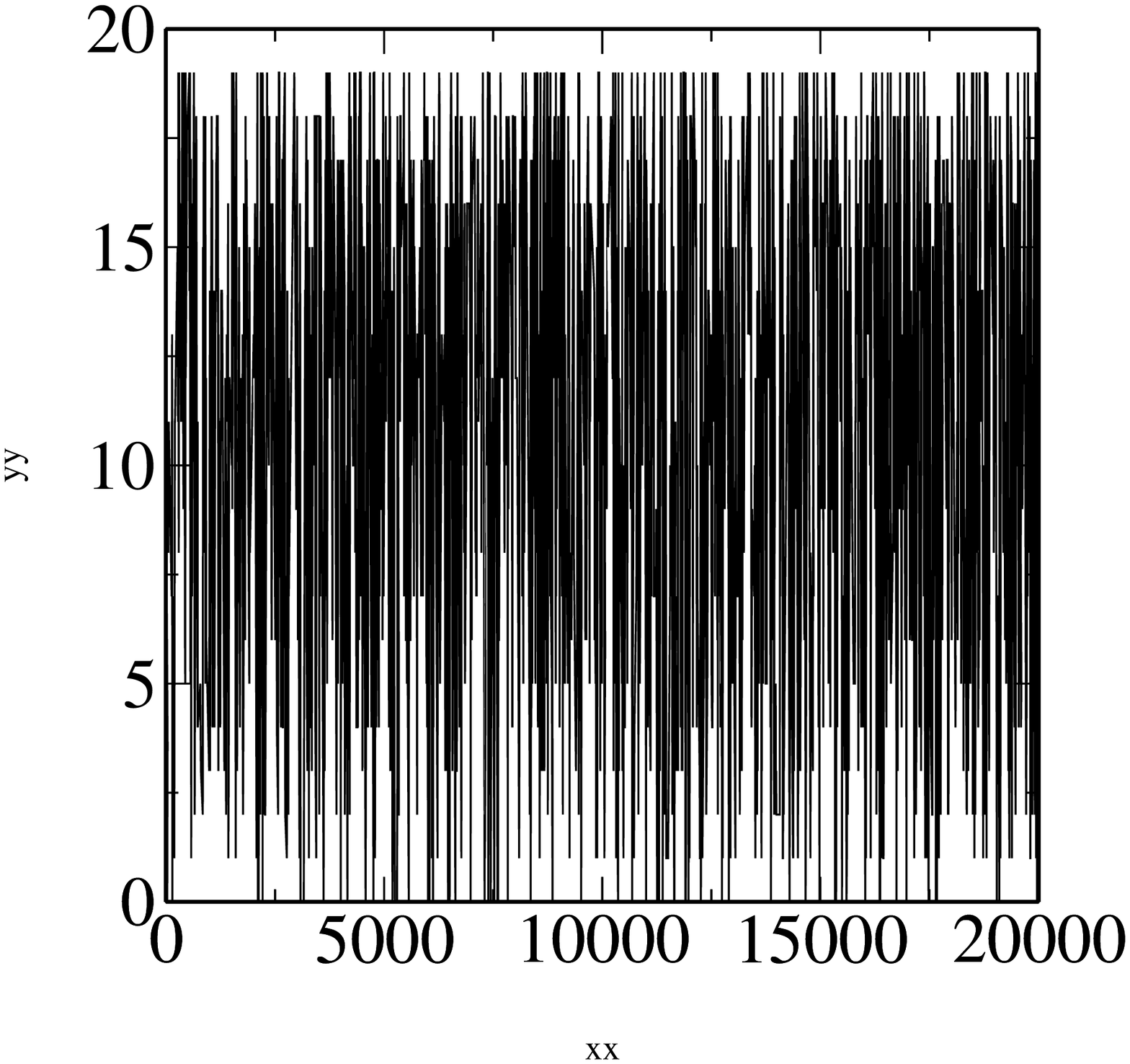}
\includegraphics*[width=4.5cm]{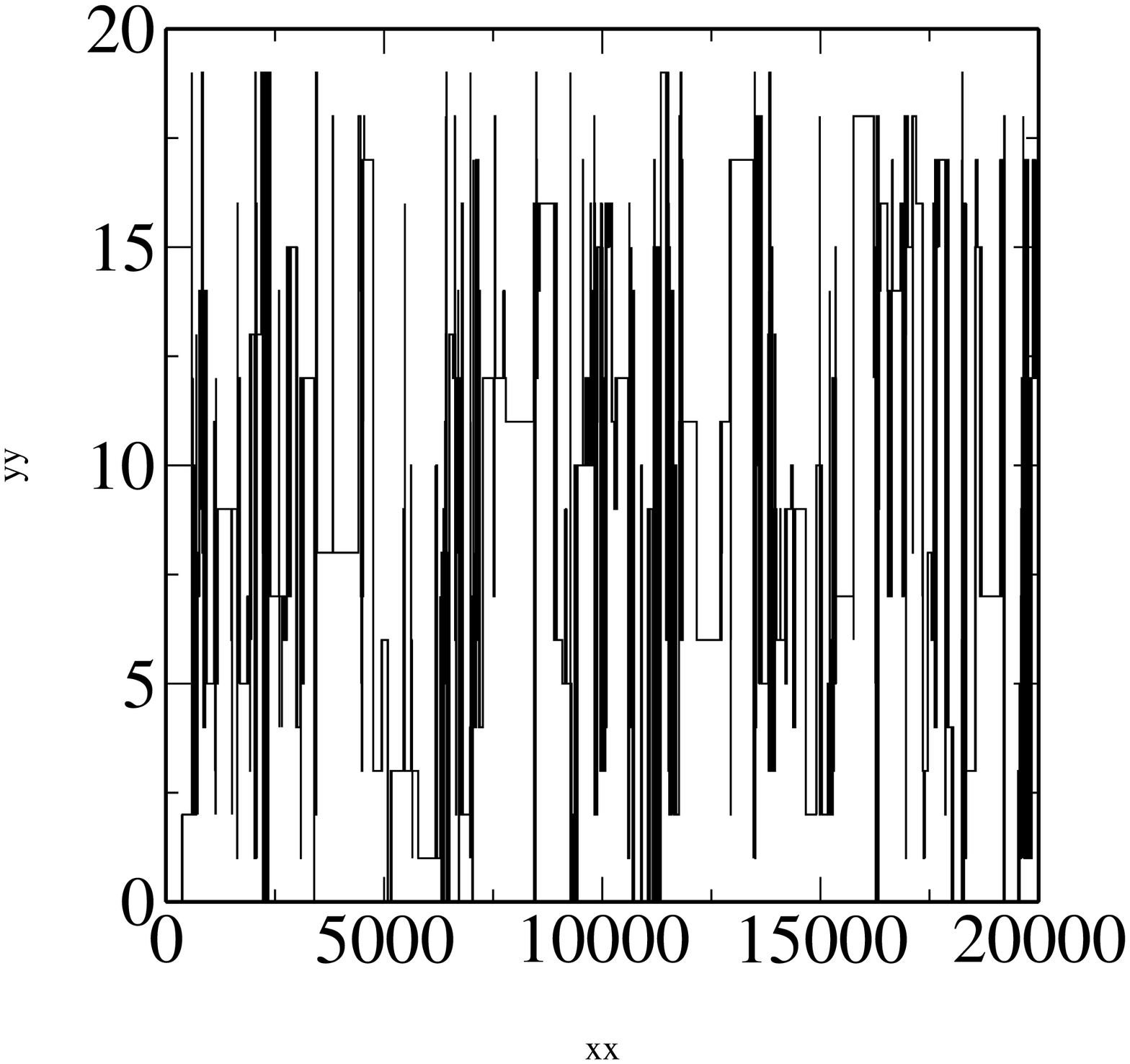}
\includegraphics*[width=4.5cm]{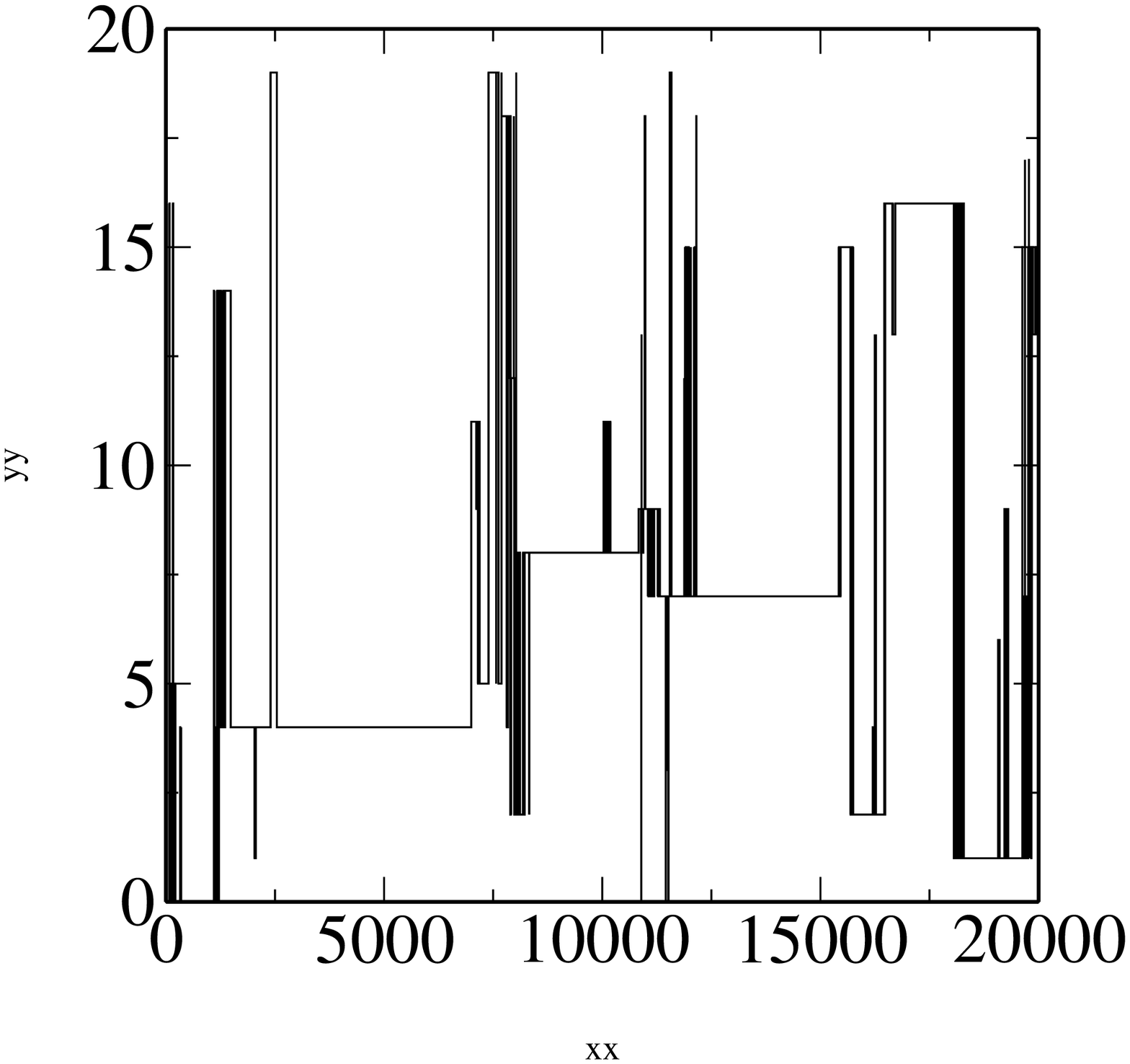}
\caption{\label{fig:periods}From left to the right: position of the node $n_{\rm max}$ with maximal number of balls as a function of time, for different densities $\rho=0.5$ (below $\rho_c=1$), $\rho=1.5$ and far in the condensed phase $\rho=2.5$. The network is a $4$-regular graph with $N=20$ nodes and the hop rate $u(m)\propto 1+3/m$. In the condensed phase, the condensate occupies a single site for a long time and then moves to another site, thus jumps in $n_{\rm max}$ occur seldom.}
\end{figure}

\begin{figure}
\center
\psfrag{xx}{$M$}
\psfrag{yy}{$\bar{T}$}
\includegraphics*[width=7cm]{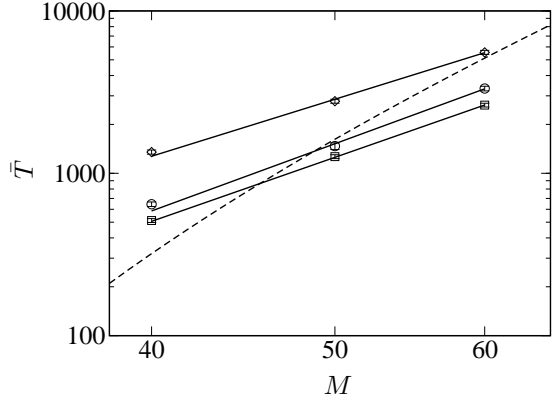}
\caption{\label{fig:thom}Average periods of time between jumps in $n_{\rm max}$ versus the density $\rho$, for homogeneous graphs. The simulation has been performed on various graphs with $N=20$ nodes and the jumping rate $u(m)=1+3/m$, the critical density $\rho_c=1$. Circles: $4$-regular graph, squares: complete graph, diamonds: one-dimensional closed chain. Solid lines are power laws $a_0 M^{a_1}$ fitted to data; for various graphs $a_1=3.6\dots 4.2$. The dashed curve giving the asymptotic formula (\ref{eq3:thom}) with arbitrary proportionality factor is far away from numerics.}
\end{figure}

Contrary to the coarsening dynamics, studies on the dynamics of an existing condensate in homogeneous systems
are rare \cite{ref:god}. 
One can ask what is the typical life-time of the condensate,
that is how much time it takes before it disappears from one site and rebuilds at another site. Let $n_{\rm max}$ be the position of the node with maximal number of balls. In figure \ref{fig:periods} we plot $n_{\rm max}$ as a function of time, for different densities $\rho$, for a regular graph.
It is clearly seen that the characteristic time between jumps is much larger in the condensed phase.
This means that the condensate, once formed, spends a lot of time without any move and then suddenly jumps to another node.
In \cite{ref:god} authors investigated this process on a complete graph.
Using a Markovian ansatz that the number of balls on the condensed site varies slowly in comparison to other $m_i$'s one can recast the original problem into a biased diffusive motion on a one-dimensional interval. The authors showed that average crossing time, i.~e. the time between melting the condensate and rebuilding it at another site, can be approximated by the formula
\bq
	\bar{T} \propto (\rho-\rho_c)^{b+1} N^b,	\label{eq3:thom}
\eq
for $b>1$. Thus for fixed size of graph and far above $\rho_c$, $\bar{T}$ grows like a $(b+1)$th power of the density of balls. For the fixed density, $\bar{T}$ grows with $N$ as $\sim N^b$. This formula holds only for quite large systems and therefore it is hard to verify in Monte Carlo simulations. In figure \ref{fig:thom} we see that for small systems the power-law dependence is rather on $M$ than on $(\rho-\rho_c)$ as it would stem from Eq.~(\ref{eq3:thom}).

Let us discuss now inhomogeneous systems.
Although the emergence of the condensate in zero-range processes has been extensively studied,
not much is known about their dynamics.
The coarsening dynamics has been examined numerically for scale-free networks in Ref.~\cite{ref:jdn}, where the jumping rate was taken to be $u(m)\propto m^\delta$ with $\delta\geq 0$.
It was observed that the dynamics is hierarchical. First, balls on the sub-network of small degrees are equilibrated, then nodes with higher degrees are equilibrated, and finally the hubs - the nodes with highest degrees. The global stationary state is reached with the time $\tau\sim N^z$, where
\bq
	z = \left\{ \begin{tabular}{cc} $1+\alpha-\delta$ & \mbox{for trees}, \\  $1-\delta$ & \mbox{for network with loops}, \\ \end{tabular} \right.
\eq
and $\alpha$ is the exponent from the cutoff scaling law $k_{\rm max}\sim N^\alpha$. Below we shall discuss the dynamics of the condensate, once it is formed. This issue has not been studied yet. Although we study only simplified models, the results will allow us to derive some conclusions about how this process looks like on S-F networks.

We shall consider the dynamics on a single-inhomogeneity network introduced earlier.
It is a very good candidate to examine how inhomogeneities influence the typical life-time of the condensate.
In order to determine this typical time $\bar{T}$ after which the condensate melts, we should first define this quantity properly.
We have seen in the figure \ref{fig:thom} that for small homogeneous systems it was impossible to reach a good agreement with theoretical predictions. One of the reasons might be that, in fact, the crossing time \cite{ref:god} has not much to do with jumps in the position $n_{\rm max}$. On the other hand, the approach presented there seems to work not only for homogeneous systems so we hope to successfully apply it to our case.
It is therefore convenient to consider the quantity $T_{mn}$ - the average time it takes to fall from $m$ to $n$ balls at the condensed site, or more precisely, the first passage time from the state with $m$ balls to the state with $n$ balls at that site. $T_{mn}$ can be easily estimated from computer simulations - one starts to count the time when $m_1$ passes through $m$, and stops when it passes through $n$ for the first time. Repeating this many times one gets the average time.

This quantity can be combined with a typical life-time $\bar{T}$ using the following picture: in the condensed phase, the node with maximal degree takes an extensive number of balls $\Delta$ while for the remaining nodes $m_i$'s fluctuate around the average number $\rho_c \ll \Delta$. We suppose these fluctuations to be much faster than the life-time of the condensate. The condensate disappears when $m_1\approx \rho_c$. Thus $\bar{T}\approx T_{mn}$ where $m\approx \Delta,n\approx \rho_c$. The averaging should actually be done over all possible value of $m,n$ with the appropriate weight. How to choose this weight and how to finally calculate $\bar{T}$ will be explained later. Now we would like to focus on the time $T_{mn}$ for given $m$ and $n$.


\begin{figure}
\center
\psfrag{xx}{$M$}
\psfrag{yy}{$T_{M0}$}
\includegraphics*[width=7cm]{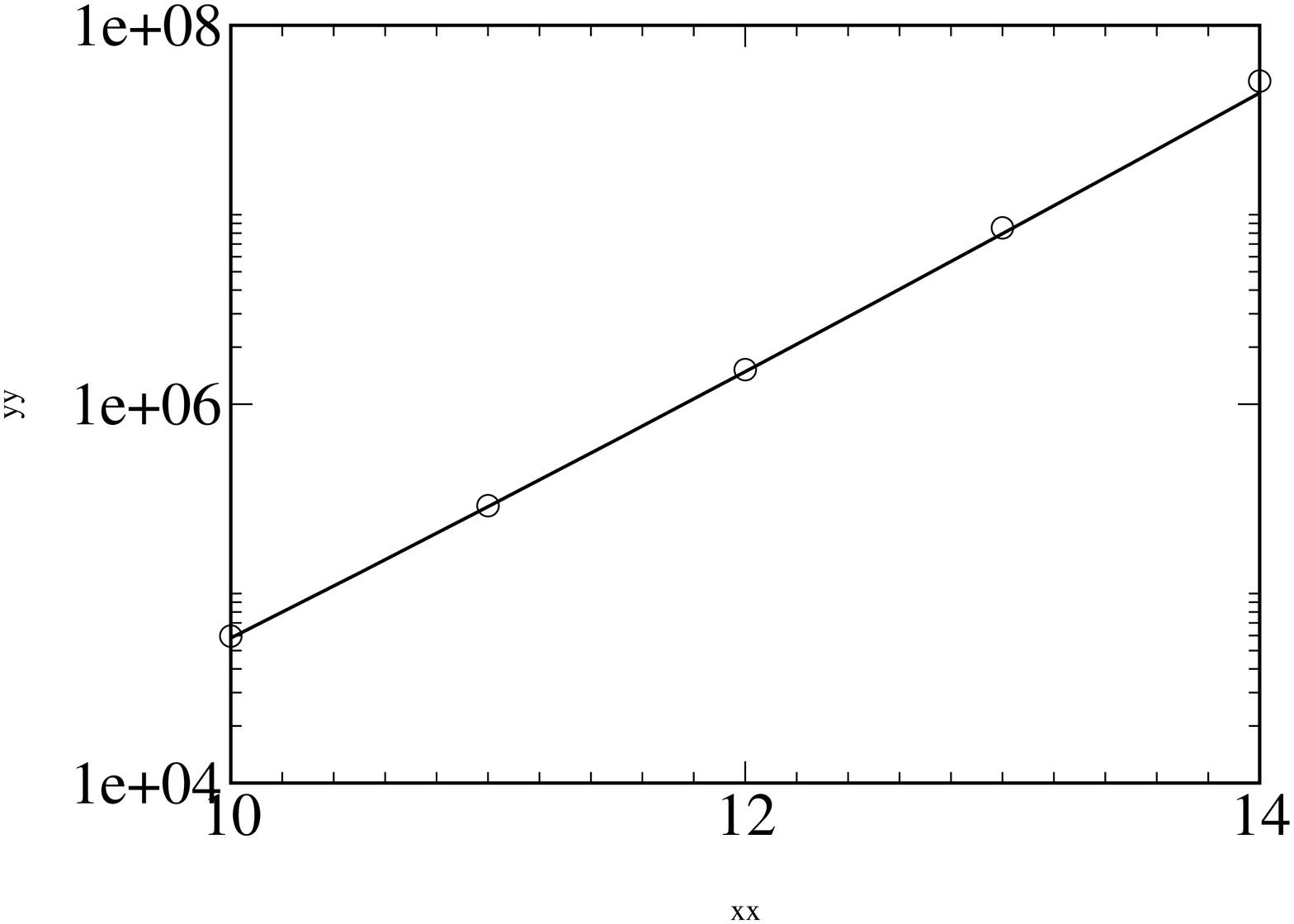}\hspace{5mm}\includegraphics*[width=7cm]{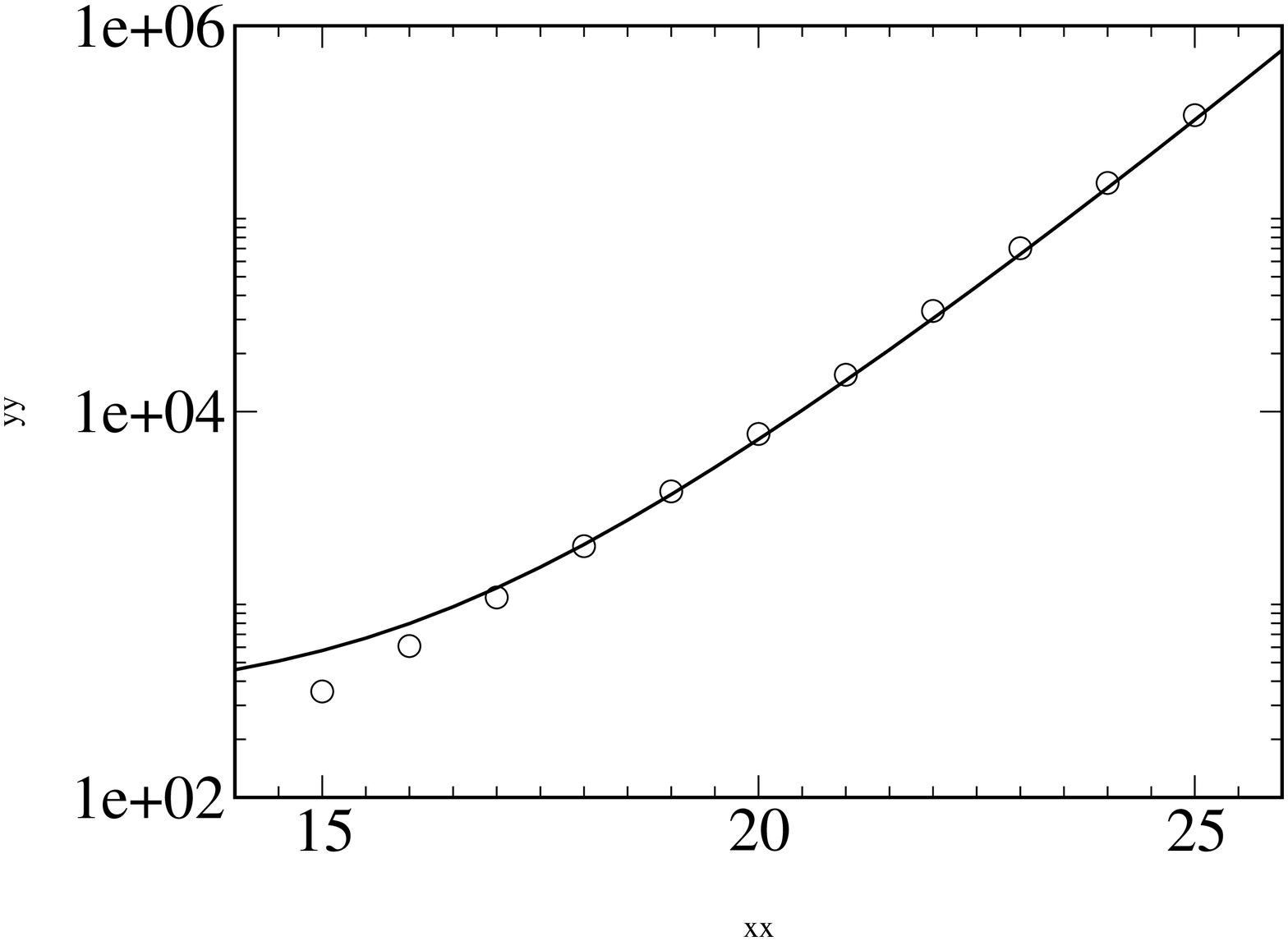}
\caption{\label{fig:star+inh-TMn}The average life-time $T_{M0}$ for a star graph with $N=10$ calculated from Eq.~(\ref{TMnstar}) (left panel) and for a single-inhomogeneity graph with $N=20,k=4,k_1=16$ from Eq.~(\ref{eq3:TMninh}) (right panel), compared to computer simulations (circles).}
\end{figure}

To calculate $T_{mn}$ we adopt the method presented in Ref.~\cite{ref:god}. We assume that the condensate can be considered as slowly changing in comparison to fast fluctuations in the bulk. Suppose that at particular time, the condensate has $m$ balls.
After one time step, the condensate can loose one ball, gain one ball or remain intact. Let us denote by $\lambda_m$ the probability
that $m\to m+1$ and by $\mu_m$ that $m\to m-1$. Assume additionally that $\mu_0=0,\lambda_M=0$ and $\mu_{m\geq 1},\lambda_{m<M}$ are greater than zero.
One can see that $T_{mn}$ has to fulfill the following equation:
\bq
	T_{mn} = 1 + \lambda_m T_{m+1,n} + \mu_m T_{m-1,n} + (1-\lambda_m-\mu_m) T_{mn},
\eq
with $T_{nn}=0$. Defining $d_m = T_{mn} - T_{m-1,n}$ we rewrite that equation in the form:
\bq
	d_m \mu_m - \lambda_m d_{m+1} = 1. \label{eq3:dm}
\eq
With zero on the right-hand side it would have a solution $d_m = \prod_{k=1}^{m-1} \mu_k /\lambda_k$.
In general, the solution has the form $d_m = c_m \prod_{k=1}^{m-1} \mu_k /\lambda_k$ with $c_m\geq 0$.
The equation (\ref{eq3:dm}) gives the recursion for $c_m$:
\bq
	c_m - c_{m+1} = \frac{1}{\mu_m} \prod_{k=1}^{m-1} \mu_k /\lambda_k. \label{eq3:cm}
\eq
The maximal number of balls at one node is $M$, so it must be $c_{M+1} = 0$. With this boundary condition, Eq.~(\ref{eq3:cm}) has the solution:
\bq
	c_m = \sum_{l=m}^{M} \frac{1}{\mu_l} \prod_{k=1}^{l-1} \frac{\lambda_k}{\mu_k}.
\eq
This leads to the following expression for $T_{mn}$:
\bq
	T_{mn} = \sum_{p=n+1}^m d_p = \sum_{p=n+1}^m \left( \prod_{k=1}^{p-1} \frac{\mu_k}{\lambda_k} \right)
	\left( \sum_{l=p}^{M} \frac{1}{\mu_l} \prod_{q=1}^{l-1} \frac{\lambda_q}{\mu_q} \right).
	\label{eq3:Tm}
\eq
In our case $\mu_m = u(m)$ since it gives the probability that $m\to m-1$ at $i$th node.
To find $\lambda_m$, let us consider a Master equation for the distribution of balls $\pi_i(m)$ at site $i$:
\bq
	\partial_t \pi_i(m) = \pi_i(m+1) \mu_{m+1} + \pi_i(m-1) \lambda_{m-1} + \pi_i(m) (1-\mu_m-\lambda_m).
\eq
In the stationary state the derivative vanishes and hence
\bq
	\pi_i(m+1)\mu_{m+1} - \pi_i(m)\lambda_m = \pi_i(m) \mu_m - \pi_i(m-1)\lambda_{m-1} = \mbox{const}. \label{eq3:pi-pi}
\eq
One sees that expressions on both sides of the last equation cannot depend on $m$. Inserting $m=1$ we see the constant is equal to zero. From Eq.~(\ref{eq3:pi-pi})
we obtain $\lambda_m$:
\bq
	\lambda_m = \mu_{m+1} \frac{\pi_i(m+1)}{\pi_i(m)}.
\eq
Inserting $\mu_m$ and $\lambda_m$ to Eq.~(\ref{eq3:Tm}), after some manipulations we obtain
\bq
	T_{mn} = \sum_{p=n+1}^{m} \frac{1}{u(p)\pi_i(p)} \sum_{l=p}^{M} \pi_i(l).
	\label{eq3:Tmngen}
\eq
We now apply the Eq.~(\ref{eq3:Tmngen}) to the case of a single-inhomogeneity graph assuming as before $u(p)=1$.
Let us start with the star graph as a special degenerate case and calculate the average transition times $T_{mn}$ for the central node on which the condensate spends almost all time. 
Using the formula (\ref{pi1inh}) with $\alpha=1/(N-1)$ for the star graph we have:
\bq
	T_{mn} = \sum_{p=n+1}^{m} \sum_{l=p}^{M} (N-1)^{l-p} \frac{(M+N-l-2)!(M-p)!}{(M+N-p-2)!(M-l)!}. \label{tmnstar}
\eq
The terms vanish for $p\to m$ if $m\gg 1$. The sum over $l$ decreases slowly with $p$ because it is a cumulative distribution for $\pi_1(l)$ (see Eq.~(\ref{eq3:Tmngen})). Thus for large $m$ the transition time is almost independent of $m$. This means that the condensate fluctuates very quickly around some value $1\ll\left<m_1\right><M$. We know from our previous considerations that $\left<m_1\right>\approx M$ and fluctuations are very small, so it is enough to concentrate on $T_{Mn}$.  Changing variables we get
\bq
	T_{Mn} = (N-2)! \sum_{r=0}^{M-n-1} \frac{r!}{(N-2+r)!} (N-1)^r \sum_{k=0}^r (N-1)^{-k} \frac{(N-2+k)!}{k!(N-2)!}.
\eq
In the last sum we can set the upper limit to infinity. We have:
\bq
	T_{Mn} \approx \left(\frac{N-1}{N-2}\right)^{N-1}(N-2)!  \sum_{r=0}^{M-n-1} \frac{r!}{(N-2+r)!} (N-1)^r.
\eq
The sum over $r$ can be done approximately by changing the variable $r\to M-n-1-r$: 
\bq
	T_{Mn} \approx \left(\frac{N-1}{N-2}\right)^{N}(N-2)! (N-1)^{M-n-1} \frac{M-n-1}{M-n-2} \frac{(M-n-1)!}{(M+N-n-3)!}.
	\label{TMnstar}
\eq
We see that, because of the factor $(N-1)^{-n}$, the time $T_{Mn}$ is very sensitive to $n$. In figure \ref{fig:star+inh-TMn} it is compared to computer simulations. This complicated formula has a simple behavior in the limit of large systems and $n=0$. For $M\to\infty$ and $N$ being fixed we get an exponential growth:
\bq
	T_{M0} \sim (N-1)^M,	\label{eq3:tm0ap1}
\eq
while for fixed density $\rho=M/N$ and $N\to\infty$ it increases faster than exponentially:
\bq
	T_{M0} \sim e^{\rho N \ln N}. \label{eq3:tm0ap2}
\eq
The approximated expressions (\ref{eq3:tm0ap1}) and (\ref{eq3:tm0ap2}) can be obtained very easily using a kind of Arrhenius law \cite{ref:god,ref:arh}, which tells that the average life-time is inversely proportional to the minimal value of the balls distribution:
\bq
	T_{M0} \sim 1/\pi_1^{\rm min},
\eq
if one thinks about the condensate's melting as of tunneling through a potential barrier. In our case the barrier $1/\pi_1(m)$ grows monotonically with $m\to 0$ so we observe that the condensate bounces from the wall at $m=0$ rather than tunnels through it. We have $\pi_1^{\rm min}\sim (N-1)^{-M}$ for fixed $N$ and large $M$ and thus we get Eq.~(\ref{eq3:tm0ap1}), while for fixed density $\rho$ the distribution $\pi_1^{\rm min}$ falls over-exponentially which results in Eq.~(\ref{eq3:tm0ap2}).

Before we comment on the exponential behavior of times $T_{mn}$, let us calculate analogous quantities for the general single-inhomogeneity graph. In the condensed state the occupation $m_1$ fluctuates quickly around the mean condensate size $\left<m_1\right>=\Delta$, with the variance estimated by Eq.~(\ref{eq3:pivar}) as $\sim N$. Although $\Delta$ is smaller than $M$ we can assume that $T_{mn}\approx T_{Mn}$ for $m>\Delta$.
This is so because the transition time between the states with high number of balls to the state with $m\approx \Delta$ must be small since the potential $1/\pi_1(m)$ decreases with $m\to \Delta$. Therefore, we can concentrate again on $T_{Mn}$ which is easier to compute. From Eqs.~(\ref{pi1inh}) and (\ref{eq3:Tmngen}) we have:
\bq
	T_{Mn} = \sum_{p=n+1}^M \sum_{l=p}^M \alpha^{p-l} \binom{M+N-l-2}{M-l} / \binom{M+N-p-2}{M-p}.
\eq
Changing variables we get:
\bq
  T_{Mn} = \sum_{p=n+1}^M \frac{(M-p)!}{(M+N-p-2)!}\sum_{q=0}^{M-p} \alpha^{-q} \frac{(M+N-p-q-2)!}{(M-p-q)!}.
\eq
The sum over $q$ can be approximated in the condensed state by an integral and evaluated by the saddle-point method.
The saddle point $q_0=\alpha (N-2)/(1-\alpha)$ is equal to $m_*$ from Eq.~(\ref{eq3:miu}) and therefore all calculations are almost identical. In this way we get:
\bq
	\sum_q \dots \approx \alpha^p \times \alpha^{\alpha\frac{N-2}{1-\alpha}-M} \frac{((N-2)/(1-\alpha))!}{(\alpha(N-2)/(1-\alpha))!}
	\sqrt{\frac{2\pi \alpha (N-2)}{(1-\alpha)^2}} .
\eq
The only dependence on $p$ is through the factor $\alpha^p$. To calculate $T_{Mn}$ it is therefore sufficient to evaluate the sum:
\bq
	\sum_{p=n+1}^M \alpha^p \frac{(M-p)!}{(M+N-p-2)!}.
\eq
Because every term is proportional to $1/\pi_1(p)$ from Eq.~(\ref{pi1inh}), in the condensed state the function under the sum has a minimum at the saddle point $p_0\approx m_*\in(1,M)$. This means that the effective contribution to the sum can be split into two terms: one for small $p$ and one for $p\approx M$. The ``small-$p$'' part can be evaluated like for static distributions in the previous section. To calculate the ``large-$p$'' part, it is sufficient to take the last two terms, namely for $p=M$ and $p=M-1$, since the large $p$ terms  decrease quickly with $p$. The complete formula for $T_{Mn}$ is given by:
\ba
	& & T_{Mn} \approx \alpha^{\alpha\frac{N-2}{1-\alpha}-M} \frac{((N-2)/(1-\alpha))!}{(\alpha(N-2)/(1-\alpha))!}
	\sqrt{\frac{2\pi \alpha (N-2)}{(1-\alpha)^2}} \nonumber \\ & & \times \left[ \frac{M!}{(M+N-2)!}  \left(\alpha\frac{M+N-2}{M}\right)^{n+1} \left( 1-\alpha\frac{M+N-2}{M}\right)^{-1} + \frac{\alpha^{M-1}(\alpha(N-1)+1)}{(N-1)!} \right]. \nonumber \\ \label{eq3:TMninh}
\ea
In Fig.~\ref{fig:star+inh-TMn} we compare the theoretical expression for $T_{M0}$ with Monte Carlo simulations. 
The agreement is the better, the larger $M$ is.
In the limit of large number of balls, $M\to\infty$, while keeping $N$ and $\alpha$ fixed, the life-time grows exponentially:
\bq
	T_{M0} \sim \left(\frac{1}{\alpha}\right)^M = \left(\frac{k_1}{k}\right)^M. \label{eq3:tm0inh}
\eq
In the limit of fixed density $\rho=M/N>\rho_c$ and for $M,N\to\infty$:
\bq
	T_{M0} \sim \exp \left[N\left( -\ln(1-\alpha)+\rho\ln(\rho/\alpha)-(1+\rho)\ln(1+\rho)\right)\right].
\eq
Let us now comment on the exponential times observed for inhomogeneous graphs. Unlike in homogeneous systems, where the life time grows like a power of $M$, in the presence of inhomogeneity it changes to the exponential behavior. This is typical for systems possessing a characteristic scale. Here it is given by the ratio $k_1/k$. This situation is to some extent similar to the relation between massless and massive interactions in particle physics. A two-point function for a massless field has a power-law decay and thus an infinite range, while for a massive field it falls off exponentially with the distance. 

$T_{M0}$ calculated above gives us some insight into the dynamics of the condensate.
Now we try to find a formula for $\bar{T}$ that is for the average time between consecutive jumps of the position $n_{\rm max}$ of the node with maximal number of balls. We do not expect that the behavior of $\bar{T}$ will be asymptotically different from $T_{mn}$ but we would like to check if we understand well the process of melting and rebuilding the condensate.

The main contribution to the average life-time of the condensate comes from situations when it occupies the node with the highest degree. The probability that at a particular time step the condensate has $m$ balls is given by $\pi_1(m)$. The condensate ends its life at a certain $m_1\equiv n$ number of balls which is no bigger than at the remaining  nodes, which means that there is at least one node $i\neq 1$ with the same or higher occupation: $m_i\geq m_1$. We call this an ``event A'' and denote the probability of its occurrence by $a(n)$. In order to calculate the average time $\bar{T}$ we have to sum over all possible $m,n$, weighted by appropriate probabilities:
\bq
	\bar{T} = \frac{1}{\sum_n a(n)} \sum_{m=0}^M \sum_{n=0}^M \pi_1(m) T_{mn} a(n), \label{avT}
\eq
where the first factor gives an appropriate normalization of $a(n)$.
We have already calculated $\pi_1(m)$ and $T_{mn}$ for the single-inhomogeneity graph. What remains is to calculate $a(n)$. We have to consider the subset of all configurations $\{m_1,\dots,m_N\}$ which favor the event A. 
One step before A happens, we have $n+1$ balls at the first node and no more than $n$ balls at other nodes. The number of balls is $M$, therefore at nodes $2,\dots,N$ there is $M-n-1$ balls in total. New configurations which lead to A are the following:
\begin{enumerate}
\item $n$ balls at the 1st node, $n+1$ balls at one node among $N-1$ remaining nodes, and $m_i\leq n$ balls on each of $N-2$ remaining nodes, that is $M-2n-1$ in total on $N-2$ nodes,
\item $n$ balls at the 1st node, $n$ balls at one node among $N-1$ nodes and $m_i\leq n$ at each of remaining nodes, with the total number of balls $M-2n$ on those nodes.
\end{enumerate}
The probability of the event A is proportional to the sum of all configurations described above:
\ba
	a(n) = \frac{1}{Z(N,M)} (N-1) \left[ \sum_{m_3=0}^n \cdots \sum_{m_N=0}^n \delta_{M,2n+1+m_3+\dots+m_N} \right. \nonumber \\
	\left. + \sum_{m_3=0}^n \cdots \sum_{m_N=0}^n \delta_{M,2n+m_3+\dots+m_N}
	\right] k_1^n k^{M-n}. \label{an}
\ea
First two factors give the normalization which corresponds to the sum over all configurations and over $N-1$ possibilities of choosing the node having exactly $n+1$ or $n$ balls. In the square bracket we have numbers of configurations with fixed amount of balls on nodes $i=3,\dots,N$. The last two terms arise from degrees of nodes: the weight $k_1^n$ for the first and $k^{M-n}$ for the rest of nodes.
Skipping the multiplicative factor and denoting the quantity in square brackets by $b(n)$ we have
\bq
	a(n) \propto b(n) \alpha^{-n}. \label{anprop}
\eq
The coefficient $b(n)$ can be expressed through the following sum:
\bq
	b(n) = \sum_{r=0}^{N-2} (-1)^r \binom{N-2}{r} \left[ \binom{N+M-4-2n-r(n+1)}{M-1-2n-r(n+1)}+ 
	\binom{N+M-3-2n-r(n+1)}{M-2n-r(n+1)} \right]. \label{bn}
\eq
This formula is obtained by using the integral representation of discrete deltas in Eq.~(\ref{an}) and by calculating each sum over $m_i$ separately.

\begin{figure}
\center
\psfrag{xx}{$M$}\psfrag{yy}{$\bar{T}$}
\includegraphics*[width=8cm]{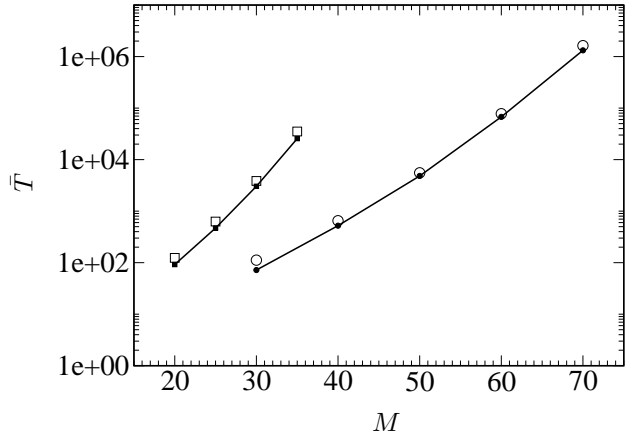}
\caption{\label{fig:T_kreg}Average life-time of the condensate on a small-inhomogeneity graph with $N=20$, $k=4$ and $k_1=8$ (circles), $k_1=12$ (squares). Empty symbols: numerical estimations, solid lines: analytical formula (\ref{avT}). Error bars are of symbols size.}
\end{figure}

We could now write in principle the formula for $\bar{T}$. It would be quite complicated so we decided not to present it here, but it can be evaluated numerically using the theoretical formulas given above. In figure \ref{fig:T_kreg} we present the comparison between average life-times computed from simulations and calculated analytically. One clearly sees that while $M$ increases, the points approach the theoretical curve but are systematically slightly above it. This means that $\bar{T}$ is a bit larger that predicted by Eq.~(\ref{avT}).

\subsection{Power-law distribution in the ZRP on inhomogeneous graphs} 
In previous sections we have noticed that there are many differences between
zero-range processes on homogeneous and inhomogeneous networks. For a homogeneous system, one can have a scale-free distribution of balls $\pi(m)$ if one tunes the hop rate $u(m)$ appropriately, while for an inhomogeneous one, distributions are in general exponential. 
In inhomogeneous systems the condensate, if exists, resides at the node with the largest degree for almost all time, and even if it
melts, it rebuilds very fast on this node, while in homogeneous systems
the condensate moves from node to node. For inhomogeneous systems 
the typical time-scale for melting the condensate grows exponentially 
with $N$ while in homogeneous ones only as a power of $N$. 

The power-law distributions at criticality are very interesting since they
are typical for systems without a characteristic scale. Inhomogeneous
systems have usually a typical scale introduced by the fact, that the flows
of balls are different between different nodes. We want to address the 
question whether it is possible to obtain a power-law distribution of balls
occupation numbers at the critical point also for inhomogeneous networks.
As we shall see, the answer to this question is in the affirmative, but it 
requires a fine-tuning of the node-degree distribution of the underlying network.
In this section we shall show how to do this. We shall also discuss some well-known 
examples of graphs including Erd\"os-R\'enyi graphs, for which the averaging over 
the ensemble of graphs leads to the particle distribution which resembles a power-law, although it is only a finite-size effect.

Unlike in previous sections, where we were interested in properties of
the ZRP on a given, fixed network, we consider now an ensemble of random graphs from Chapter 2.
The graphs are defined by 
specifying a desired degree distribution $\Pi(k)$ in the thermodynamical limit.
They can be generated by the Monte Carlo algorithm described in Sec. 2.1.5. In
this case, the probability $P(k_1,\dots,k_N)\equiv P(\vec{k})$ of having
a network with a sequence of degrees $k_1,\dots,k_N$ factorizes for $N\to\infty$: 
\bq 
	P(\vec{k}) = \Pi(k_1)\cdots \Pi(k_N).	\label{eq3:pifac} 
\eq 
This assumption means that we consider only
uncorrelated networks. It is approximately fulfilled for finite-size
graphs if $\Pi(k)$ falls quickly with $k$, as it results from 
the equivalence between canonical and grand-canonical partition
function for networks \cite{ref:snd-stat-mech} discussed previously,
since in the grand-canonical ensemble there is no constraint on the sum of
degrees. The factorization breaks down for scale-free networks.
Particularly strong deviations from the factorization are observed for
$\Pi(k) \sim k^{-\gamma}$ with $2<\gamma\le 3$, where finite-size
effects are especially strong \cite{ref:extr2}. Below we shall discuss
networks which are free of these problems and for which the
factorization works fine.

Let us recall the partition function for the ZRP on a given graph: 
\bq
	Z(N,M,\vec{k}) = \sum_{m_1,\dots,m_N=0}^M \delta_{\sum_i m_i,M} \prod_{i=1}^N f(m_i) k_i^{m_i}.	\label{eq3:part} 
\eq 
We are now interested in the behavior of the ZRP on a ``typical'' network,
taken from the ensemble of graphs with distribution of degrees given by Eq.~(\ref{eq3:pifac}). 
We want to average over all possible degree sequences
in the given ensemble. We thus define a canonical partition function:
\bq 
	Z(N,L,M) =\sum_{k_1\dots k_N} P(\vec{k}) Z(N,M,\vec{k}), \label{eq3:canon} 
\eq 
where $L$ is the total number of edges in the graph which, as in the canonical partition function for networks, is assumed to be fixed.
The dependence on $L$ we pull into the probability $P(\vec{k})$.

To simplify calculations, we set $u(m)=1$ as before, since we expect the effect of network inhomogeneity to be stronger than the effect
coming from the dependence of $u(m)$ on $m$. The canonical partition function (\ref{eq3:canon}) assumes now the form: 
\bq 
	Z(N,L,M) = \sum_{\vec{m}} \delta_{\sum_i m_i, M} \prod_{i=1}^N \mu(m_i), \label{eq3:zz}
\eq 
where $\mu(m)$ is the $m$th moment of the degree distribution $\Pi(k)$: 
\bq
	\mu(m) = \sum_{k=1}^\infty \Pi(k) k^m. \label{eq3:mom} 
\eq 
This partition function has exactly the form of the partition function 
(\ref{partsimple}) for homogeneous ZRPs, which we have discussed before. 
This shows that averaging over uncorrelated networks is equivalent
to considering another ZRP, for a homogeneous system. The averaging
smears the inhomogeneity and restores the symmetry with respect to the
permutation of occupation numbers. Instead of $\tilde{f}_i(m)$, distinct for different 
nodes, we have only one $f(m)\equiv \mu(m)$, identical for all nodes. Moreover, 
we will see that if all the moments $\mu(m)$ exist, the system
is self-averaging in the sense that for large $N$ a single network 
taken from the given ensemble reproduces the limiting distribution of balls.

\begin{figure} \center \psfrag{xx}{$k/\phi$}
\psfrag{yy}{$\Pi(k/\phi)/\phi$} 
\includegraphics*[width=8cm]{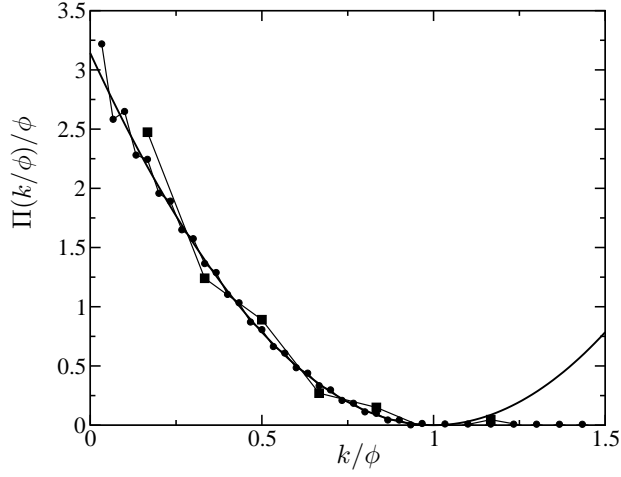}
\caption{\label{fig:fig1}$\Pi(k)$ calculated from the exact formula (\ref{eq3:piviaMz}) (points) and approximated one (\ref{eq3:piqgen})
(thick line), for $\mathcal{N}=2\pi$ and $b=3$. Squares: $\phi=6$, circles: $\phi=30$. The approximate solution diverges for $x>1$ and has
to be cut. For $0<x\leq 1$ the approximation is the better, the larger is $\phi$.} 
\end{figure}

\begin{figure} \center \psfrag{xx}{$m$} \psfrag{yy}{$\mu(m)$}
\includegraphics*[width=8cm]{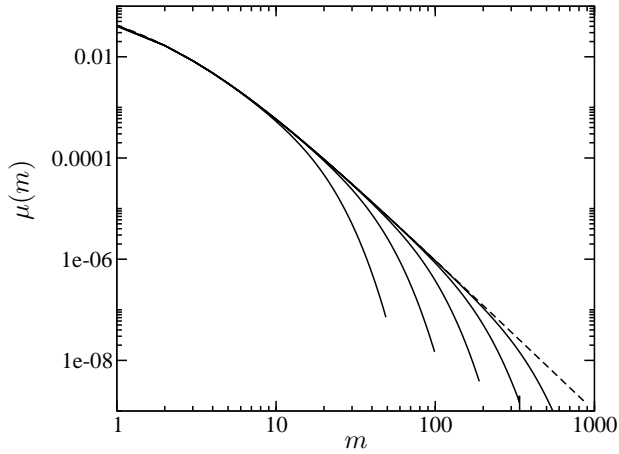} 
\caption{\label{fig:fig2}Desired (dotted line) versus real $\mu(m)$ for finite networks (solid lines)
obtained from Eqs.~(\ref{eq3:piqgen}) and (\ref{eq3:mom}), for $b=3$. Lines from left to right: $\phi=5,10,20,40,80$, which corresponds to
$\bar{k}=1.67, 2.89, 5.4, 10.4, 20.4$ from Eq.~(\ref{eq3:phifor3}). These plots approximate also $\pi(m)$ at the critical point, for infinite networks. The parameter $\phi$ grows almost linearly with $\bar{k}$.} 
\end{figure}

After these preliminaries we are now ready to attack the main question, namely
how to choose $\Pi(k)$ in order to obtain a scale-free distribution of balls occupation
numbers: $\pi(m)\sim m^{-b}$, at the critical point. From section 3.2.2 
and Eq.~(\ref{eq3:zz}) we see that $\mu(m)$ should behave as $m^{-b}$ for large $m$. Thus we are looking for 
the degree distribution $\Pi(k)$ which gives the following moments (\ref{eq3:mom}):
\bq
	\mu(m) = \frac{\Gamma(m+1)}{\Gamma(m+1+b)} \phi^m.	\label{eq3:mugamma} 
\eq 
This particular form of $\mu(m)$ is well suited for analytical calculations, but of course we expect a similar behavior for any other
$\mu(m)$ having the asymptotic behavior $\sim m^{-b}$. 
The exponential term $\phi^m$ does not change the $\pi(m)$ at the critical point $\rho_c$, since for conserved number of balls it
gives only an overall factor $\phi^M$ in $Z(N,L,M)$. We shall use the freedom of
choice of $\phi$ to adjust the mean value of the distribution $\Pi(k)$ to
the average degree $\bar{k}=2L/N$ which is fixed in the ensemble with given $N,L$.
Introducing a generating function
\bq 
	M(z) = \sum_{m=0}^\infty \mu(m) \frac{z^m}{m!} = \sum_{m=0}^\infty \frac{(z\phi)^m}{\Gamma(m+1+b)}, \label{eq3:mz} 
\eq 
we can recover $\Pi(k)$ for $k>0$ as a Fourier coefficient by means of the inverse transform: 
\bq 
	\Pi(k) = \mathcal{N} \frac{1}{2\pi} \int_{-\pi}^\pi \dd z\, e^{izk} M(-iz),	\label{eq3:piviaMz} 
\eq 
where $\mathcal{N}$ gives appropriate normalization, since $M(-iz)$ is by definition equal to $\sum_k \Pi(k)
\exp(-i z k)$. The integral in Eq.~(\ref{eq3:piviaMz}) is in general hard to calculate and express through special functions like sine and
cosine integral. However, the function $M(-iz)$ falls to zero with $z\to\pm\infty$ sufficiently fast and thus for $\phi\gg 1$ we can extend
the limits of integration to $\pm\infty$. Then equation (\ref{eq3:piviaMz}) becomes a Fourier Transform of the function
$M(-iz)$ from Eq.~(\ref{MZgen}), introduced in section 3.1.2. Hence we know that it can be written as an infinite series expansion (\ref{fftsimpl}) (see
also \cite{ref:bwis}). This complicated expression simplifies very much in the present case. Changing variables $k\to x\phi$ we have 
\bq 
	\Pi(x\phi) = \frac{\mathcal{N}}{2\pi\phi} \int_{-\infty}^\infty \dd z\, e^{izx} \sum_{m=0}^\infty \frac{(-iz)^m}{\Gamma(m+1+b)}. 
\eq 
and we can now apply the results of section 3.1.2. According to Eq.~(\ref{fftsimpl}), the last integral gives 
\bq 
	\frac{\mathcal{N}}{\phi} \sum_{k=0}^\infty \frac{(-x)^k}{k!\Gamma(b-k)}, 
\eq 
and hence 
\bq 
	\Pi(k)=  \frac{\mathcal{N}}{\phi} \sum_{q=0}^\infty \frac{(-k/\phi)^q}{q!\Gamma(b-q)} = (\phi-k)^{b-1} \frac{\mathcal{N}}{\Gamma(b)\phi^b}.   \label{eq3:piqgen} 
\eq 
In figure \ref{fig:fig1} we show $\Pi(k)$ calculated from the exact equation (\ref{eq3:piviaMz}) and from the approximated one (\ref{eq3:piqgen}). Because the probabilistic interpretation of $\Pi(k)$ requires that it must be non-negative, the solution (\ref{eq3:piqgen}) is physical only for $k\leq \phi$. So we have to set $\Pi(k)=0$ above $\phi$. Thus $\left\lfloor\phi\right\rfloor$ can be interpreted as the maximal degree existing in the network. One must be, however, aware that even the true $\Pi(k)$ calculated directly from Eq.~(\ref{eq3:piviaMz}) can be negative above $\phi$ and that cutting the negative part leads to some discrepancy between the desired $\mu(m)$ from Eq.~(\ref{eq3:mugamma}) and that obtained when (\ref{eq3:piqgen}) is inserted into Eq.~(\ref{eq3:mom}). Hence we must make sure if we really have $\pi(m)\sim m^{-b}$ at the critical point. The answer is that for any finite network we can always choose $\phi$ so that the discrepancy between the power-law behavior and the real $\mu(m)$ becomes negligible. In figure \ref{fig:fig2} we plot the desired function $\mu(m)$ and we compare it to that calculated  for various $\phi$ from Eqs.~(\ref{eq3:mom}) and (\ref{eq3:piqgen}). As $\phi$ increases, the plots tend to the power law. We see that in order to get the correct ball distribution in the thermodynamic limit we have to scale $\bar{k}$ to have $\lim_{N\to\infty} \bar{k} = \infty$. Such networks with $\Pi(k)$ given by Eq.~(\ref{eq3:piqgen}) are neither sparse nor very dense since $\bar{k}$ scales with a power of $N$ less than one.

The parameter $\phi$ is related to the average degree by the formula: $\bar{k}=\sum_{k=1}^\phi \Pi(k) k$. The normalization $\mathcal{N}$ must
be chosen so that $\sum_{k=1}^\phi \Pi(k)=1$. The sum goes from one because there can be no isolated nodes ($k=0$) on the graph.
When $b=2,3,4$, one is able to find closed formulas for the normalized degree distribution $\Pi(k)$. For instance, for $b=3$ we have 
\bq 
	\Pi(k) = \frac{(\phi-k)^2}{\phi(\phi-1)(2\phi-1)}, 
\eq 
for $0<k\leq\phi$ and zero for $k=0$ and $k>\phi$, with $\phi$ given by the following expression: 
\bq 
	\phi = \left(-1+4\bar{k}+\sqrt{1-16\bar{k}+16\bar{k}^2}\right)/2. \label{eq3:phifor3} 
\eq 
In general, for large $\phi$, the relation between $\phi$ and $\bar{k}$ is almost linear: 
\bq 
	\bar{k} = \frac{\sum_{k=1}^\phi (k-\phi)^{b-1} k}{\sum_{k=1}^\phi (k-\phi)^{b-1}}
	\approx \frac{\int_0^\phi (\phi-k) k^{b-1} \dd k}{\int_0^\phi k^{b-1} \dd k} = \frac{\phi}{b+1}. 
\eq 
Because $\phi$ grows with $\bar{k}$, one should take graphs large enough to minimize finite-size corrections. In other
words, the ratio $\phi/N$ should be much smaller that 1.

We performed Monte Carlo simulations of the ZRP on random networks with the degree distribution (\ref{eq3:piqgen}) to check whether one indeed obtains a power law in the distribution of balls $\pi(m)$. The simulations were made as follows. First we generated a connected graph from the ensemble of random graphs with the degree distribution from Eq. (\ref{eq3:piqgen}), using the Monte Carlo algorithm described in previous chapter. The graph had no degree-degree correlations, except of those introduced by finite-size effects. On that graph we simulated the zero-range process starting from a uniform distribution of balls. We collected a histogram of $\pi(m)$ and repeated the simulation for other networks from the ensemble. In total, we generated over 50 networks for the given set of parameters $N,M,L$.

The crucial point is to ensure that those graphs are connected. The Monte Carlo algorithm presented before generates in principle
graphs which may have disconnected parts. But we know (see e.g. \cite{ref:barab}) that for random graphs there exists a critical average
degree $\bar{k}_c$ (a percolation threshold), above which a single component is always formed in the limit of $N\to\infty$. In our
simulations we always checked that we were above $\bar{k}_c$ and that the graph we used was connected. 
We also simulated tree graphs which are by definition connected. For trees, however, finite-size effects are stronger than for graphs.

\begin{figure} 
\center 
\psfrag{xx}{$m$} \psfrag{yy}{$\pi(m)$}
\includegraphics*[width=8cm,bb=1 46 702 584]{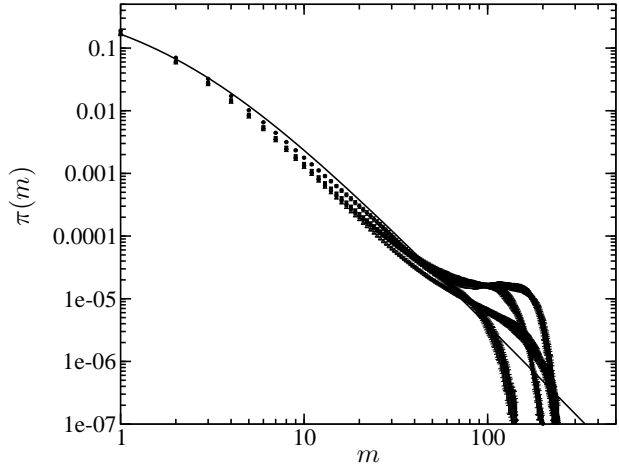} 
\caption{\label{fig:fig3}The distribution of balls in the ensemble of graphs with degree distribution $\Pi(k)\sim(\phi-k)^2$. Solid line: theoretical $\pi(m)\sim m^{-3}$ at the critical point $\rho_c=1$ and for infinite system. Circles: for trees with $N=M=400$, averaged over 50 graphs, $\phi\approx 6$. Squares: for simple graphs with $N=M=400$, $\bar{k}=8$, $\phi\approx 30$, diamonds: as before but $M=300$, triangles: $N=800,M=600$.} 
\end{figure}

In figure \ref{fig:fig3} we compare a theoretical distribution at the critical point $\rho_c=1$ for $b=3$, with experimental ones obtained
by numerical simulations. The agreement is not perfect. Finite-size effects are strong. But we see an apparent power law in the  
distribution of balls. The curves shown in figure \ref{fig:fig2} would suggest that for $\bar{k}=8$, the power law should continue up to $m$ of order $100$. In figure \ref{fig:fig3} we see that it deviates already before, probably
because it is not exactly at the critical point. Indeed, we observe a reminiscence of the condensation suggesting that the system is already off the transition point. These deviations are induced by the fact that for any finite graph there is a condensation on the most inhomogeneous node \cite{ref:kulki}. This gives a peak in $\pi(m)$. Because the condensate takes an extensive number of balls, it effectively increases the critical density for the rest of the system, so we are slightly below $\rho_c$.

\begin{figure} 
\center \psfrag{xx}{$m$} \psfrag{yy}{$\pi(m)$}
\includegraphics*[width=8cm]{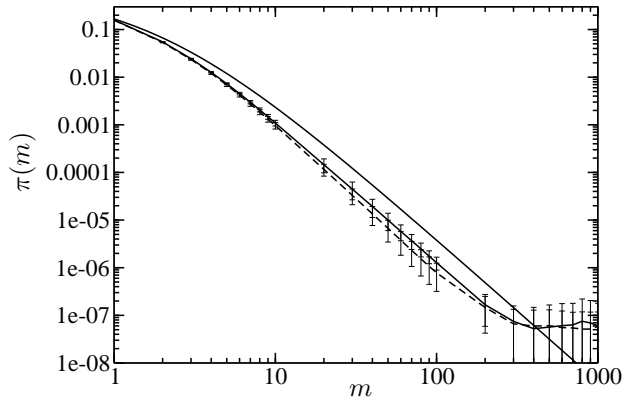}
\caption{\label{fig:selfav}Demonstration of self-averaging: $\pi(m)$ for a single network with degree distribution $\Pi(k)\sim(\phi-k)^2$, for
two different sizes $N=M=5000$ (thin solid line) and $10000$ (dashed line) and $\bar{k}=16$. For each case four networks were generated to
estimate error bars. Only a few experimental points are shown for brevity. The thick solid line gives the asymptotic distribution.}
\end{figure}

In figure \ref{fig:selfav} we present results for large networks, but without averaging over the ensemble. We again get a power law which
indicates that a self-averaging takes place. As before, the experimental line does not agree ideally with the theoretical one, $\pi(m)$, but the power-law behavior is clear.

\begin{figure} \center \psfrag{xx}{$m$} \psfrag{yy}{$\pi(m)$}
\includegraphics*[width=8cm,bb=1 46 702 584]{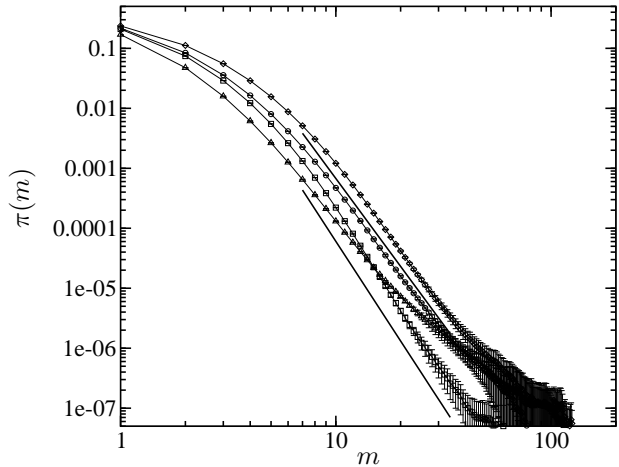}
\caption{\label{fig:fig4}The experimental distribution of balls averaged over 200 graphs taken from the ensemble of random trees and ER graphs. Circles: ER graphs $N=400,M=300,\bar{k}=8$, squares: as before but $N=800,M=450$, diamonds: ER with $N=M=400, \bar{k}=16$, triangles: trees with $N=800,M=300$. Thick lines show power laws $\pi(m)\sim m^{-b}$ with $b=4.87$ (upper line) and $5.52$ (lower line).} 
\end{figure}

The argumentation presented above suggests that one has to fine-tune the degree distribution in order to obtain the scale-free distribution of the number of balls. What happens if one takes different distributions? We have performed the ZRP also on some other networks and surprisingly found that $\pi(m)$ seems to be also heavy-tailed. In figure \ref{fig:fig4} we show results of numerical experiments for random trees \cite{ref:bbjk} and ER random graphs. Random trees are equilibrated trees with weights $p(k)=1$ and have been already mentioned in Sec. 2.1.4. They can be generated using the Monte Carlo algorithm given in Sec. 2.1.5. The degree distribution for random trees reads
\bq
	\Pi(k) = \frac{e^{-1}}{(k-1)!},	\label{eq3:piRT} 
\eq 
for $k>0$ and $\Pi(0)=0$. For ER graphs, $\Pi(k)$ is approximately Poissonian  as we know from Eq.~(\ref{eq2:poiss}). In figure \ref{fig:fig4} we see the results of measuring $\pi(m)$ on networks of size of order few hundreds.  The distribution $\pi(m)$ seems to follow a power law, in a certain range. In order to check whether this range increases in the large $N$
limit, one would have to perform a systematic analysis for networks of growing sizes.

However, here we prefer to present some theoretical discussion of whether it can be a power-law or rather some other distribution. Let us
calculate the theoretical distribution $\pi(m)=\mu(m)$ at the critical point, in the thermodynamical limit. For random trees, the generating
function $M(z)$ defined in Eq.~(\ref{eq3:mz}) has a closed form: $M(z)=\exp(z+e^z)$, as follows from Eq.~(\ref{eq3:piRT}). The function
$\mu(m)$ is given by the inverse Laplace transform: 
\bq 
	\mu(m) = \frac{m!}{2\pi i} \oint M(z) z^{-m-1} \dd z, 
\eq 
which can be evaluated by the saddle-point integration around $z_0\approx \ln(m/\ln m)$:
\bq 
	\ln \mu(m) = m(\ln m - \ln\ln m) + O(m), 
\eq 
and hence $\mu(m)$ grows over-exponentially for large $m$. The hop rate $u(m)$ for an equivalent homogeneous ZRP decays fast with $m$. This
means that the condensation happens for any density of balls, and the bulk distribution falls faster than any power-law. Similarly, one can
estimate that for random ER graphs $M(z) \propto \exp(\bar{k} e^z)-1$ and hence the leading term in $\ln \mu(m)$ is also $m\ln m$, so 
we again have the condensation.

\begin{figure} \center \psfrag{xx}{$m$} \psfrag{yy}{$\pi(m)$}
\includegraphics*[width=8cm]{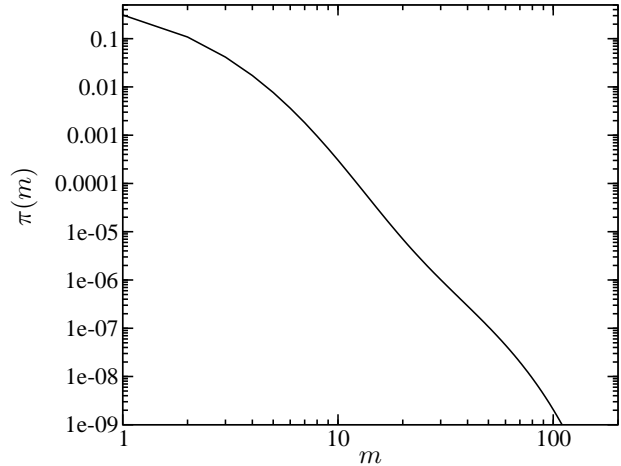}
\caption{\label{fig:fig5}The distribution $\pi(m)$ obtained from Eq.~(\ref{eq3:mom}) with cut Poissonian degree distribution. The almost straight line on the log-log plot explains partially the quasi-power-law observed in numerical experiments in the figure \ref{fig:fig4}.} 
\end{figure}

It is clear from the above arguments that in the limit $M\to\infty$ one cannot obtain power-law distribution of balls for maximally random
graphs like ER graphs or random trees. If it is not a power law, how the behavior in Fig.~\ref{fig:fig4} can be explained? For finite systems, we observe that the power law goes only over one-two decades, so it can be easily confused with another function. Such a quasi-power-law behavior is presented in Fig.~\ref{fig:fig5}, where we have calculated $\pi(m)$ for finite-size ER graph by means of Eq.~(\ref{eq3:mom}) multiplied by a factor $\exp(-m \beta)$. In order to mimic such finite-size effects we have assumed that the degree distribution $\Pi(k)$ had a cutoff at some $k_{\rm max}$ calculated from the condition that in $n=200$ samples of graphs of size $N=400$ it
should be around one node with degree $k_{\rm max}$: $nN \Pi(k_{\rm max}) \approx 1$, and hence $k_{\rm max}\approx 22$. The factor $\beta$ has been chosen
to get an almost straight line on the plot. Normally this is done by the factor $Z(N-1,M-m)/Z(N,M)$ in the formula for distribution of balls, and in real simulations by taking an appropriate value of $M$, which brings the system to the critical density $\rho_c$.

To summarize the discussion of this subsection, we have shown that tuning the node-degree distribution $\Pi(k)$ on a random network, on which the zero-range process is defined, one can obtain the power law in the balls occupation distribution $\pi(m)$. This makes the system scale-free at the critical point, in contrast to the previously discussed single-inhomogeneity graphs. 
The key point is that although degrees of nodes differ, their distribution is so tuned that averaging over distributions of balls for nodes with different degrees gives exactly the power law. On every single node $i$, however, the distribution of balls is not a power law, but it falls exponentially as $(k_i/k_{\rm max})^m$. The only exception is the node with maximal degree $k_{\rm max}$, where the condensation takes place.

This is not the case for maximal random graphs like ER graphs, where the degree distribution is concentrated around $\bar{k}$ and the ZRP behaves like for a homogeneous system with constant hop rates, having an over-exponential decay in $\pi(m)$ for small $m$ and a condensation peak at large $m$. The situation is similar to that for S-F networks, where the node-degree fluctuations are strong enough to produce a node of degree much larger than other degrees. This node attracts the condensate \cite{ref:jdn}.

\chapter{Conclusions and outlook} 
Complex networks have been widely
studied in recent years. Being a discipline on the interface of
physics, chemistry, biology, social and computer sciences, and others,
it applies a variety of methods. Most people try to understand
observed properties of networks by introducing simplified models and
then by making computer simulations in order to compare results to
real-world data. Some of them use a multitude of so-called mean field
approaches, when the quantity of interest is assumed to evolve in an
averaged field of all interactions. This, however, can rarely allow one to
examine such effects like phase transitions or condensation, and the
results can be only qualitative. Moreover, some problems may be
ill-posed when one does not specify what the word ``averaged'' means. In
this thesis we have tried to present a consistent theory of statistical
mechanics of complex networks, where all problems can be formulated in
terms of some averaged quantities over a well-defined statistical
ensemble. The starting point of the formulation is the ensemble of
Erd\"{o}s-R\'{e}nyi graphs, where all graphs have the same statistical
weight. But we have seen that ER graphs weakly reproduce features
observed in real-world networks. Therefore, we have assigned different
statistical weights to graphs from the same set, enhancing the probability
of occurrence of graphs of a certain type. For instance, by assigning to
each node a functional weight $p(k)$ depending on its degree $k$, we
have been able to obtain any desired degree distribution, either for simple
or degenerated graphs, for trees or graphs with cycles, and for causal
as well as for equilibrated networks. In particular, we can reproduce for
equilibrated networks the scale-free degree distribution, one of the most
important properties of real networks. We have shown also how the approach
via statistical ensemble can be used to calculate degree-degree
correlations or the assortativity coefficient. We have pointed out that the same 
method can be used for growing networks. We have discussed how to reformulate models of
preferential attachment in the language of network's ensembles and how to relate 
them to the rate-equation approach, which is a very powerful analytical method.

At the end of discussion devoted to statistical ensembles of graphs we have presented
a comparison between growing and equilibrated networks. We have shown how to choose 
functional weights in both ensembles, in order to obtain the same power-law degree
distributions and not to introduce node-node correlations. 
Then we have focused on some global properties like the
assortativity or the diameter. We have found that both types of
networks are disassortative but that the degree-degree correlation function $\eps(k,q)$ exhibits
different behavior for these graphs. We have observed a similar difference for the
diameter, which scales like $\ln N$ for growing unweighted networks, also for
growing trees, thus indicating the small-world behavior, while for equilibrated unweighted trees it grows like
$\sim N^{1/2}$. In other words, we have explicitly shown that graphs in the
two ensembles, despite having identical degree distributions, may have completely different geometrical
properties. In this particular case, the origin of the
differences, shortly speaking, comes from the fact that the set of causal graphs forms
only a small subset of all possible graphs in the statistical ensemble, and the
properties of that subset are quite different to those observed as
``typical'' for the whole set.

Further differences between various graphs with the same degree
distribution have been discussed in Chapter 3. Using analytical and numerical methods 
we have tested theoretical predictions for the position of the cutoff known from the literature
on the subject. We have encountered an unexpected
difference between the values of the exponent $\alpha$, describing the scaling of the cutoff $k_{max}$ with the
network size, for causal and equilibrated trees. We have pointed out that the two estimates of $k_{max}$
for simple equilibrated graphs found in the literature seem to be
inconsistent in light of results presented in this thesis. As a by-product of
this analysis we have developed a method of calculating the cutoff function $w(x)$, 
which allows one to treat many models of growing networks in a unified, standard fashion.

In the second part of Chapter 3 we have discussed dynamics on networks. We have
studied the zero-range process, being just the balls-in-boxes model with
a certain type of local dynamics. We have shown that for
inhomogeneous systems, that is when node degrees differ much from each other,
static and dynamical properties of the system are different than those for
homogeneous systems studied in the past. For instance, when the
inhomogeneity is strong enough, it triggers 
the condensation on the most inhomogeneous node. The critical density of balls,
above which the condensation takes place, depends on $k_1/k$, where $k_1$ is the
largest degree and $k$ is the typical degree in the network.
In particular, on S-F networks the ZRP always develops the condensation.
Another interesting effect of inhomogeneity is a qualitative change in the behavior
of a typical life-time of condensate, which grows exponentially or faster, in contrast
to homogeneous systems where it grows only as a power of the system size.
We have seen also that the effect of inhomogeneity can be weakened for some node-degree distributions.
In particular, we have found a special form of the distribution $\Pi(k)\sim
(\phi-k)^{b-1}$, for which the system of balls behaves very much like on a homogeneous
networks at the critical point, where the distribution of balls occupation number is scale-free: $\pi(m)\sim m^{-b}$.

At the end, let us say a few words about possible directions of further studies, and concepts
which can be an interesting continuation of ideas discussed in this thesis.
Among many interesting ideas, it is of the primary interest of the
author to study dynamical processes taking place on dynamically
rearranged networks. Suppose that the ZRP can interact with the topology
of the underlying network and change it while the occupation of nodes is
changing. A slow change in the network's structure should be
well approximated by averaging of the ZRP over an ensemble of static
networks as it was done in Sec. 3.2.4. If the network evolves quickly in
comparison to the characteristic time of the ZRP, its evolution can be viewed as
a sequence of rewirings as those described in section devoted to Monte Carlo simulations, but with
additional weights for nodes, depending on the state of the ZRP. It is very
interesting to study what happens in between, that is when the two 
characteristic time scales are comparable.

Another class of problems where this kind of the two-fold evolution becomes important
is related to neural networks.
If one couples the evolution of neuron's states to the evolution of
connections between them, one observes a self-organized criticality that
produces a S-F network and a small-world \cite{ref:rew3}. The question is whether
one can mimic this behavior using a simpler model, or to predict it analytically in
the framework described here.

There are also many other questions, for instance if one can use the
approach via moments of the distribution $\Pi(k)$ to estimate
the cutoff in some other models of growing networks, especially with
degree-degree correlations, or how the properties of causal networks
change when, after a certain time, we allow for some rewirings that
homogenize the network. As the example of Watts and Strogatz's
small-world model shows, an interesting behavior is possible. We hope to
address these and other problems in the future investigations.

\end{document}